\scrollmode
\documentclass[12pt,twoside]{book}
\usepackage{a4wide,amsmath,amssymb,array,cite,eepic,epsfig,  
  fancyheadings,multirow,verbatim,axodraw,graphicx}   
\advance \headheight by 3.truept

\newcommand{\postscript}[2]{\setlength{\epsfxsize}{#2\hsize}
   \centerline{\epsfbox{#1}}}

\def\stone{$\tilde{t}_1$}
\def\sttwo{$\tilde{t}_2$}

\def\stauone{$\tilde{\tau}_1$}
\def\stautwo{$\tilde{\tau}_2$}

\def\snone{$\tilde{\chi}_1^0$}
\def\sntwo{$\tilde{\chi}_2^0$}
\def\snthree{$\tilde{\chi}_3^0$}
\def\snfour{$\tilde{\chi}_4^0$}

\def\scone{$\tilde{\chi}_1^{\pm}$}
\def\sctwo{$\tilde{\chi}_2^{\pm}$}

\def\beq{\begin{equation}}
\def\eeq{\end{equation}}
\def\beqa{\begin{eqnarray}}
\def\eeqa{\end{eqnarray}}

\begin{document}
\pagenumbering{arabic}


\pagestyle{empty}

\begin{center}
{\LARGE Technische Universit\"at M\"unchen\\[2mm]
und\\[2mm]
Universit\'e de Montpellier II\\[2mm] 
(Frankreich)}

\bigskip
\bigskip
\bigskip
\bigskip
\bigskip

{\Large\bf Super-heavy $X$ particle decay\\[3mm]
 and\\[5mm]
Ultra-high Energy Cosmic Rays} \\[8mm] 
\bigskip
{\Large Cyrille Barbot} \\[4mm]
\end{center}

\bigskip
\bigskip
\bigskip
\bigskip

{\large Vollst\"andiger Abdruck der von der Fakult\"at f\"ur Physik der
Technischen Universit\"at M\"unchen zur Erlangung des akademischen
Grades eines Doktors der Naturwissenschaften genehmigten Dissertation.}

\bigskip
\bigskip

\bigskip
\bigskip

{\large Vorsitzender: ~~~~~~~~~~~~~~Univ.-Prof. Dr. Lothar Oberauer}\\

\bigskip

{\large Pr\"ufer der Dissertation:}

{
\setlength{\parindent}{6cm}
\indent 
{\large 1. Univ.-Prof. Dr. Manuel Drees}\\
\indent 
{\large 2. Univ.-Prof. Dr. Manfred Lindner}\\
\indent
{\large 3. Prof. Dr. Abdelhak Djouadi,~-~Frankreich}\\
\indent
{\large 4. Prof. Dr. Jean Orloff,~-~Frankreich}\\
}

\bigskip
\bigskip
\bigskip
\bigskip

{\large Die Dissertation wurde am 12/06/2003 bei der Technische
  Universit\"at M\"unchen eingereicht und durch die Fakult\"at f\"ur
  Physik am 18/07/2003 angenommen.}


\clearpage \pagestyle{empty}~
\clearpage

\addcontentsline{toc}{chapter}{Acknowledgements}
\pagestyle{empty}
\begin{flushright}

``Chercheur, trouveras-tu ce qu'ils n'ont pas trouv\'e ?\\
Songeur, r\^everas-tu plus loin qu'ils n'ont r\^ev\'e ?''\\

\bigskip
\bigskip

``Esprit, fais ton sillon, homme, fais ta besogne.\\
Ne va pas au-del\`a. Cherche Dieu. Mais tiens toi,\\
Pour le voir, dans l'amour, et non pas dans l'effroi.''\\

\bigskip
\bigskip

\end{flushright}

\setlength{\parindent}{5.5cm}
\indent ``\^Ame ! \^etre, c'est aimer.\newline
\setlength{\parindent}{10.5cm}
\indent Il est.\newline
\setlength{\parindent}{12cm}
\indent C'est l'\^etre extr\^eme.\newline
\setlength{\parindent}{5.5cm}
\indent Dieu, c'est le jour sans borne et sans fin qui dit : J'aime.''\\
\setlength{\parindent}{0.5cm}
\begin{flushright}

Victor HUGO

\bigskip
\bigskip
\bigskip
\bigskip
\bigskip
\bigskip

I dedicate this thesis to all curious people,\\
who would like to understand better all the\\
incredible features of Nature, especially\\
the craziest ones...\\

\bigskip

And to my parents.

\end{flushright}

\clearpage

\begin{center}
\section*{Acknowledgments}
\end{center}

\bigskip

First I would like to thank the Technische Universit\"at of M\"unchen
(TUM) and its personnel for the very pleasant environment they offered
me during these three years and even more for the possibility and the
funds they have given me to pursue my studies and research in
theoretical particle physics. A very special thank to Karin Ramm, the
secretary of the t30 institute, whose help was very precious to me
from the beginning to the end of this thesis.

Many thanks to all my colleagues for their presence, friendship,
encouragements and their daily support. A special thank to Benedikt
Gassmaier who had to bear me three years long in the same office, and to
Prof. A. Buras for the excellent idea he had once to buy a ping-pong
table for the institute!

Many thanks also to my housemates Bernd Stegmann, Bernhard Pedrotti,
Matthias Wilke, Peter Behl, and Roger Abou-Jaoud\'e, and to all my
friends, who strongly contributed in making my permanency in Germany a
very pleasant and joyful experience. I'll never thank enough my
parents and my family; without their support and encouragements, I
certainly would not have been able to defend a PhD thesis today.

Thanks to the university of Montpellier II and its personnel for their
strong support in the difficult way to go through when one wants to
get a thesis in co-tutella. A special thank to Prof. Abdelhak Djouadi,
who accepted to be my French advisor, and to Josette Cellier and
Florence Picone, for their precious help and their incredible patience
with me...

I thank the French doctoral school of ``Mati\`ere condens\'ee'' and
its director Prof. Francis Larch\'e, as well as the German SFB
program, which partly funded this work.

I thank all referees and members of the jury who accepted to read this
thesis and to attend to its defense, especially Profs. Manfred Lindner
and Jean Orloff, who both helped me a lot when accepting this task at
the last moment.

Finally, a VERY special thank to Prof. Manuel Drees, who offered me a
second chance in physics, and was a very competent and patient
advisor, from whom I learned a lot, not only in physics.




\pagestyle{empty}
\addcontentsline{toc}{chapter}{Abstract}
\begin{center}
{\Large \bf Abstract}
\end{center} 
\bigskip

In this thesis, I describe in great detail the physics of the decay of
any Super-Heavy $X$ particle (with masses up to the grand unification
scale $\sim 10^{16}$ GeV and possibly beyond), and the computer code I
developed to model this process - which currently is the most
complete available one. The general framework for this work is the
Minimal Supersymmetric Standard Model (MSSM). The results are
presented in the form of fragmentation functions of any (s)particle of
the MSSM into any final stable particle (proton, photon, electron,
three types of neutrino, lightest superparticle LSP) at a virtuality
$Q = M_X$, over a scaled energy range $x \equiv 2 E / M_X \in
[10^{-13}, 1]$. At very low $x$ values, color coherence effects have
been taken into account through the Modified Leading Log Approximation
(MLLA). The whole process is explicitely shown to conserve energy with
a numerical accuracy up to a few part per mille, which allows to make
quantitative predictions for any $N$-body decay mode of any $X$
particle. I then apply the results to the old - and yet unsolved -
problem of Ultra High Energy Cosmic Rays (UHECRs). In particular, I
provide quantitative predictions of generic ``top-down'' models for
the neutrino and neutralino fluxes which could be observed in the next
generation of detectors.


\pagestyle{empty}
\addcontentsline{toc}{chapter}{Zusammenfassung}
\begin{center}
{\Large \bf Zusammenfassung}
\end{center} 
\bigskip

In dieser Doktorarbeit betrachte ich in Detail die Physik des Zerfalls
beliebiger, superschwerer Teilchen $X$ (mit einer Masse bis zur Skala
der grossen Vereinheitlichung $\sim 10^{16}$ GeV und m\"oglicherweise
jenseits). Weiterhin wird das von mir entwickelte Programm - momentan
das kompletteste verf\"ugbar - zur numerischen Simulation dieses
Prozess vorgestellt. Der allgemeine Rahmen f\"ur diese Arbeit ist das
Minimale Supersymmetrischen Standard Modell (MSSM). Die Ergebnisse
werden als Fragmentierungsfunktionen von beliebigen MSSM
(Super)teilchen in verschiedene (stabile) Endzust\"ande wie Protonen,
Photonen, Elektronen, die drei Typen von Neutrinos, und das leichteste
Superteilchen LSP) mit Virtualit\"at $Q = M_X$ repr\"asentiert, \"uber
eine Energieabstand $x \equiv 2 E / M_X \in [10^{-13}, 1]$. F\"ur sehr
kleine $x$ Werte wurden QCD Farbeneffekte durch ``Modified Leading Log
Approximation'' (MLLA) betrachtet.  W\"ahrend der kompletten
numerischen Simulation dieser Multi Teilchen Kaskaden konnte zum
ersten Mal Energieerhaltung mit einer numerischen Genauigkeit auf dem
permille Niveau erzielt werden. Mit dieser Pr\"azision werden zu
beliebigen $X$ Zerfallsmode gute quantitative Voraussagen erm\"oglicht.
In einem zweiten Teil dieser Arbeit habe ich diese Ergebnisse in
Zusammenhang mit dem so genannten ``Ultrahoch Energetische Kosmische
Strahlungen'' (UHECRs) Problem angewandt.


\addcontentsline{toc}{chapter}{R\'esum\'e}
\begin{center}
{\Large \bf R\'esum\'e}
\end{center} 
\bigskip

Dans cette th\`ese, je d\'ecris en d\'etail la d\'esint\'egration
d'une particule supermassive - not\'ee $X$ -, dont la masse est de
l'ordre de l'\'echelle de grande unification $\sim 10^{16}$ GeV ou
au-del\`a, ind\'ependamment de tout mod\`ele particulier d\'ecrivant
cette particule ; je d\'ecris \'egalement le programme que j'ai
d\'evelopp\'e - actuellement le plus complet dans le domaine - pour
mod\'eliser ce processus. J'ai trait\'e l'ensemble du probl\`eme dans
le cadre du Mod\`ele Standard Supersym\'etrique Minimal (MSSM). Les
r\'esultats sont pr\'esent\'es sous la forme de fonctions de
fragmentation d'une quelconque particule du MSSM vers les particules
stables finales (proton, photon, \'electron, un des trois types de
neutrinos, ou enfin la particule supersym\'etrique la plus l\'eg\`ere,
appel\'ee LSP), \`a une virtualit\'e $Q = M_X$, sur un intervalle
d'\'energie d\'efini par $x \equiv 2 E / M_X \in [10^{-13}, 1]$. Dans
le domaine des faibles valeurs de $x$, j'ai pris en compte les effets
de coh\'erence de couleur en incluant une correction \`a l'ordre
dominant appel\'ee MLLA (``Modified Leading Log Approximation'').
L'ensemble du programme conserve explicitement l'\'energie avec une
pr\'ecision num\'erique de l'ordre de quelques pour mille ; cela
permet d'utiliser ces r\'esultats pour faire des pr\'edictions
quantitatives sur le spectre final d'une quelconque d\'esint\'egration
\`a $N$ corps, quel que soit le type de particule $X$ consid\'er\'e.
J'ai ensuite appliqu\'e ces r\'esultats au probl\`eme - encore non
r\'esolu - des rayons cosmiques \`a ultra-haute \'energie (UHECRs).


\setcounter{tocdepth}{3}
\tableofcontents


\pagestyle{fancyplain}
\renewcommand{\chaptermark}[1]{\markboth{\thechapter.  #1}{}}
\renewcommand{\sectionmark}[1]{\markright{\thesection\  #1}}
\rhead[\leftmark]{\fancyplain{} {\bf \thepage}}
\lhead[\bf \thepage]{\rightmark}
\cfoot{}

\chapter{Introduction}
\setcounter{footnote}{1}

Although they obviously have never been observed, many different types
of super-heavy (SH) particles (with masses up to the grand unification
scale, at $10^{16}$ GeV and even beyond) are predicted to exist in a
number of theoretical models, e.g. grand unified \cite{gut} and string
models. But even without calling upon these particular theories, the
existence of such SH particles is quite natural; indeed, it is known
that the Standard Model of particle physics (SM) cannot be the
fundamental theory, but only an effective theory at low energy (say,
up to the TeV region); thus one should find one (or more) fundamental
energy scale(s) at higher energies, and there are reasons to believe
that some (super-heavy) particle(s) would be associated to this new
scale. A general overview on the weaknesses of the SM can be found for
example in \cite{reviewMartin}, and a list of different SH candidates
appears in \cite{reviewSigl,Cronin}.

If $X$ particles exist, they should have been produced in large
quantities during the first phases of the universe, especially during
or immediately after inflation \cite{creat}. Their decay could have
had a strong influence on the particle production in the early
universe; this is certainly true for the decay of the inflatons
themselves. Moreover, the decay of such particles has been proposed as
a (``top-down'') alternative solution for the ultra-high energy cosmic
ray (UHECR) problem. Indeed, if $X$ particles have survived until our
epoch\footnote{At first sight, this assumption seems to be rather
  extreme, but many propositions have been made in the literature for
  explaining such a long lifetime; for example, the $X$ particles
  could be protected from decay by some unknown symmetry, which would
  only be broken by non-renormalizable operators of high orders
  occurring in the Lagrangian; or they could be ``trapped" into very
  stable objects called topological defects (TDs), and released when
  the TDs happen to radiate (For a review, see\cite{reviewSigl}).},
their decay could explain the existence of particles carrying energy
up to $10^{20}$ eV, which have been observed in different cosmic ray
experiments over the past 30 years
\cite{volcano,Suga:jh,yakutsk,haverah,agasa:1,hires,largest,newhires}
and still remain one of the greatest mysteries in astrophysics.

Because of the energy scales considered in these models, it is clear
that we will need theories going beyond the SM. Up to now, one of the
most promising class of models able to cure the most dangerous
aspects of the SM are the so-called {\it supersymmetric} (often
surnamed ``SUSY'') theories. Without going into any detail, we
just note here that the so-called Minimal Supersymmetric Standard
Model (MSSM) offers two beautiful and very useful features:
\begin{itemize}
\item[1)] a solution at all orders to the so-called ``hierarchy
  problem'' occurring in the SM. It allows us to consider safely energy
  scales larger than the TeV and thus the very existence of SH
  particles.
\item[2)] the impressive unification of all gauge couplings of the SM
  at a ``grand unification'' (GUT) scale of order $10^{16}$ GeV. This
  offers us a natural scale for the mass of our $X$ particles.
\end{itemize}

That is the reason why I will work in this whole thesis within the
framework of the MSSM. For an excellent review on this subject, see
\cite{reviewMartin}. For a more theoretical introduction to
Supersymmetry, see e.g. \cite{Wess}.

In order to protect the unification of couplings mentioned above -
which occurs {\it naturally} in the MSSM -, we are driven to formulate
the so-called ``desert hypothesis'', which consists in assuming that
there is no ``new physics'', thus no new energy scale, between the TeV
region and the GUT scale. Within this assumption, the only available
particle content is the one of the MSSM, and it becomes reasonable to
assume that $X$ will decay only into some ``light'' particles of the
MSSM, independently of the particular model that one considers for
$X$. The primary decay products will initiate parton cascades, the
development of which can be predicted from the ``known'' interactions
contained in the MSSM. For studying in detail the predictions of
these models, a new code taking into account the full complexity of
the decay cascade of SH particles was still needed.

The program SHdecay\footnote{SHdecay is a public code and can be downloaded
 from:\\ http://www1.physik.tu-muenchen.de/$\sim$barbot/} has been designed
for this purpose. It allows to compute the spectra of the final stable
decay products of any $N$-body decay of $X$, independently of the
model describing the nature of $X$; the only fundamental assumption
behind this work is the one stated above: whatever the actual $X$
decay modes are, $X$ will only decay into known particles of the MSSM.

It is useful to note here that, although there could be a lot of other
applications for this work, historically the main reason for this
computation has always been the possibility of explaining the origin of the
UHECRs through the so-called ``top-down'' models mentioned above. I
make no exception to this rule and will essentially apply our results
to this problem.

The remaining of this work will be organized as follows: in
chapter~\ref{chap:decay}, I give all the details on the physics of a SH
particle decay, and describe the calculation of the spectrum of {\it
 stable} particles (protons, electrons, photons, three kinds of
neutrinos, and possibly the lightest superparticles called LSPs)
produced in these decays, in a pure ``particle physics oriented''
approach. Chapter~\ref{chap:code} is aimed to be a ``user guide''
for the particular code I developed - called ``SHdecay''-, including
all features detailed in chapter~\ref{chap:decay}.

I then turn to particular applications of this work in the general
framework of UHECRs (chapter~\ref{chap:applications}). After a brief
introduction, I give quantitative predictions for the fluxes of
neutrinos and neutralinos in the context of ``top-down'' models.

Chapter~\ref{chap:conclusion} offers a general summary of this work and
gives some perspectives on how to pursue it.

Finally, a series of Appendices gives some theoretical and technical
information which are useful for a better understanding of this work.


\chapter{Decay of a super-heavy $X$ particle}
\label{chap:decay}

\section{Physics background}
\label{sec:intro_decay}

Before going into technical details, I briefly outline the physics
involved in the decay of a SH particle; it is summarized in
fig.~\ref{X_decay}. By assumption its primary decay is into 2 or more
particles of the MSSM. These primary decay products will generally not
be on--shell; instead, they have very large (time--like) virtualities,
of order $M_X$. Each particle produced in the primary decay will
therefore initiate a parton shower. The basic mechanism driving the
shower development is the splitting of a virtual particle into two
other particles with (much) smaller virtualities; the dynamics of this
process is described by a set of splitting functions (SFs). As long as
the virtuality is larger than the typical sparticle mass scale $M_{\rm
 SUSY}$, all MSSM particles participate in this shower. At virtuality
$M_{\rm SUSY} \sim 1$ TeV the breaking of both supersymmetry and of
$SU(2) \times U(1)_Y$ gauge invariance becomes important. All the
massive superparticles that have been produced at this stage can now
be considered to be on--shell, and will decay into Standard Model (SM)
particles and the only (possibly) stable sparticle, the LSP. The same
is true for the heavy SM particles, i.e. the top quarks and the
massive bosons. However, the lighter quarks and gluons will continue a
perturbative parton shower until they have reached either their
on--shell mass scale or the typical scale of hadronization $Q_{\rm
 had} \sim 1$ GeV. At this stage, strong interactions become
non--perturbative, forcing partons to hadronize into colorless mesons
or baryons. Finally, the unstable hadrons and leptons will also decay,
and only the stable particles will remain. The spectra of these
particles constitute the result of our calculation, which can give for
example the spectrum of Ultra-High Energy Cosmic Rays at the location
of $X$ decay, as it was mentioned in the Introduction\footnote{Of
 course, the spectrum on Earth might be modified considerably due to
 propagation through the (extra)galactic medium \cite{reviewSigl}; we
 will come back to these issues in chapter~\ref{chap:applications}.}.

Technically the shower development is described through fragmentation
functions (FFs). The dependence of these functions on the virtuality
is governed by the DGLAP evolution equations \cite{AP} extended to
include the complete spectrum of the MSSM. All splitting functions
needed in this calculation are collected in Appendix~\ref{app:SFs}.
We numerically solved the evolution equations for the FFs of any
particle of the MSSM into any other. At scale $M_{\rm SUSY}$ we
applied unitary transformations to the FFs of the unbroken fields
(``current eigenstates'') in order to obtain those of the physical
particles (``mass eigenstates''); details are given in
Appendix~\ref{app:transfo}. We then model the decays of all particles
and superparticles with mass $\sim M_{\rm SUSY}$, using the public
code ISASUSY \cite{Isasusy} to compute the branching ratios for all
allowed decays, for a given set of SUSY parameters. If R-parity is
conserved, we obtain the final spectrum of the stable LSP at this
step; the rest of the available energy is distributed between the SM
particles. After a second perturbative cascade down to virtuality
$\sim \max(m_q,Q_{\rm had})$, the quarks and gluons will hadronize, as
stated before. This non--perturbative phenomenon is parameterized in
terms of ``input'' FFs. We use the results of ref.\cite{Poetter},
which are based on fits to LEP data. We paid special attention to the
conservation of energy; this was not possible in previous studies,
because of the incomplete treatment of the decays of particles with
mass of order $M_{\rm SUSY}$. We are able to check energy conservation
at each step of the calculation, up to a numerical accuracy of a few
per mille. A brief summary of these results has appeared in \cite{bd1},
and the more complete analysis presented here were published in \cite{bd2}.

\vspace*{5mm}
\begin{center} 
\begin{figure}
\label{X_decay}
\begin{picture}(-400,300)(0,-190)
\SetOffset(500,0)
\SetPFont{Helvetica}{24}
\PText(-405,4)(0)[]{X}
\DashLine(-410,0)(-440,0){3} \Text(-425,7)[]{$\tilde{q}_L$} 
\DashLine(-400,0)(-370,0){3} \Text(-385,7)[]{$\tilde{q}_L$} 
\Line(-370,1)(-340,26)
\Line(-370,0)(-340,25) \Text(-360,25)[]{$\tilde{g}$} 
\Line(-340,26)(-310,46)
\Line(-340,25)(-310,45) \Text(-330,45)[]{$\tilde{g}$}
\Gluon(-340,25)(-310,10){-3}{4}\Text(-320,25)[]{$g$} 
\DashLine(-310,45)(-280,55){3}  \Text(-295,60)[]{$\tilde{q}_L$} 
\GCirc(-280,55){3}{0}
\Line(-310,45)(-280,35) \Text(-295,30)[]{$q_L$} 
\DashLine(-280,-120)(-280,110){4} \Text(-280,-130)[c]{1 TeV}
\Text(-280,-145)[c]{(SUSY}
\Text(-280,-160)[c]{+ $SU(2)\otimes U(1)$}
\Text(-280,-175)[c] {breaking)}

\Line(-280,55)(-250,80) \Text(-265,80)[]{$q$} 
\Line(-250,80)(-190,70) \Text(-220,68)[]{$q$} 
\Gluon(-250,80)(-220,90){3}{4} \Text(-235,97)[]{$g$} 
\Line(-220,90)(-190,100) \Text(-205,103)[]{$q$} 
\Line(-220,90)(-190,80) \Text(-205,80)[]{$q$} 

\Line(-280,56)(-250,46)
\Line(-280,55)(-250,45)
\Text(-265,42)[]{$\tilde{\chi}_2^0$}
\GCirc(-250,45){3}{0}
\Line(-250,45)(-220,55) \Text(-235,57)[]{$q$}
\Line(-250,46)(-170,36)
\Line(-250,45)(-170,35) \Text(-166,35)[l]{$\tilde{\chi}_1^0$}
\Line(-250,45)(-220,25) \Text(-235,27)[]{$q$} 
\Line(-220,55)(-190,65) 
\Gluon(-220,55)(-190,45){-3}{4} 
\DashLine(-190,-120)(-190,110){4} \Text(-190,-130)[c]{1 GeV}
\Text(-190,-145)[c]{(hadronization)}

\Line(-370,0)(-340,-25) \Text(-360,-18)[]{$q_L$} 
\Photon(-340,-25)(-310,-5){3}{5} \Text(-330,-5)[]{$B$} 
\DashLine(-310,-5)(-280,5){3} \Text(-295,8)[]{$\tilde{q}_R$}
\GCirc(-280,5){3}{0}
\DashLine(-310,-5)(-280,-15){3} \Text(-295,-18)[]{$\tilde{q}_R$}
\GCirc(-280,-15){3}{0}
\Line(-340,-25)(-310,-45) \Text(-330,-40)[]{$q$} 
\Photon(-310,-45)(-280,-25){3}{5} \Text(-308,-31)[]{$W$} 
\GCirc(-280,-25){3}{0}
\Line(-280,-25)(-190,-5) \Text(-240,-8)[]{$\tau$} 
\GCirc(-190,-5){3}{0}
\Line(-190,-5)(-160,5) \Text(-175,9)[]{$a_1^{-}$}
\GCirc(-160,5){3}{0}
\Line(-160,5)(-130,15) \Text(-145,17)[]{$\rho^{-}$}
\GCirc(-130,15){3}{0}
\Line(-130,15)(-100,15) \Text(-115,11)[]{$\pi^{-}$}
\GCirc(-100,15){3}{0}
\Line(-100,15)(-70,15) \Text(-66,15)[l]{$\nu_\mu$} 
\Line(-100,15)(-70,5) \Text(-78,2)[]{$\mu^{-}$} 
\GCirc(-70,5){3}{0}
\Line(-70,5)(-50,5) \Text(-46,5)[l]{$\nu_\mu$} 
\Line(-70,5)(-50,-5) \Text(-46,-7)[l]{$\nu_e$} 
\Line(-70,5)(-50,-15) \Text(-46,-16)[l]{$e^{-}$} 

\Line(-130,15)(-100,35) \Text(-115,35)[]{$\pi^0$} 
\GCirc(-100,35){3}{0}
\Photon(-100,35)(-70,35){3}{5} \Text(-66,35)[]{$\gamma$} 
\Photon(-100,35)(-70,55){3}{5} \Text(-66,57)[]{$\gamma$} 

\Line(-160,5)(-130,-5) \Text(-145,-8)[]{$\pi^0$}
\GCirc(-130,-5){3}{0}
\Photon(-130,-5)(-100,-25){3}{5} \Text(-96,-27)[]{$\gamma$} 
\Photon(-130,-5)(-100,-5){3}{5} \Text(-96,-3)[]{$\gamma$} 
\Line(-190,-5)(-160,-15) \Text(-175,-16)[]{$\nu_\tau$} 
\Line(-280,-25)(-100,-45) \Text(-96,-45)[l]{$\nu_\tau$} 

\Line(-310,-45)(-280,-65) \Text(-300,-62)[]{$q$} 
\Line(-280,-65)(-250,-55) \Text(-265,-53)[]{$q$} 
\Gluon(-280,-65)(-250,-75){-3}{4} \Text(-265,-80)[]{$g$}
\Gluon(-250,-75)(-220,-95){-3}{4} \Text(-240,-95)[]{$g$} 
\Line(-220,-95)(-190,-105) \Text(-205,-105)[]{$q$} 
\Line(-220,-95)(-190,-85) \Text(-205,-85)[]{$q$}

\Gluon(-250,-75)(-220,-65){3}{4} \Text(-240,-62)[]{$g$} 
\Line(-220,-65)(-190,-55)  \Text(-205,-55)[]{$q$}
\Line(-220,-65)(-190,-75)  \Text(-205,-75)[]{$q$}

\GOval(-190,65)(20,7)(0){0} 
\GCirc(-170,65){3}{0}
\Text(-175,65)[r]{$n$}
\Line(-170,65)(-130,80) \Text(-126,82)[l]{$p$}
\Line(-170,65)(-130,65) \Text(-126,65)[l]{$e^{-}$}
\Line(-170,65)(-130,50) \Text(-126,48)[l]{$\nu_e$} 

\GOval(-190,-80)(15,7)(0){0} 
\GCirc(-168,-83){3}{0}
\Text(-170,-80)[r]{$\pi^0$}
\Photon(-168,-83)(-130,-65){3}{5} \Text(-126,-63)[]{$\gamma$} 
\Photon(-168,-83)(-130,-95){3}{5} \Text(-126,-98)[]{$\gamma$} 

\end{picture} 
\caption{Schematic MSSM cascade for an initial squark with a
virtuality $Q \simeq M_X$. The full circles indicate decays of massive
particles, in distinction to fragmentation vertices. The two vertical
dashed lines separate different epochs of the evolution of the
cascade: at virtuality $Q > M_{\rm SUSY}$, all MSSM particles can be
produced in fragmentation processes. Particles with mass of order
$M_{\rm SUSY}$ decay at the first vertical line. For $M_{\rm SUSY} > Q
> Q_{\rm had}$ light QCD degrees of freedom still contribute to the
perturbative evolution of the cascade. At the second vertical line,
all partons hadronize, and unstable hadrons and leptons decay. See the
text for further details.}
\end{figure}
\end{center}
\vspace*{5mm}

\section{Technical aspects of the calculation}
\label{sec:technical}

In this section we describe how to calculate the spectra of stable
particles produced in $X$ decays: protons, electrons, photons, the
three types of neutrinos and LSPs, and their antiparticles. Note that
at most one out of the many particles produced in a typical $X$ decay
will be observed on Earth. This means that we cannot possibly measure
any correlation between different particles in the shower; the energy
spectra of the final stable particles are indeed the only measurable
quantities. These spectra are given by the differential decay rates $d
\Gamma_X / d E_P$, where $P$ labels the stable particle we are
interested in. This is a well--known problem in QCD, where parton
showers were first studied. The resulting spectrum can be written in
the form \cite{QCDrev}
\beq \label{def_ff}
\frac {d \Gamma_X} {d x_P} = \sum_I \frac {d \Gamma(X \rightarrow I)}
{d x_I} \otimes D^P_I(\frac{x_P}{x_I}, M_X^2),
\eeq
where $I$ labels the MSSM particles into which $X$ can decay, and we
have introduced the scaled energy variable $x = 2 E / M_X$. $d
\Gamma(X \rightarrow I) / d x_I$ depends on the phase space in a
particular decay mode; for a two--body decay, $d \Gamma(X \rightarrow
I) / d x_I \propto \delta(1-x_I)$. The convolution is defined as
\beq \label{conv}
f(z) \otimes g(x/z) = \int_{x}^{1} f(z) g \left( \frac {x} {z} \right)
\, \frac{dz}{z}.
\eeq
All the nontrivial physics is now contained in the fragmentation
functions (FFs) $D^P_I(z,Q^2)$. They encode the probability for a
particle $P$ to originate from the shower initiated by another
particle $I$, where the latter has been produced with initial
virtuality $Q$. This implies the ``boundary conditions''
\beq \label{boundary}
D^J_I(z,m_J^2) = \delta_I^J \cdot \delta(1-z),
\eeq
which simply say that an on--shell particle cannot participate in the
shower any more. As already explained in the Introduction, for $Q >
M_{\rm SUSY}$ all MSSM particles $J$ are active in the shower, and
thus have to be included in the list of ``fragmentation products''.

The first calculations of this kind \cite{Hill} used simple scaling
fragmentation functions to describe the transition from partons to
hadrons. Later analyses \cite{Birkel,Berezinsky:2000} used Monte Carlo
programs to describe the cascade. However, since we can only expect to
see a single particle from any given cascade, we only need to know the
one--particle inclusive decay spectrum of $X$. This is encoded in
fragmentation functions; the evolution of the cascade corresponds to
the scale dependence of these FFs, which is described by generalized
DGLAP equations \cite{AP}.

In the next two subsections we discuss these evolution equations, and
their solution, in more detail. We first only include strong
(SUSY--QCD) interactions. However, at energies above $10^{20}$ eV all
gauge interactions are of comparable strength. The same is true for
interactions due to the Yukawa coupling of the top quark, and possibly
also for those of the bottom quark and tau lepton. In a second step we
therefore extend the evolution equations to include these six
different interactions\footnote{Earlier analyses using this technique
 only included (SUSY) QCD \cite{Rubin:1999, Coriano:2001,
  Sarkar:2001, FodorKatz}, or at best a partial treatment of
 electroweak interactions \cite{ToldraLSP, Berezinsky:2002}.}. We
then describe the decays of heavy (s)particles, which happen at
virtuality $Q = M_{\rm SUSY} \sim 1$ TeV. At $Q < M_{\rm SUSY}$ only
QCD interactions need to be included, greatly simplifying the
treatment of the evolution equations in this domain. Finally, we
describe the non-perturbative hadronization, and the weak decays of
unstable hadrons and leptons.

We will show that the inclusion of electroweak gauge interactions in
the shower gives rise to a significant flux of very energetic photons
and leptons, beyond the highest proton energies. Moreover, we
carefully model decays of all unstable particles. As a result, we are
for the first time able to fully account for the energy released in
$X$ decay. We cover all possible primary $X$ decay modes, i.e. our
results should be applicable to all models where physics at energies
below $M_X$ is described by the MSSM.

The remainder of this chapter is organized as follows. In sec.~2 we
describe the technical aspects of the calculation. The derivation and
solution of the evolution equations is outlined. We also check that
our final results are not sensitive to the necessary extrapolation of
the input FFs. Numerical results are presented in sec.~3. We give the
energy fractions carried by the seven stable particles for any primary
$X$ decay product, and study the dependence of our results on the SUSY
parameters. We finally describe our implementation of color coherence
effects at small $x$ using the modified leading log approximation
(MLLA). Sec.~4 is devoted to a brief summary and conclusions.
Technical details are delegated to a series of Appendices, giving the
complete list of splitting functions (Appendix~\ref{app:SFs}), the
unitary transformations from the interaction states to the physical
states (Appendix~\ref{app:transfo}), our treatment of 2-- and 3--body
decays (Appendix~\ref{app:decays}), parameterizations of the input FFs
(Appendix~\ref{app:FFs}), and finally a complete set of FFs obtained with
our program for a given set of SUSY parameters (corresponding to a
gaugino--like LSP with a low value of $\tan \beta \sim 3.6$ and
$M_{\rm SUSY} \sim 500$ GeV. See Appendix~\ref{app:curves}).

\subsection{Evolution equations in QCD and SUSY--QCD}

For convenience, we review here the DGLAP evolution equations in
ordinary QCD. As already noted, the FF $D_{p}^P (x,Q^2)$ of a parton
(quark or gluon) $p$ into a particle (parton or hadron) $P$ describes
the probability of fragmentation of $p$ into $P$ carrying energy $E_P
= xE_p$ at a virtuality scale $Q$. If $P$ is itself a parton, the FF
has to obey the boundary condition (\ref{boundary}). However, if $P$
is a hadron, the $x-$dependence of the FF cannot be computed
perturbatively; it is usually derived from fits to experimental
data. Perturbation theory does predict the dependence of the FFs on
the virtuality $Q$: it is described by a set of coupled
integro--differential equations. In leading order (LO), these QCD
DGLAP evolution equations can be written as \cite{QCDrev}:
\beqa \label{e1}
\frac {dD_{q_i}^P (x,Q^2)} {d\log(Q^2)}  &=&
\frac {\alpha_S(Q^2)} {2\pi} \left\{ P_{gq}(z) \otimes
  D_{g}^{P}(\frac{x}{z},Q^2) + P_{qq}(z) \otimes D_{q_i}^P
(\frac{x}{z},Q^2) \right\}\,, \nonumber \\
\frac {dD_g^P(x,Q^2)} {d\log(Q^2)}
&=& \frac {\alpha_{S}(Q^2)} {2\pi} \left\{ P_{gg}(z) \otimes
  D_{g}^P(\frac{x}{z},Q^2) + \sum_{i=1}^{2F} P_{qg}(x) \otimes
D_{q_i}^P (\frac{x}{z}, Q^2) \right\}\,, \eeqa
where $\alpha_S$ is the running QCD coupling constant, $F$ is the
number of active flavors (i.e. the number of Dirac quarks whose mass
is lower than $Q$), and $i$ labels the quarks and
antiquarks.\footnote{Note that the DGLAP equations given here are the
{\it time--like} ones, which describe the evolution of fragmentation
functions. In leading order they differ from the space--like DGLAP
equation (describing the evolution of distribution functions of
partons inside hadrons) only through a transposition of the matrix of
the splitting functions.} The convolution has been defined in
eq.(\ref{conv}). The physical content of these equations can be
understood as follows. A virtual quark $q_i$ can reduce its
virtuality by emitting a gluon; the final state then contains a quark
and a gluon. Either of these partons (with reduced virtuality) can
fragment into the desired particle $P$; this explains the occurrence of
two terms in the first eq.(\ref{e1}). Analogously, a gluon can either
split into two gluons, or into a quark--antiquark pair, giving rise to
the two terms in the second eq.(\ref{e1}).

These partonic branching processes are described by the splitting
functions (SFs) $P_{p_2 p_1}(x)$, for parton $p_1$ splitting into
parton $p_2$, where $x = E_{p_2} / E_{p_1}$. As already noted, in pure
QCD there are only three such processes: gluon emission off a quark or
gluon, and gluon splitting into a $q \bar q$ pair. The first of these
processes gives rise to both SFs appearing in the first eq.(\ref{e1});
momentum conservation then implies $P_{qq}(x) = P_{gq}(1-x)$, for $x
\neq 1$. Similarly, $P_{gg}(x) = P_{gg}(1-x)$ and $P_{qg}(x) =
P_{qg}(1-x)$ follows from the symmetry of the final states resulting
from the splitting of a gluon. Special care must be taken as $x
\rightarrow 1$. Here one encounters infrared singularities, which
cancel against virtual quantum corrections. The physical result of
this cancellation is that the energy of the fragmenting parton $p$ is
conserved, which requires 
\beq \label{momcons}
\sum_P \int_0^1 x D_p^P(x,Q^2) = 1 \,\,\, \forall p, Q^2.
\eeq
This can be ensured, if
\beq \label{sumrule}
\int_0^1 dx\, x \sum_{p'} P_{p'p}(x) = 0 \,\,\, \forall p.
\eeq
Note that these integrals must give zero (rather than one), since
eqs.(\ref{e1}) only describe the {\em change} of the FFs. The explicit
form of the QCD SFs is \cite{AP}:
\beqa
\label{e3}
P_{qq}(x) &=& \frac{4}{3} \left( \frac {1+x^2} {1-x} \right )_+\,,
\nonumber \\
P_{gq}(x) &=& \frac{4}{3} \frac {1 + (1-x)^2} {x}\,, 
\nonumber \\
P_{qg}(x) &=& \frac{1}{2} \left[ (1-x)^2 + x^2 \right] \,,
\nonumber \\
P_{gg}(x) &=& 6 \left[ \frac {1-x} {x} + x(1-x) + \frac {x} {(1-x)_+}
+ \delta(1-x) \left( \frac{11}{12} - \frac{F}{18} \right) \right] \,.
\eeqa
The ``+'' distribution, which results from the cancellation of $x
\rightarrow 1$ divergences as outlined above, is defined as:
\beq \label{e4} 
\int_{0}^{1} f(x) g(x)_+ dx = \int_{0}^{1} [f(x) - f(1)] g(x),
\eeq
while $g(x)_+ = g(x)$ for $x \neq 1$. Finally, the scale dependence of
$\alpha_S(Q^2)$ is described by the following solution of the relevant
renormalization group equation (RGE):
\beq \label{alphas}
\alpha_S(Q^2) = \frac{2\pi B}{\log\frac{Q^2}{\Lambda^2}} \,,
\eeq
where $\Lambda \sim 200$ MeV is the QCD scale parameter, and $B =
6/(33 - 2F)$. 

Note that eqs.(\ref{e1}) list different FFs for all (anti)quark flavors
$q_i$. At first sight it thus seems that one has to deal with a system
of $2F+1$ coupled equations. In practice the situation can be
simplified considerably by using the linearity of the evolution
equations. This implies 
\beq \label{split}
D_p^P(x,Q^2) = \sum_{p'} \tilde D_p^{p'}(z, Q^2, Q_0^2) \otimes
D_{p'}^P(\frac{x}{z}, Q_0^2),
\eeq
where the generalized FFs $\tilde D_p^{p'}$ again obey the evolution
equations (\ref{e1}). Moreover, they satisfy the boundary conditions
$D_p^{p'}(x,Q_0^2, Q_0^2) = \delta(1-x) \delta_p^{p'}$ at some
convenient value of $Q_0 < Q$. The $\tilde D$ thus describe the purely
perturbative evolution of the shower between virtualities $Q$ and
$Q_0$. This ansatz simplifies our task, since all quark flavors have
exactly the same strong interactions, i.e. we can use the {\em same}
$\tilde D_{q_i}^{p'}$ for all quarks $q_i$ with $m_{q_i} < Q_0$.
Moreover, we only have to distinguish three different cases for $p': \
q_i, q_j$ with $j \neq i$, and $g$. All flavor dependence is then
described by the $D_{p'}^P(x, Q_0^2)$; for sufficiently small $Q_0$,
these can be taken directly from fits to experimental data. If we make
the additional simplifying assumption that all quarks and antiquarks
are produced with equal probability in primary $X$ decays, we
effectively only have to introduce two generalized FFs $\tilde D$ for
a given particle $P$, one for the fragmentation of gluons and one for
the fragmentation of any quark. In other words, in pure QCD we only
need to solve a system of two coupled equations.

The introduction of squarks $\tilde q_i$ and gluinos $\tilde g$,
i.e. the extension to SUSY--QCD, requires the introduction of FFs
$D_{\tilde q_i}^P, \ D_{\tilde g}^P$. This gives rise to new SFs,
describing the emission of a gluon by a squark or gluino, as well as
splittings of the type $q_i \rightarrow \tilde q_i \tilde g, \ \tilde
q_i \rightarrow q_i \tilde g$ and $\tilde g \rightarrow \tilde q_i
\bar q_i$. We thus see that any of the four types of partons $(q_i,
\tilde q_i, g, \tilde g)$ can split into any (other) parton. The
complete set of evolution equations thus contains 16 SFs \cite{Jones},
which we collect in Appendix~\ref{app:SFs}. The presence of new particles with
$SU(3)$ interactions also modifies the running of $\alpha_S$. One can
still use eq.(\ref{alphas}), but now $B_{SUSY} = 2/(9 - F)$.

\subsection{Evolution equations in the MSSM}

We now extend our discussion of the evolution equations to the full
MSSM. We already saw in the Introduction that superparticles can only
be active in the shower evolution at virtualities $Q > M_{\rm SUSY}
\sim 1$ TeV. This means that the supersymmetric part of the shower
evolution can be described in terms of generalized FFs $\tilde D_I^J$
satisfying the boundary condition 
\beq \label{boundary_MSSM}
\tilde D_I^J(x,M^2_{\rm SUSY},M^2_{\rm SUSY}) = \delta_I^J \delta(1-x),
\eeq
where $I$ and $J$ label any (s)particle contained in the MSSM. Note
that eq.(\ref{boundary_MSSM}) differs from eq.(\ref{boundary}) since
the former is valid for {\em all} particles in the MSSM, including
light partons. According to the discussion following eq.(\ref{split})
we only have to consider those particles to be distinct that have
different interactions. We include all gauge interactions in this part
of the shower evolution, as well as the Yukawa interactions of third
generation (s)fermions and Higgs bosons, but we ignore first and
second generation Yukawa couplings, as well as all interactions
between different generations. This immediately implies that we do not
need to distinguish between first and second generation particles.
Moreover, we ignore CP violation, which means that we need not
distinguish between particles and antiparticles. 

Finally, the electroweak $SU(2)$ symmetry can be taken to be exact at
virtuality $Q > M_{\rm SUSY}$, i.e. we need not distinguish between
members of the same $SU(2)$ multiplet\footnote{This is analogous to
  ordinary QCD, where one does not need to introduce different FFs for
  quarks with different colors. Our assumption implies that $X$ is an
  $SU(2)$ singlet. Had we allowed \cite{Berezinsky:2002} $X$ to
  transform nontrivially under $SU(2)$, the $SU(2)$ splitting
  functions would have to be modified \cite{CCC}.}. Altogether we
therefore need to treat 30 distinct particles: six quarks $q_L, u_R,
d_R, t_L, t_R, b_R$, four leptons $l_L, e_R, \tau_L, \tau_R$, three
gauge bosons $B, W, g$, two Higgs bosons $H_1, H_2$, and all their
superpartners; $H_1$ couples to down--type quarks and leptons, while
$H_2$ couples to up--type quarks.  Note that a ``particle'' often
really describes the contribution of several particles which are
indistinguishable by our criteria. For example, the ``quark'' $u_R$
stands for all charge$-2/3$ right--handed quarks and antiquarks of the
two first generations, i.e.  $u_R, \, c_R$ and their antiparticles
$\overline{u}_R, \, \overline{c}_R$. This can be expressed formally as
$D_{u_R}^P = \left( D_{u_R}^P + D_{c_R}^P + D_{\overline{u}_R}^P +
  D_{\overline{c}_R}^P \right) / 4$, where in our approximation the
four terms in the sum are all identical to each other after the final
state $P$ has been summed over particle and antiparticle.\footnote{A
  consistent interpretation of, e.g., $u_R$ as a ``particle'' requires
  that $u_R$ stands for the {\em average} of $u_R, \, c_R$ etc. when
  $u_R$ appears as {\em lower} index of a generalized FF, as described
  in the text.  However, $u_R$ stands for the {\em sum} of $u_R, \,
  c_R$ etc. when $u_R$ is an {\em upper} index of a $\tilde D$. With
  this definition, we have $\tilde D_{u_R}^{u_R}(x,M^2_{\rm SUSY}) =
  \delta(1-x)$. This interpretation also fixes certain multiplicity
  factors in the DGLAP equations, as detailed in
  Appendix~\ref{app:SFs}. This treatment is only possible if $X$ has
  equal branching ratio into $u_R, \, c_R$ etc. However, we expect the
  differences between decays into first or second generation quarks to
  be very small even in models where these branching ratios are not
  the same.} Similarly, $q_L$ stands as initial particle for an
average over the two $SU(2)$ quark doublets of the two first
generations $(u_L,d_L)$ and $(c_L,s_L)$, and their antiparticles. Note
that all group indices of the particle in question are summed over. In
the usual case of QCD this only includes summation over color indices,
but in our case it includes summation over $SU(2)$ indices, since
$SU(2)$ is (effectively) conserved at energies above $M_{\rm SUSY}$.

Let us first discuss the scale dependence of the six coupling
constants that can affect the shower evolution significantly at scales
$Q > M_{\rm SUSY}$. These are the three gauge couplings $g_Y$, $g_2$
and $g_S$, which are related to the corresponding ``fine structure
constants'' through $\alpha_i \equiv g_i^2 / (4\pi), \ i \in \left\{
Y, 2, S \right\}$. Moreover, the third generation Yukawa couplings are
proportional to the masses of third generation quarks or leptons:
\beqa \label{e6}
y_t &=& \frac{g\,m_t}{\sqrt{2}\,m_W\sin{\beta}}\,,
\nonumber \\
y_b &=& \frac{g\,m_b}{\sqrt{2}\,m_W\cos{\beta}}\,,
\nonumber \\
y_\tau &=& \frac{g\,m_\tau}{\sqrt{2}\,m_W\cos{\beta}}\,,
\eeqa
where $\tan\beta \equiv \langle H_2^0 \rangle / \langle H_1^0
\rangle$. The couplings $y_b$ and $y_\tau$ are only significant if
$\tan\beta \gg 1$. Note that in many models, values $\tan\beta \simeq
m_t(m_t) / m_b(m_t) \simeq 60$ are possible, in which case $y_b$ and
$y_\tau$ are comparable in magnitude to $g_S$ and $g_2$, respectively.
The LO RGEs for these six MSSM couplings are \cite{susyrge}:
\beqa \label{e7}
\frac{dg_Y}{dt} &=& 11\,\frac{g_Y^3}{16\pi^2}\,,
\nonumber \\
\frac{dg_2}{dt} &=& \frac{g_2^3}{16\pi^2}\,,
\nonumber \\
\frac{dg_S}{dt} &=& -3\,\frac{g_S^3}{16\pi^2}\,,
\nonumber \\
\frac{dy_t}{dt} &=& \frac {y_t} {16\pi^2} \left( 6y_t^2 + y_b^2 -
\frac {13} {9} g_Y^2 -3 g_2^2 -\frac {16} {3} g_S^2 \right) \,,
\nonumber \\
\frac{dy_b}{dt} &=& \frac {y_b} {16\pi^2} \left( 6y_b^2 + y_t^2 +
y_\tau^2 - \frac{7}{9} g_Y^2 - 3g_2^2 - \frac{16}{3} g_S^2 \right)\,,
\nonumber \\
\frac{dy_\tau}{dt} &=& \frac {y_\tau} {16\pi^2} \left( 3y_b^2 +
4y_\tau^2 - 3g_Y^2 - 3g_2^2\right)\,,
\eeqa
where $t = \log \frac{Q}{Q_0}$ parameterizes the logarithm of the
virtuality, and $Q_0$ is an arbitrary scale where the numerical values
of these couplings constants are ``known'' (in case of the Yukawa
couplings, up to the dependence on $\tan\beta$). As well known
\cite{gut}, given their values measured at $Q_0 \simeq 100$
GeV eqs.(\ref{e7}) predict the three gauge couplings to unify at scale
$M_{\rm GUT} \simeq 2 \cdot 10^{16}$ GeV, i.e. $g^2_S(M_{\rm GUT}) =
g^2_2(M_{\rm GUT}) = 5 g^2_Y(M_{\rm GUT}) / 3 \simeq 0.52$, where
the Clebsch--Gordon factor of 5/3 is predicted by most simple unified
groups, e.g. $SU(5)$ or $SO(10)$. We solved these equations by the
Runge--Kutta method; of course, the RGEs for the gauge couplings can
trivially be solved analytically, but the additional numerical effort
required by including eqs.(\ref{e7}) in the set of coupled
differential equations that need to be solved numerically is
negligible.

\setcounter{footnote}{0}
The main numerical effort lies in the solution of the system of 30
coupled DGLAP equations, which are of the form:
\beq
\label{e8}
\frac {d \tilde D_I^J} {d\log(Q^2)} (x,Q^2,M^2_{\rm SUSY}) = \sum_K \frac {\alpha_{KI}
(Q^2)} {2\pi} P_{KI}(z) \otimes \tilde D_K^J(\frac{x}{z} ,Q^2,M^2_{\rm SUSY})\,,
\eeq
where $I,J,K$ run over all the 30 particles, and $\alpha_{KI}(Q^2) =
g_{KI}^2 / 4\pi$ is the (running) coupling constant associated
with the corresponding vertex; note that at this stage we are using
interaction (or current) eigenstates to describe the spectrum.
Generically denoting particles with spin 1, 1/2 and 0 as $V, \ F$ and
$S$ (for vector, fermion and scalar), we have to consider\footnote{We
do not need to consider $S \rightarrow SS$, since the corresponding
dimensionful coupling is ${\cal O}(M_{\rm SUSY}) \ll Q$ in this
domain, i.e. these processes are much slower than the relevant time
scale $1/Q$.} branching processes of the kind $V \rightarrow VV, \ V
\rightarrow FF, \ V \rightarrow SS, \ F \rightarrow FV, \ F
\rightarrow FS, \ S \rightarrow SV$ and $S \rightarrow F F$. All these
branching processes already occur in SUSY--QCD. The splitting
functions can thus essentially be read off from the results of
ref.\cite{Jones}, after correcting for different group [color and/or
$SU(2)$] and multiplicity factors. The coefficients of the
$\delta(1-x)$ terms in diagonal SFs can be fixed using the momentum
conservation constraint in the form (\ref{sumrule}); note that these
constraints have to be satisfied for each of the six interactions
separately. The explicit form of the complete set of MSSM SFs
$P_{KI}(x)$ is given in Appendix~\ref{app:SFs}.

We solved these equations numerically using the Runge--Kutta
method. To that end the FFs were represented as cubic splines, using
50 points which were distributed equally on a logarithmic scale in $x$
for $10^{-7} \leq x \leq 0.5$, and 50 additional points distributed
equally in $\log(1-x)$ for $0.5 \leq x \leq 1-10^{-7}$. Starting from
the boundary conditions\footnote{Technically, these $\delta-$functions
are represented by narrow Gaussians centered at $x=1$, normalized to
give unity after integration over $x \leq 1$.} (\ref{boundary_MSSM}),
we arrive at the $30\times 30$ generalized fragmentation functions at
virtuality $Q=M_X$. Here we assume that the evolution equations
describe the perturbative cascade at these energies correctly. We will
comment on the limitations of our treatment at the end of this
Section.

\subsection{Evolution of the cascade below $Q = 1$ TeV}
\label{subsec:non_pert}
\setcounter{footnote}{0}

Here we would like to describe the physics at scales at and below
$M_{\rm SUSY}$: the breaking of both supersymmetry and $SU(2)\otimes
U(1)$ symmetry, the decay of unstable (s)particles with masses of
order $M_{\rm SUSY}$, the pure QCD shower evolution down to $Q_{\rm
had}$, the non--perturbative hadronization of quarks and gluons, and
finally the weak decays of unstable leptons and hadrons. For
simplicity we assume that all superparticles, the top quark as well as
the $W, \ Z$ and Higgs bosons all decouple from the shower and decay
at the same scale $M_{\rm SUSY} \simeq 1$ TeV. The fragmentation of
$b$ and $c$ quarks is treated using the boundary condition
(\ref{boundary}) at their respective mass scales of 5 and 1.5 GeV,
while the nonperturbative hadronization of all other partons takes
place at $Q_{\rm had} = 1$ GeV.

At $Q = M_{\rm SUSY}$ we break both Supersymmetry and $SU(2) \otimes
U(1)$. All (s)particles acquire their masses in this process, and in
many cases mix to give the mass eigenstates. This means that we have
to switch from a description of the particle spectrum in terms of
current eigenstates to a description in terms of physical mass
eigenstates. This is accomplished by unitary transformations of the
type\footnote{Note that the squares of the coefficients $c_{SJ}$
appear in eq.(\ref{trafo}), since the FFs describe probabilities,
which are related to the square of the wave functions of the particles
in question.}
\beq \label{trafo}
\tilde D^S_I = \sum_J |c_{SJ}|^2 \tilde D^J_I\,.
\eeq
Unitarity requires $\sum_S |c_{SJ}|^2 = \sum_J |c_{SJ}|^2 = 1$, if the
current state $J$ has the same number of degrees of freedom as the
physical state $S$. This is often not the case in the usual
convention; then some care has to be taken in writing down the
$|c_{SJ}|^2$, see Appendix~\ref{app:transfo}. We use the following
physical particles: $u\,, d\,, b\,, s\,, c\,, t$ quarks and $e\,,
\mu\,, \tau$ leptons now have both left-- and right--handed
components, i.e. they have twice as many degrees of freedom as the
corresponding states with fixed chirality. The neutrinos remain
unchanged, since we ignore the interactions of right--handed
neutrinos. The gluons also remain unchanged, since $SU(3)$ remains
exact below $M_{\rm SUSY}$. The electroweak gauge sector of the SM is
described by $W := W^+ + W^-$, Z and $\gamma$; note that the massive
gauge bosons absorb the Goldstone modes of the Higgs sector, and hence
receive corresponding contributions in eq.(\ref{trafo}). The Higgs
sector consists of two charged Higgs bosons $H^{\pm}$ (described by $H
= H^+ + H^-$) and the three neutral ones $H^{0}$, $h^{0}$ and $A^{0}$;
the neutral Higgs bosons are described by real fields, which contain a
single degree of freedom. In the SUSY part of the spectrum, the gluino
$\tilde g$ as well as the first and second generation sfermions
$\tilde{u}_{L,R}$, $\tilde{d}_{L,R}$, $\tilde{s}_{L,R}$,
$\tilde{c}_{L,R}$ and sneutrinos remain unchanged (but $\tilde u_L$
and $\tilde d_L$, etc., are now distinguishable). The $SU(2)$ singlets
and doublets of third generation charged sfermions mix to form mass
eigenstates \stone, \sttwo, $\tilde b_1, \ \tilde{b}_2$, \stauone,
\stautwo. Similarly, the two Dirac charginos \scone\, and \sctwo are
mixtures of charged higgsinos and winos, and the four Majorana
neutralinos \snone, \sntwo, \snthree, \snfour, in order of increasing
masses, are mixtures of neutral higgsinos, winos and binos.

The numerical values of many of the $c_{SJ}$ depend on the parameters
describing the breaking of supersymmetry. We choose four different
sets of parameters, which describe typical regions of the parameter
space, in order to study the impact of the details of SUSY breaking on
the final spectra. We take two fairly extreme values of $\tan(\beta) =
3.6$ and $48$, and two sets of dimensionful parameters corresponding
to higgsino--like and gaugino--like states \scone, \snone\ and \sntwo.
We used the software ISASUSY \cite{Isasusy} to compute the mass
spectrum and the mixing angles of the sparticles and Higgses for a
given set of SUSY parameters.

Having computed the spectrum of physical (massive) particles, we have
to treat the decay of all unstable particles with mass near $M_{\rm
SUSY}$. Since we assumed $R-$parity to be conserved, the lightest
supersymmetric particle (LSP) is stable. In our four scenarios (as in
most of parameter space) the LSP is the lightest neutralino \snone.
The end products of these decays are thus light SM particles and
LSPs. Note that decays of heavy sparticles often proceed via a
cascade, where the LSP is produced only in the second, third or even
fourth step, e.g. $\tilde g \rightarrow \bar u \tilde u_L \rightarrow
\bar u d \tilde \chi_1^+ \rightarrow \bar u d e^+ \nu_e \tilde
\chi_1^0$. In order to model these decays we again use ISASUSY, which
computes the branching ratios for all allowed tree--level 2-- and
3--body decay modes of the unstable sparticles, of the top quark and
of the Higgs bosons. Together with the known branching ratios of the
$W$ and $Z$ bosons, this allows us to compute the spectra of the SM
particles and the LSP after all decays, by convoluting the spectra of
the decaying particles with the energy distributions calculated for
2-- or 3--body decays. The total generalized FF of any MSSM current
eigenstate $I$ into a light or stable physical particle $s$ (quark,
gluon, lepton, photon or LSP) is then
\beq \label{decay}
\tilde D_I^s = \tilde D_I^{S=s} + \sum_{S \neq s} \tilde D_I^S \otimes
\tilde P_{sS},
\eeq
where $\tilde P_{sS}$ describes the spectrum of $s$ in the decay $S
\rightarrow s$. We compute these spectra from phase space, including
all mass effects, but we didn't include the matrix elements. The
spectra for each decay mode of the heavy particle $S$ are normalized
to give the correct branching ratio, as computed by ISAJET. As far as
LSPs are concerned, eq.(\ref{decay}) already gives the final result,
i.e. $D_I^{\rm LSP} = \tilde D_I^{\rm LSP}$. If $s$ is a lepton or
photon, eq.(\ref{decay}) describes the FF at all virtualities between
$M_{\rm SUSY}$ and $m_b = 5$ GeV.

As we will see shortly, in some cases two--body decays can lead to
sharp edges in the FFs at intermediate values of $x$. This can happen
if the primary decay product is a massive particle with only weak
interactions. In that case a substantial fraction of the initial
$\delta-$peak at $x=1$ survives even after the evolution; convolution
of this $\delta-$peak with a two--body decay distribution leads to a
flat $x$ distribution of the decay products between some $x_{\rm min}$
and $x_{\rm max}$. An accurate description of these contributions to
the FFs sometimes requires the introduction of additional points near
$x_{\rm min}$ and/or $x_{\rm max}$ in the splines describing these
FFs.

The perturbative evolution in the QCD sector does not stop at $M_{\rm
SUSY}$, but continues until virtuality $Q_0 = \max(m_q, Q_{\rm
had})$. This part can be treated by introducing generalized FFs
$\tilde D_p^{p'}$ as in eq.(\ref{split}), where $(p, p') \in
\{u,d,s,c,b,g\}$ are light QCD partons. We use once more the DGLAP
evolution equations, but this time for pure QCD, evolving these
generalized FFs between $Q_0$ and $M_{\rm SUSY}$. The generalized
partonic FFs between $Q_0$ and $M_X$ can then be computed through one
more convolution:
\beq \label{dofq0}
\tilde D_I^p(x,M_X^2,Q_0^2) = \sum_{p'} \tilde D_I^{p'}(z,M_X^2,M_{SUSY}^2)
\otimes \tilde D_{p'}^p(\frac{x}{z},M_{SUSY}^2,Q_0^2) \,.
\eeq

The total partonic FFs at $M_X$ can finally be computed through
eq.(\ref{split}) by using known ``input FFs''. They describe the
non--perturbative hadronization of quarks and gluons into mesons and
baryons, which happens at $Q = Q_0$. These FFs $D_i^h(x,Q_0^2)$, where
$i \in \{u,d,s,c,b,g\}$ and $h$ represents a hadron, can be obtained
directly from a fit to (e.g.) LEP data. We used the results of
\cite{Poetter}, where the FFs of a quark or gluon into protons,
neutrons, pions and kaons (or more exactly the sum over particles and
antiparticles) are parameterized in the form
$Nx^{\alpha}(1-x)^{\beta}$.

The original form \cite{Poetter} of these functions is only valid down
to $x = 0.1$. Kinematic and color coherence effects, which are not
included in the usual DGLAP framework, become important
\cite{Basics_of_QCD}) at $x \leq \sqrt{(Q/Q_{\rm had})} \sim 0.1$,
where in the second step we have used the LEP energy scale $Q \sim
100$ GeV. For $Q \sim M_X \sim 10^{16}$ GeV these effects become large
only for $x \leq 10^{-8}$; they can thus safely be ignored for many
(but not all; see below) applications. In \cite{bd1} we
therefore chose a rather simple extrapolation of the functions given
in \cite{Poetter} towards small $x$. Our default choice was a $N
x^{-\alpha'}$ parameterization; $N$ and $\alpha'$ were computed by
requiring the continuity of this parameterization with the FFs of
\cite{Poetter} at some $x_0 \simeq 0.1$, energy conservation, and, as
additional constraint, an identical power law behavior at small $x$
(i.e. identical $\alpha'$) for all the FFs of a given quark into the
different hadrons. This last assumption was motivated by the fact that
we obtain such an identical power law at small $x$ during the
perturbative part of the cascade, and by the well accepted LPHD
hypothesis (Local Parton-Hadron Duality) \cite{LPHD}, which
postulates a local proportionality in phase space between the spectra
of partons and hadrons. We chose different $x_0$ for each initial
parton in such a way that we obtain $\alpha'$ between 0 and 2; the
upper bound on $\alpha'$ follows from energy conservation (the energy
integral $\int_0^1dx\,xD(x)$ has to be finite).

\begin{figure}
\setlength{\unitlength}{1cm}
\input{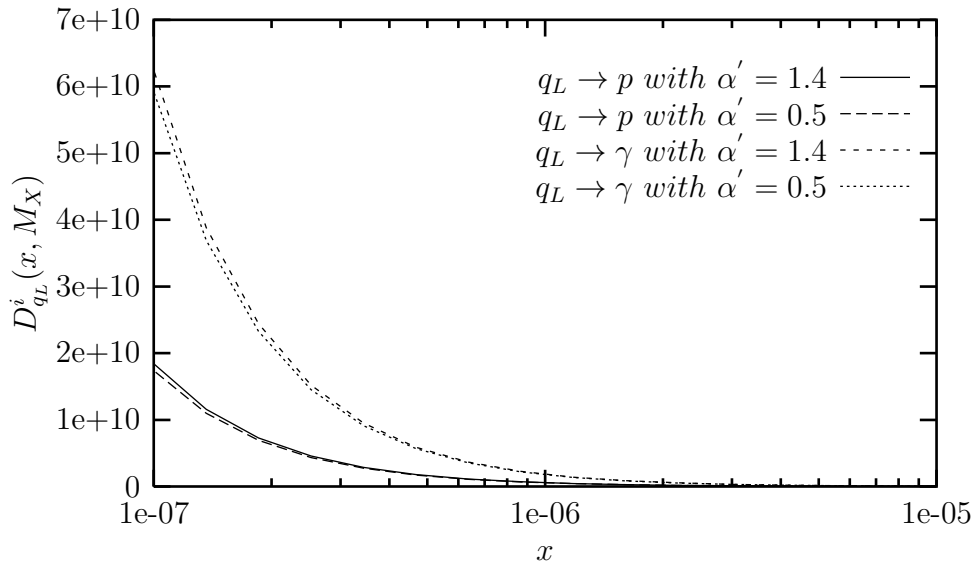}
\caption{Effect of varying the low$-x$ extrapolation of the input FFs
on the final FFs $D_{q_L}^p$ and $D_{q_L}^\gamma$. See the text for
further explanations.}
\label{smallx}
\end{figure}

In order to check the consistency of this parameterization, we used
another functional form with three free parameters: $D(x) =
ax^{-\alpha'} + b\log{x} + c\,, a > 0$. This allowed us to freely
choose $\alpha'$, keeping the same assumptions about continuity etc.
as above. This enabled us to compare two extreme values of $\alpha'$,
namely 0.5 and 1.4. The first is the smallest value compatible with $a
> 0$, while the second approximates the small$-x$ behavior of the
perturbative QCD evolution between 1 GeV and 1 TeV; requiring $\alpha'
< 1.4$ thus ensures that this perturbative evolution dominates the
behavior of the FFs at small $x$. Note also that the perhaps most
plausible value, $\alpha' \sim 1$ (which corresponds to a flat
distribution of particles in rapidity when perturbative effects are
ignored) is comfortably bracketed by these limiting values. In
fig.~\ref{smallx} we plot the final result at small $x$ for different
FFs with these two extreme parameterizations, {\it after} convolution
with the perturbative FFs. As can be seen, the effect of varying
$\alpha'$ is very small once energy conservation is imposed. This
indicates that our final results are not sensitive to the necessary
small$-x$ extrapolation of the input FFs.\footnote{However, the
original FFs of ref.\cite{Poetter} should {\em not} be used on the
whole range [$10^{-7},1$], since they violate energy conservation
badly, leading to over--production of particles at small $x$.} The
main uncertainty at moderately small $x$ ($10^{-5} \leq x \leq 0.1$)
will then come from perturbative higher order corrections, which might
be quite significant in this range.

Unfortunately, we were not able to perform a complete NLO analysis,
for the following reasons. Beyond leading order the SFs for
space--like and time--like processes are no longer identical
\cite{Furmanski}. Already at next--to--leading order (NLO) the
time--like SFs have a rather bad behavior at small $x$, with a
negative leading term $-\frac{40}{9}\frac{1}{x}$ in $P_{qq}$. This
term is tempered in the final spectra (which have to be positive) by
the convolution occurring in the DGLAP equations, as well as by the
convolution of the FFs with NLO ``coefficient functions'' which modify
the basic relation (\ref{def_ff}) once higher order corrections are
included. Note that the FFs, SFs and NLO coefficient functions are
scheme dependent; worse, the coefficient functions are also
process--dependent, i.e. they will depend on the spins of $X$ and its
primary decay products. NLO results are known for the classical
processes occurring in pure (non--supersymmetric) QCD, but they are not
available for most of the processes we are interested in. Moreover, in
cases where they are known, these coefficient functions often contain
the most important part of the NLO correction, rendering useless any
attempt to give a partial result by only including NLO terms in the
SFs. We conclude that it might be possible and interesting to carry
out a full NLO analysis in the pure QCD case, but this is not possible
in the more interesting supersymmetric case using available
results. Note that part of the perturbative NLO effects are absorbed
in the input FFs, through their fit to experimental data. At very
small $x$, NLO effects just give the leading ``color coherence''
corrections, which are re-summed analytically in the MLLA formula, as
will be discussed in Sec.~3.4.

Finally, having computed the spectra of long--lived hadrons and
leptons, we still need to treat weak decays of unstable particles, in
order to obtain the final spectra of protons, electrons, photons and
the three types of neutrinos. This is again done using the formalism
of eq.(\ref{decay}). We limit ourselves to 2-- and 3--body decays,
considering the 4--body decays of the $\tau$ to be cascades of 2--body
decays. As before, we compute the decay functions $P_{sH}$ for $H
\rightarrow s$ decays from phase space only, and we ignore decays with
branching ratio smaller than 1\%. We then renormalize the branching
ratios of the decays we do include, so that we maintain energy
conservation. We also explicitly treated the leptonic part of the
semi--leptonic decays of $b-$ and $c-$flavored hadrons, which are
evidently not included in the FFs of \cite{Poetter}. We used the
Peterson parameterization for non--perturbative heavy quark
fragmentation \cite{Peterson}, and then treated the semi--leptonic
decays in the spectator model (i.e. using the same spectra as for free
quark decays, with $m_c = 1.5$ GeV and $m_b = 4.5$ GeV). Details of
our treatment of decays are given in Appendix~\ref{app:decays}.

\section{Results and analysis}
\label{sec:results}

\subsection{General features of the final fluxes}

A fairly complete set of results of our code for a given set of SUSY
parameters is given in Appendix~\ref{app:curves}. Here we assumed
similar masses for all sfermions, higgsinos, heavy Higgs bosons and
gluinos, $m_{\tilde f} \simeq m_A \simeq m_{\tilde g} \simeq \mu
\simeq 500$ GeV; this leads to a gaugino--like LSP, since we assume
``gaugino mass unification'', i.e. $6 m_{\tilde B} \simeq 3 m_{\tilde
 W} \simeq m_{\tilde g}$. We also choose a small value for the ratio
of vevs, $\tan\beta = 3.6$. We see that the final spectra depend
sensitively on the primary $X$ decay products \cite{bd1}, especially
in the large $x$ region. This strong dependence on the unknown primary
$X$ decay mode(s) should be kept in mind when one is trying to
quantitatively test ``top--down'' models (see
chapter~\ref{chap:applications}). Nevertheless, we can make a few
general statements about these results. To that end we first analyze
ratios of FFs of the different stable particles divided by the FF of
the same initial particle into protons. Recall that these FFs directly
represent the flux at source if $X$ undergoes two--body decay.

\begin{figure}
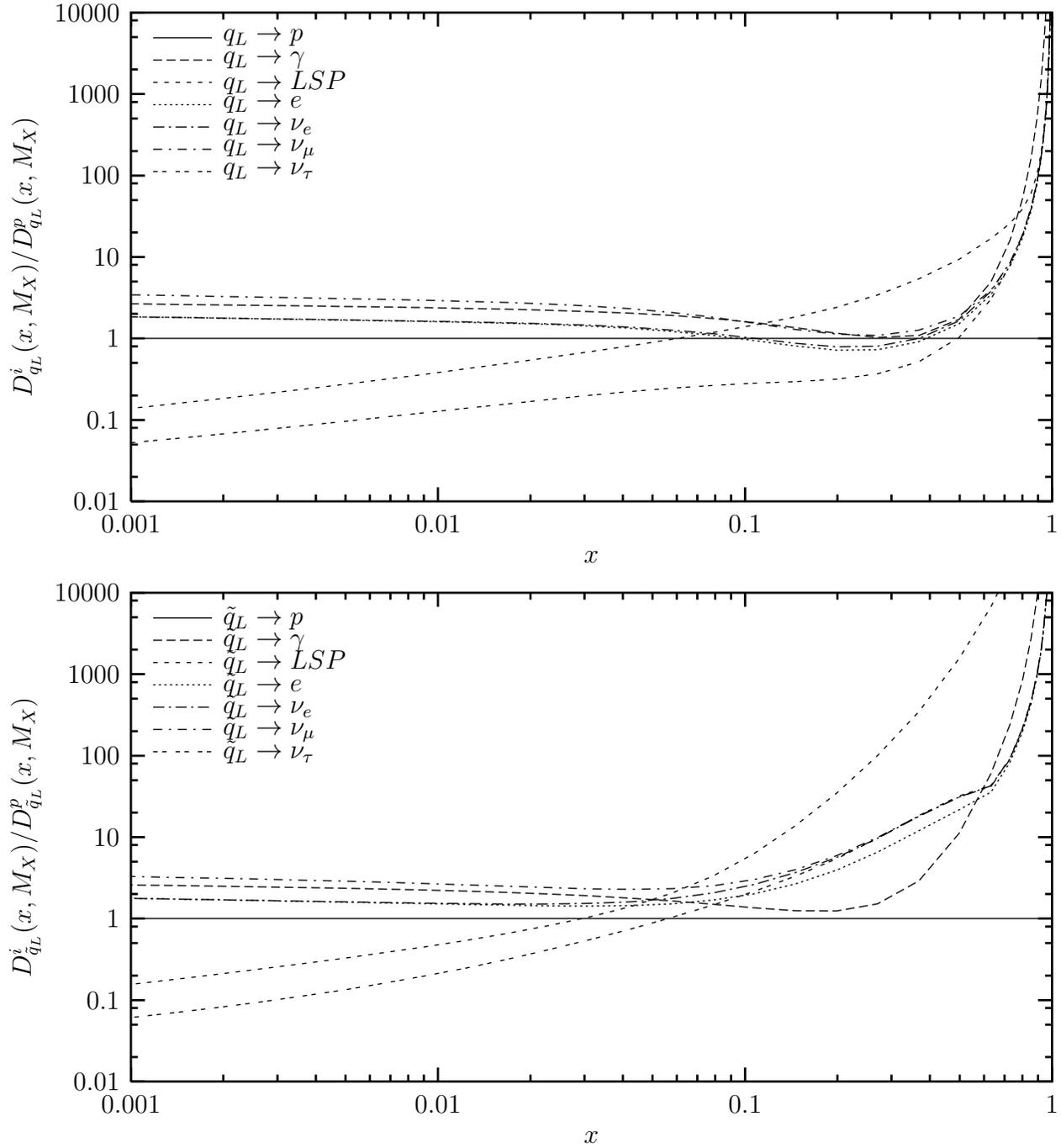

\input{Figures/Low_G_ratio_uL.tex}
\input{Figures/Low_G_ratio_uL_.tex}
\caption{Ratios of FFs $D_I^h/D_I^p$ for different stable particles $h$,
for an initial first or second generation $SU(2)$ doublet quark, $I =
q_L$, (top) or squark, $I = \tilde q_L$ (bottom).} 
\label{ratio_q}
\end{figure}

\begin{figure}[t]
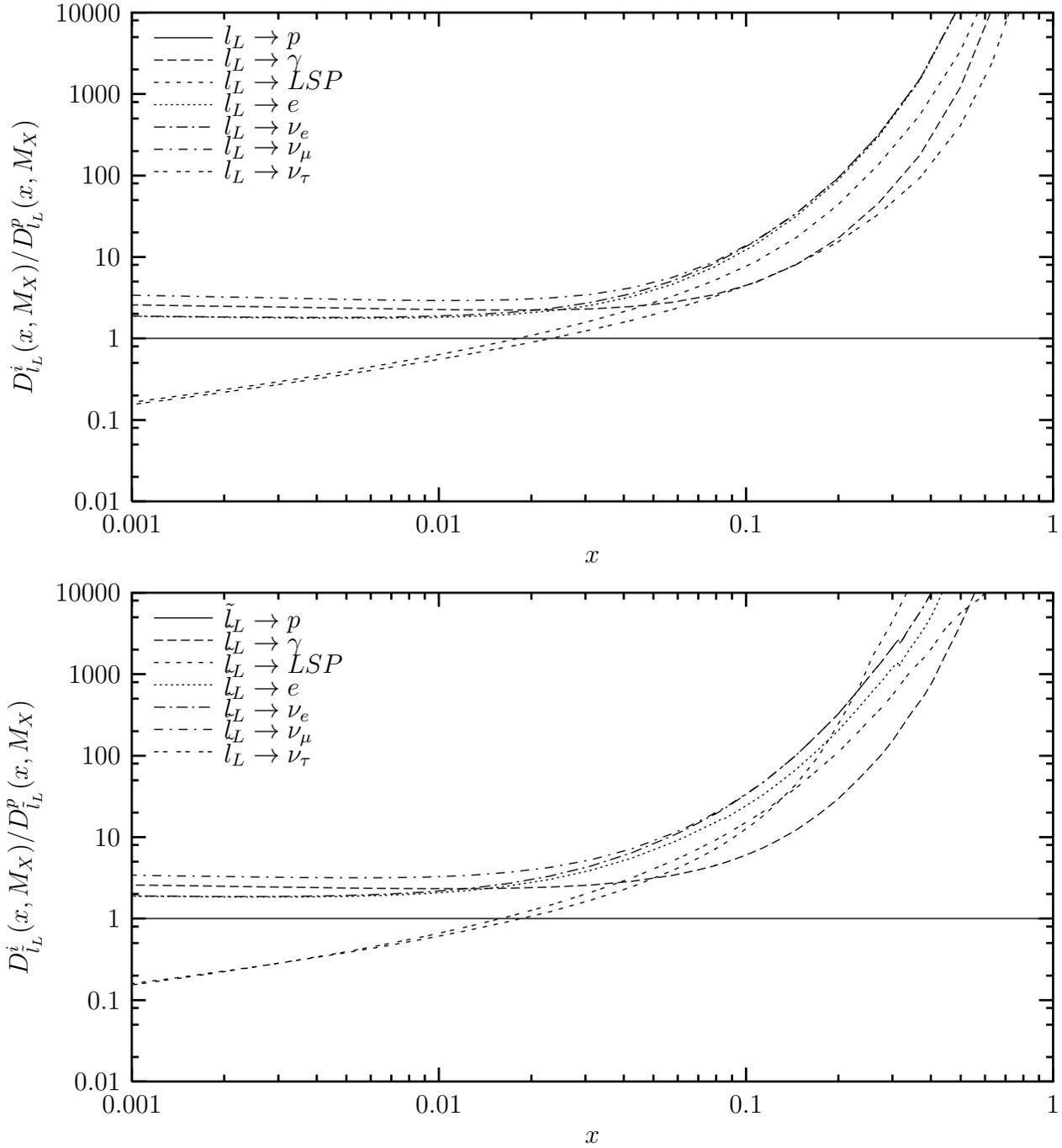

\input{Figures/Low_G_ratio_eL.tex}
\input{Figures/Low_G_ratio_eL_.tex}
\caption{As in fig.~\ref{ratio_q}, but for initial first or second
generation $SU(2)$ doublet lepton, $I = l_L$, (top) or slepton, $I =
\tilde l_L$ (bottom).}
\label{ratio_l}
\end{figure}

Taking the ratios of the different FFs renders some features more
evident, as can be seen from figs.~\ref{ratio_q} and \ref{ratio_l}.
First of all, in the low $x$ region most FFs show the same power law
behavior, and the ratios become quite independent of the initial
particle. The exceptions are the FFs into the LSP and $\nu_\tau$. This
comes from the fact that the LSP flux as well as most of the
$\nu_\tau$ are produced in the perturbative cascade above 1 TeV and in
the following decays of the heavy particles of the spectrum; they
receive no contribution from the decays of light hadrons, although the
$\nu_\tau$ flux receives a minor contribution from the decay of
$b-$flavored hadrons. In contrast, at low $x$ the fluxes of $\nu_e, \
\nu_\mu, \ e$ and $\gamma$ all dominantly originate from the decays of
light hadrons, in particular of charged or neutral pions; we saw in
fig.~\ref{smallx} that the shape of the light hadron spectrum at small
$x$ is essentially determined by the perturbative QCD evolution,
i.e. is independent of the initial particle $I$. In the region $x
\leq 0.01$ we thus predict FFs into $\nu_\mu$ and $\gamma$ to be
approximately 3 to 4 times larger than the FF into protons, while the
FFs into electrons and $\nu_e$ are around twice the FF into
protons. The FFs into LSP and $\nu_\tau$ are five to 20 times
smaller than the one into protons. Note that the LSP flux at small
$x$ from an initial particle is almost the same as that from its
superpartner. It is determined completely by the MSSM cascade, i.e. by
the supersymmetric DGLAP equations, and is almost independent of
details of the supersymmetric spectrum. However, even at $x = 0.01$
the FF into the LSP does retain some sensitivity to the start of the
cascade, i.e. to the initial particle $I$ and hence to the primary $X$
decay mode(s).

At larger values of $x$ the ratios of the FFs depend more and more
strongly on the initial particle. As $x \rightarrow 1$ the proton flux
is always orders of magnitude smaller than the fluxes of all other
stable particles. One reason is that the proton is a composite
particle, i.e. its FF contains a convolution with a non--perturbative
factor which falls as a power of $1-x$ at large $x$. Even before this
convolution the flux of partons (quarks and gluons) that can give rise
to protons is suppressed at large $x$ due to copious emission of
(soft) gluons, whereas the FFs into leptons, photons and LSPs can
remain large at large $x$. If the progenitor $I$ of the cascade is a
strongly interacting superparticle, at large $x$ the FF into the LSP
always dominates over the other FFs. For an initial quark or gluon,
the flux of $\gamma$ (which is the second after LSP for a squark or
gluino) will dominate at large $x$. On the contrary, in the case of an
initial lepton, $W$, $B$ or $H_i$, the strongest fluxes will be
leptonic ones, the exact order depending of the initial particle.
Moreover, for an initial (s)lepton, the fluxes will be significantly
higher at high $x$ (and hence smaller at low $x$, because of energy
conservation) than for strongly interacting (super)particles or Higgs
bosons. Finally, an initial $B$ or $\tilde B$ has a $\delta-$peak at
$x=1$ (not visible in the figures) in $D_B^\gamma$ and $D_{\tilde
 B}^{\rm LSP}$, respectively, in addition to a smooth component that
vanishes as $x \rightarrow 1$. This behavior reflects the inability of
$B$ or $\tilde B$ to radiate a boson, i.e. there are no splitting
processes $B \rightarrow B + X$ or $\tilde B \rightarrow \tilde B +
X$.

\clearpage

\clearpage

\subsection{Energy distribution between the final stable particles}

In the following tables we show the total energy carried per each type
of particle at the end of the cascade, depending on the progenitor of
the cascade, for the same set of SUSY parameters as in Sec.~3.1. As
stated earlier, we are able to verify energy conservation up to at
most a few per mille at each step of the cascade, including its very
end. We see that the ``lost'' energy is somewhat larger for
(s)quarks, gluons and gluinos than for (s)leptons. This is due to
numerical artefacts. The biggest numerical uncertainties arise from
the Runge--Kutta method.\footnote{For practical reasons, we used a
  fixed virtuality step in this algorithm, which we had to keep
  reasonably large, the whole program being already quite
  time--consuming (see chapter~\ref{chap:code}). In the worst
  cases, our choice of the virtuality step leads to errors of the
  order of a few per mille; such a precision is certainly sufficient
  for our purposes.}

Note that even for an initial quark or gluon, more than 35\% of the
energy is carried by the electromagnetic channels (electrons plus
photons), while neutrinos carry about 40\%; in this case most of these
fluxes originate from the decays of light hadrons, chiefly pions. The
corresponding numbers for superparticles are slightly smaller, the
difference being made up by the increased energy fraction carried by the
LSP (at large $x$); an initial $SU(2)$ singlet squark leads to a
higher energy fraction in LSPs, since $SU(2)$ singlet sfermions
usually decay directly into the LSP, which is Bino--like for our
choice of parameters, whereas $SU(2)$ doublet sfermions preferentially
decay via a cascade involving $\tilde \chi_2^0$ or $\tilde
\chi_1^\pm$. 

Lepton--induced showers have a far smaller photon component, but now
an even larger fraction of the energy is carried by electrons and/or
neutrinos, while protons carry at most 2\% of the primary's energy. In
this case the difference between an initial particle and its
superpartner is much larger than in case of strongly interacting
particles, since a much higher fraction of an initial slepton's energy
goes into LSPs, due to the reduced perturbative shower and shorter
superparticle decay cascades. This also explains why more than 70\% of
the energy of an initial $B$ ($\tilde B$) goes into photons (LSPs).
The energy fractions for an initial $SU(2)$ gauge or Higgs boson
resemble those for a quark (with the exception of an increased
$\nu_\tau$ component, which is however washed out by neutrino
oscillations), although the shapes of the corresponding FFs differ
quite dramatically. The energy fraction carried by protons is always
quite small. Pions are created much more abundantly in the
non--perturbative hadronization, and decay into leptons (2/3) and
photons (1/3). As noted earlier, this explains the regularity and the
features of the small $x$ behavior.

\begin{table}[t]
\fontsize{10pt}{1ex}
\begin{center}
\begin{tabular}{|c||c|c|c|c|c|c|c|c|c|c|c|c|}\hline
init part $\,\,\rightarrow$
&\rule[-3mm]{0mm}{8mm} $q_L$ & $\tilde{q}_L$ & $u_R$ & $\tilde{u}_R$ &
$d_R$ & $\tilde{d}_R$ & $t_L$ & $\tilde{t}_L$ & $t_R$ & $\tilde{t}_R$
& $b_R$ & $\tilde{b}_R$ \\  
energy [\%] $\,\,\downarrow$&&&&&&&&&&&& \\ \hline
$p$ &\rule[-3mm]{0mm}{8mm} $10.0$ & $8.3$ & $9.1$ & $7.0$ & $11.5$ & $8.4$ & $9.3$ &
$8.0$ & $8.8$ & $7.8$ & $10.3$ & $8.1$ \\ \hline
$\gamma$ & \rule[-3mm]{0mm}{8mm} $22.9$ & $19.1$ & $25.2$ & $19.1$ & $24.1$ &
$18.0$ & $20.5$ & $17.8$ & $22.0$ & $19.0$ & $22.0$ & $18.0$ \\ \hline
$LSP$ & \rule[-3mm]{0mm}{8mm} $5.8$ & $17.8$ & $6.4$ & $28.8$ & $6.1$ & $29.1$ &
$5.9$ & $17.3$ & $5.6$ & $19.0$ & $4.9$ & $19.1$ \\ \hline
$e$ & \rule[-3mm]{0mm}{8mm} $15.7$ & $14.0$ & $15.5$ & $11.7$ & $14.9$ & $11.3$ &
$16.5$ & $14.5$ & $16.4$ & $13.9$ & $16.3$ & $14.1$ \\ \hline
$\nu_e$ & \rule[-3mm]{0mm}{8mm} $15.6$ & $14.0$ & $15.2$ & $11.5$ & $14.7$ &
$11.2$ & $16.4$ & $14.5$ & $16.2$ & $13.8$ & $16.1$ & $13.9$ \\ \hline
$\nu_\mu$ & \rule[-3mm]{0mm}{8mm} $28.0$ & $24.2$ & $27.5$ & $20.8$ & $27.8$ &
$20.9$ & $27.5$ & $24.1$ & $26.9$ & $23.2$ & $27.9$ & $23.5$ \\ \hline 
$\nu_\tau$ & \rule[-3mm]{0mm}{8mm} $1.3$ & $1.8$ & $0.4$ & $0.4$ & $0.3$ & $0.4$ &
$3.0$ & $2.9$ & $3.4$ & $2.6$ & $1.6$ & $2.3$ \\ \hline \hline
sum & \rule[-3mm]{0mm}{8mm} $99.2$ & $99.2$ & $99.3$ & $99.3$ & $99.2$ & $99.2$ &
$99.2$ & $99.2$ & $99.2$ & $99.2$ & $99.1$ & $99.2$ \\ \hline
\end{tabular}
\caption{Energy fractions $\int_0^1 dx \, x D_I^p(x,M_X^2)$ carried by
the stable particles $p$ at the end of the cascade, for initial
(s)quarks of the 1st/2nd and 3rd generations.} 
\label{nrj_quarks}
\end{center}
\end{table}

\begin{table}[b]
\begin{center}
\begin{tabular}{|c||c|c|c|c|c|c|c|c|}\hline
initial (s)particle $\,\,\rightarrow$
&\rule[-3mm]{0mm}{8mm} $l_L$ & $\tilde{l}_L$ & $e_R$ & $\tilde{e}_R$ &
$\tau_L$ & $\tilde{\tau}_L$ & $\tau_R$ & $\tilde{\tau}_R$ \\
energy fraction (in \%)$\,\, \downarrow$&&&&&&&& \\ \hline
$p$ & \rule[-3mm]{0mm}{8mm} $1.2$ & $2.2$ & $0.1$ & $0.1$ & $1.2$ & $2.1$ & $0.1$ & $1.0$ \\ \hline
$\gamma$ & \rule[-3mm]{0mm}{8mm} $4.5$ & $6.4$ & $6.1$ & $5.1$ & $10.4$ & $9.6$ & $20.0$ & $11.3$ \\ \hline
$LSP$ & \rule[-3mm]{0mm}{8mm} $2.6$ & $28.5$ & $2.0$ & $47.6$ & $2.7$ & $30.5$ & $1.8$ & $36.6$ \\ \hline
$e$ & \rule[-3mm]{0mm}{8mm} $29.6$ & $19.2$ & $60.2$ & $31.0$ & $9.1$ & $8.9$ &
$14.3$ & $7.1$ \\ \hline
$\nu_e$ & \rule[-3mm]{0mm}{8mm} $29.6$ & $19.1$ & $15.2$ & $7.9$ & $9.1$ & $8.8$ & $14.1$ & $6.9$ \\ \hline
$\nu_\mu$ & \rule[-3mm]{0mm}{8mm} $31.1$ & $21.9$ & $15.3$ & $8.0$ & $12.9$ &
$12.8$ & $19.5$ & $9.8$ \\ \hline
$\nu_\tau$ & \rule[-3mm]{0mm}{8mm} $1.1$ & $2.2$ & $0.1$ & $0.1$ & $54.3$ & $27.1$ & $30.0$ & $27.1$ \\ \hline \hline
sum\rule[-3mm]{0mm}{8mm} & $99.8$ & $99.7$ & $99.8$ & $99.8$ & $99.8$ & $99.7$ & $99.8$ & $99.7$ \\ \hline
\end{tabular}
\caption{Energy fractions carried by the stable particles at the end
of the cascade, for initial (s)leptons of the 1st/2nd and 3rd generations.}
\label{nrj_leptons}
\end{center}
\end{table}
\clearpage

\begin{center}
\begin{table}[t]
\begin{center}
\begin{tabular}{|c||c|c|c|c|c|c|c|c|c|c|}\hline
init part $\,\,\rightarrow$
& \rule[-3mm]{0mm}{8mm} $B$ & $\tilde{B}$ & $W$ & $\tilde{W}$ & $g$ &
$\tilde{g}$ & $H_1$ & $\tilde{H}_1$ & $H_2$ & $\tilde{H}_2$ \\
energy [\%] $\,\,\downarrow$ &&&&&&&&&& \\ \hline
$p$ & \rule[-3mm]{0mm}{8mm} $1.8$ & $1.5$ & $7.3$ & $6.1$ & $9.8$ & $9.1$ & $8.5$ & $7.0$ & $8.0$ & $5.6$ \\ \hline
$\gamma$ & \rule[-3mm]{0mm}{8mm} $71.6$ & $4.1$ & $17.8$ & $14.2$ & $22.5$ &
$20.7$ & $19.4$ & $16.7$ & $18.9$ & $14.1$ \\ \hline
$LSP$ & \rule[-3mm]{0mm}{8mm} $4.2$ & $76.9$ & $7.0$ & $24.5$ & $8.4$ & $14.0$ & $4.9$ & $18.6$ & $4.9$ & $31.2$ \\ \hline
$e$ & \rule[-3mm]{0mm}{8mm} $7.2$ & $5.7$ & $17.0$ & $14.0$ & $15.2$ & $14.4$ &
$17.2$ & $14.6$ & $17.1$ & $12.4$ \\ \hline
$\nu_e$ & \rule[-3mm]{0mm}{8mm} $5.2$ & $4.0$ & $17.4$ & $14.1$ & $15.0$ & $14.2$ & $17.2$ & $14.6$ & $17.5$ & $12.5$ \\ \hline
$\nu_\mu$ & \rule[-3mm]{0mm}{8mm} $7.6$ & $5.9$ & $26.4$ & $21.6$ & $27.1$ &
$25.4$ & $27.2$ & $22.9$ & $27.1$ & $19.3$ \\ \hline
$\nu_\tau$ & \rule[-3mm]{0mm}{8mm} $2.1$ & $1.7$ & $6.5$ & $4.9$ & $1.0$ & $1.2$ &
$5.1$ & $4.4$ & $6.1$ & $4.2$ \\ \hline \hline
sum & \rule[-3mm]{0mm}{8mm} $99.8$ & $99.8$ & $99.4$ & $99.4$ & $99.1$ & $98.9$ & $99.5$ & $98.9$ & $99.5$ & $99.3$ \\ \hline
\end{tabular}
\caption{Energy fractions carried by the stable particles at the end
 of the cascade, for initial bosons and bosinos.}
\label{nrj_bosons}
\end{center}
\end{table}
\end{center}

\subsection{Dependence on SUSY parameters}
\label{subsec:SUSYdep}

As stated in \cite{bd1}, the general features of our results
described above depend very little on the set of SUSY parameters we
are using. Here we give a more precise analysis of the influence of
different parameters describing the SUSY spectrum. As usual we present
our results as $x^3 \cdot D_I^p(x,M_X^2)$. The multiplication with
the third power of the energy leads to an approximately flat cosmic
ray spectrum for $E \leq 10^{10}$ GeV \cite{reviewSigl}. In our case
it suppresses the small$-x$ region, leading to maxima in the curves at
$x$ between 0.1 and 1.

We first studied the dependence of our results on the overall SUSY
mass scale, by comparing results for two different ISASUSY input mass
scales for scalars and gluinos: $M_{SUSY} \sim 500$ GeV and $1$
TeV. As expected, this change has almost no impact on the final
results, since the details of the decay chains of heavy (s)particles
will depend mostly on the relative ordering of the (s)particle spectrum
(e.g. allowing or preventing some decay modes), rather than on their
absolute mass scale. Moreover, a factor 2 or 3 in the scale where the
MSSM evolution is terminated does not change the FFs much, since the
DGLAP equations describe an evolution which is only logarithmic in the
virtuality.

\begin{figure}[ht]
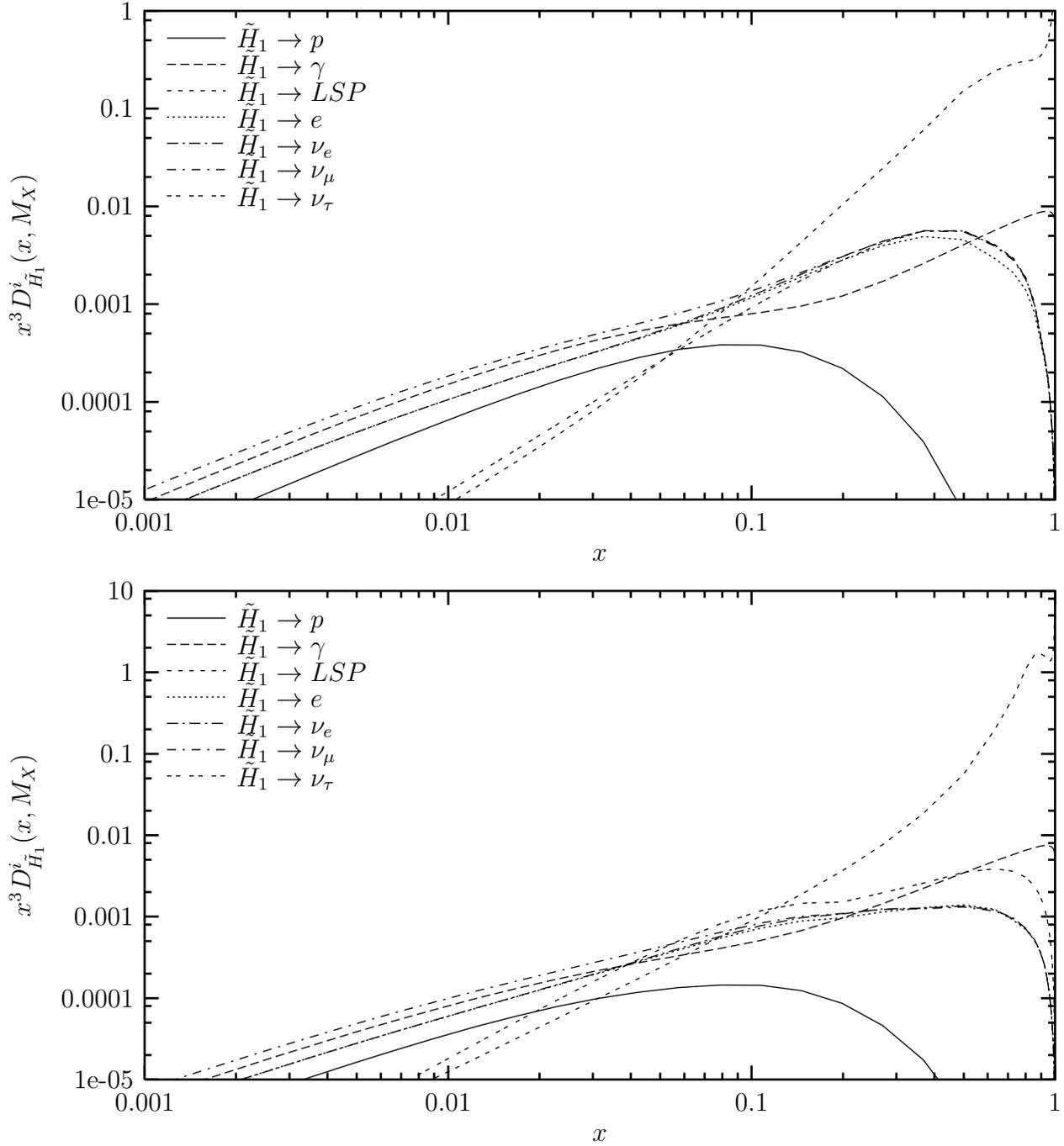

\input{Figures/Low_H_H1_.tex}
\input{Figures/Big_H_H1_.tex}
\caption{FFs into the final stable particles for an initial
$\tilde{H}_1$ for $\tan\beta = 3.6$ (top) or 48 (bottom).}
\label{LB1}
\end{figure}

\clearpage

Next we compared two rather extreme values of $\tan \beta$, namely 3.6
and 48, leaving all dimensionful parameters at the weak scale
unchanged. Once again the effect is very small. The only visible
difference occurs for initial $H_1$ and $\tilde{H}_1$, where the
increase of $\tan \beta$ produces more $\nu_\tau$ at large $x$, as can
be seen in fig.~\ref{LB1}. However, flavor oscillations will
essentially average the three neutrino fluxes between source and
detector, so we expect very little direct dependence of measurable
quantities on $\tan\beta$. The main remaining effect is an increase of
the overall multiplicity by $\sim 30\%$ for an initial $H_1$ or
$\tilde H_1$ in case of large $\tan\beta$, due to the increased shower
activity from the much larger bottom Yukawa coupling. However, the
situation could be different in more constrained models, where the
spectrum is described by a few soft breaking parameters specified at
some high energy scale. In this case a change of $\tan\beta$ generally
changes the sparticle and Higgs spectrum, and can also greatly modify
some branching ratios.

In order to get a feeling for how the various FFs depend on the
relative ordering of the dimensionful parameters describing the SUSY
spectrum, we investigated two rather extreme cases. They resemble two
qualitatively different regions of parameter space in the minimal
supergravity (mSUGRA or CMSSM) model where the thermal LSP relic
density is acceptably small \cite{msugra}.\footnote{In our case $X$
particles could contribute significantly to the Dark Matter; in this
scenario, which is realized only for a small region of the total
allowed $M_X, \tau_X$ plane, the upper bound on the LSP relic density
would have to be tightened accordingly, but the allowed regions of
parameter space would be qualitatively the same.} In the first
scenario the LSP $\tilde \chi_1^0$ has small mass splitting to the
lightest stau, \stauone. We took the following values for the relevant
soft breaking parameters: $m_{\tilde q} \simeq m_{\tilde g} = 1$ TeV
for all squarks, $m_{\tilde l_L} = 250$ GeV for all $SU(2)$ doublet
sleptons, $m_{\tilde l_R} \simeq 200$ GeV for $l = e, \, \mu$ but
reduced $m_{\tilde \tau_R}$ so that $m_{\tilde \tau_1} = m_{\tilde
\chi_1^0} + 13$ GeV $= 163$ GeV; note that in mSUGRA one needs large
mass splitting between squarks and sleptons if the LSP mass is to be
close to the \stauone\ mass. The physical sfermion masses receive
additional contributions from $SU(2) \times U(1)_Y$ symmetry breaking,
and, in case of the third generation, from mixing between singlet and
doublet sfermions; in case of $\tilde t$, contributions $+m_t^2$ to
the diagonal entries of the mass matrix also have to be added. Our
choice $\mu = 1$ TeV together with the assumption of gaugino mass
unification ensures that the LSP is an almost pure bino.

In contrast, in the second scenario we took $\mu = -100$ GeV,
$m_{\tilde g} = 800$ GeV, so that the LSP is dominated by its higgsino
components, although the bino component still contributes $\sim
20\%$. In this scenario we took $m_{\tilde q} = 1.5$ TeV for all
squarks and $m_{\tilde l} = 1.2$ TeV for all sleptons, since in mSUGRA
large scalar masses are required if the LSP is to have a large
higgsino component. We took CP--odd Higgs boson mass $m_A = 1$ TeV in
both cases, and $\tan\beta = 3.6$; we just saw that the latter choice
is not important for us. In the following we will refer to these two
choices as the ``gaugino'' and ``higgsino'' set of parameters,
respectively.

\begin{figure}
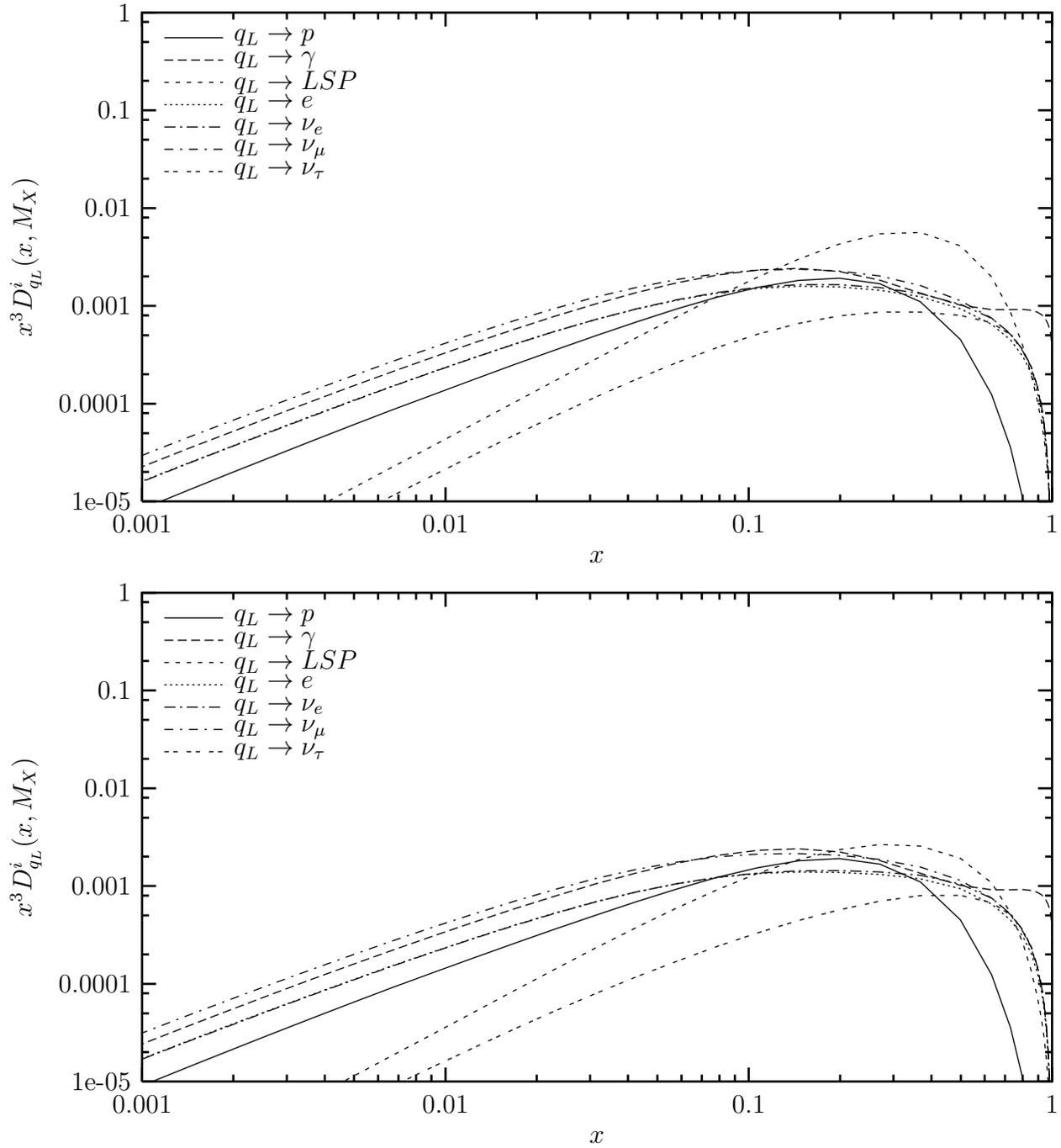

\input{Figures/Low_Ga_uL.tex}
\input{Figures/Low_Hi_uL.tex}
\caption{FFs into the final stable particles for an initial
$SU(2)$ doublet quark of the first or second generation $q_L$, for
the gaugino (top) and higgsino (bottom) set of parameters.}
\label{GH_uL}
\end{figure}

In Fig.~\ref{GH_uL} we compare the FFs of an initial first or second
generation $SU(2)$ doublet quark $q_L$ for these two scenarios. The
main difference occurs in the FF into the LSP, which is significantly
softer for the higgsino set. The reason is that most heavy
superparticles (sfermions and gluinos) preferentially decay into
gaugino--like charginos and neutralinos, which have much larger
couplings to most squarks than the higgsino--like states do. These
gaugino--like states are the lighter two neutralinos and lighter
chargino in case of the gaugino set, but they are the heavier $\tilde
\chi$ states for the higgsino set. The supersymmetric decay chains
therefore tend to be longer for the higgsino set, which means that
less energy goes into the LSP produced at the very end of each chain.

\begin{figure}
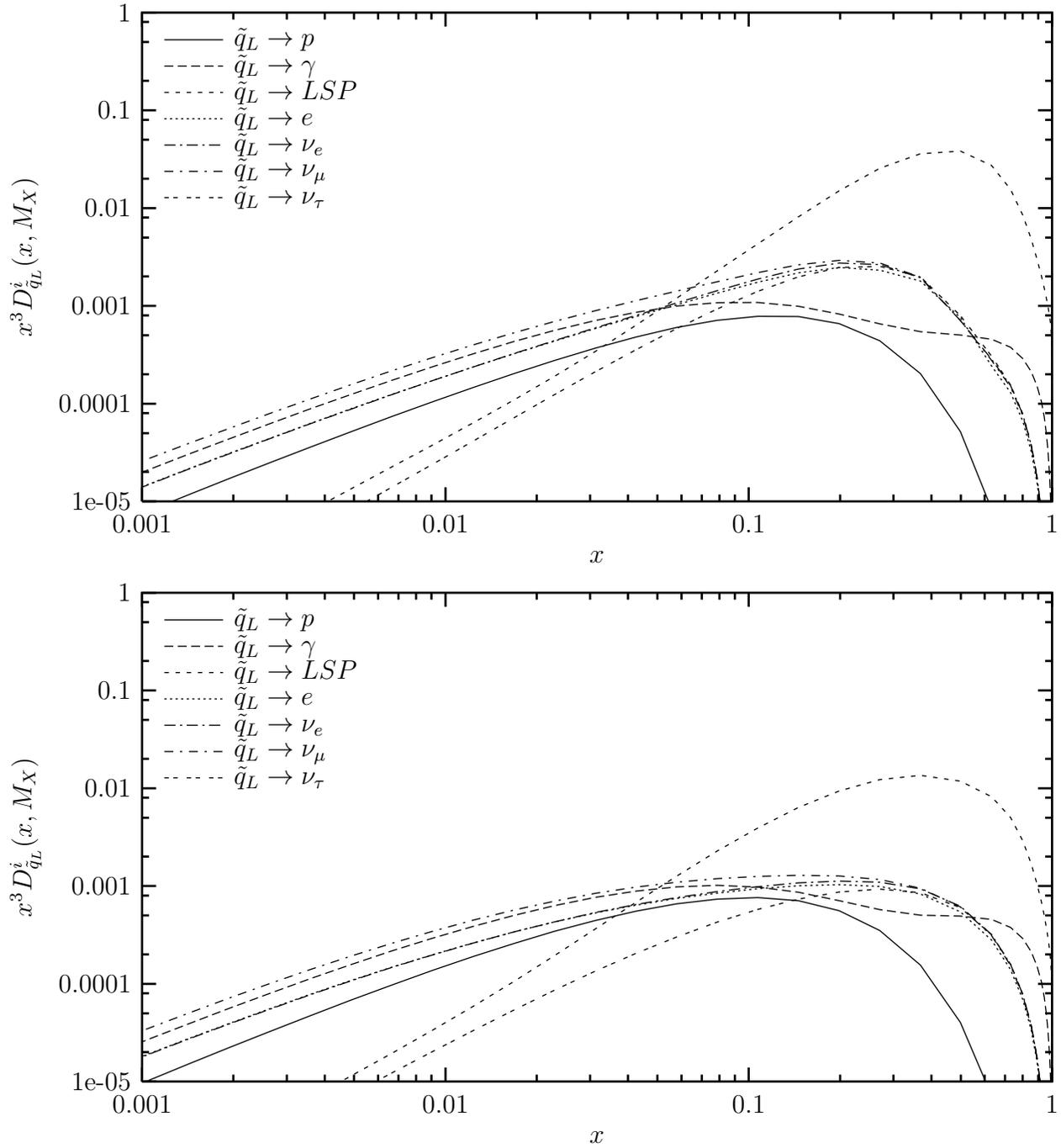

\input{Figures/Low_Ga_uL_.tex}
\input{Figures/Low_Hi_uL_.tex}
\caption{FFs into the final stable particles for an initial
$SU(2)$ doublet squark of the first or second generation $\tilde q_L$,
for the gaugino (top) and higgsino (bottom) set of parameters.}
\label{GH_uL_}
\end{figure}

Fig.~\ref{GH_uL_} shows the same comparison for an initial first or
second generation $SU(2)$ doublet squark $\tilde q_L$. Not
surprisingly, the FFs of a squark are more sensitive to details of the
sparticle spectrum than those of a quark. In particular, in addition
to the reduced FF into the LSP, we now also see that the FFs into
neutrinos and electrons are suppressed for the higgsino set relative to
the gaugino set. This is partly again due to the longer decay chains,
which pushes these FFs towards smaller $x$ where the $x^3$
normalization factor suppresses them more strongly, and partly because
the branching ratios for leptonic decays of the $SU(2)$ gaugino--like
$\tilde \chi$ states are smaller here than for the gaugino set, which
implies that fewer leptons are produced in sparticle decays. On the
other hand, the longer decay chains and larger hadronic branching
ratios for $\tilde \chi$ decays are characteristic of the higgsino set
lead to an increase of the total multiplicity of 25\% or so, as can be
seen from the FFs at small $x$; of course, in this region the ratios
of these FFs again approach their universal values, as discussed in
Sec.~3.1.

\begin{figure}
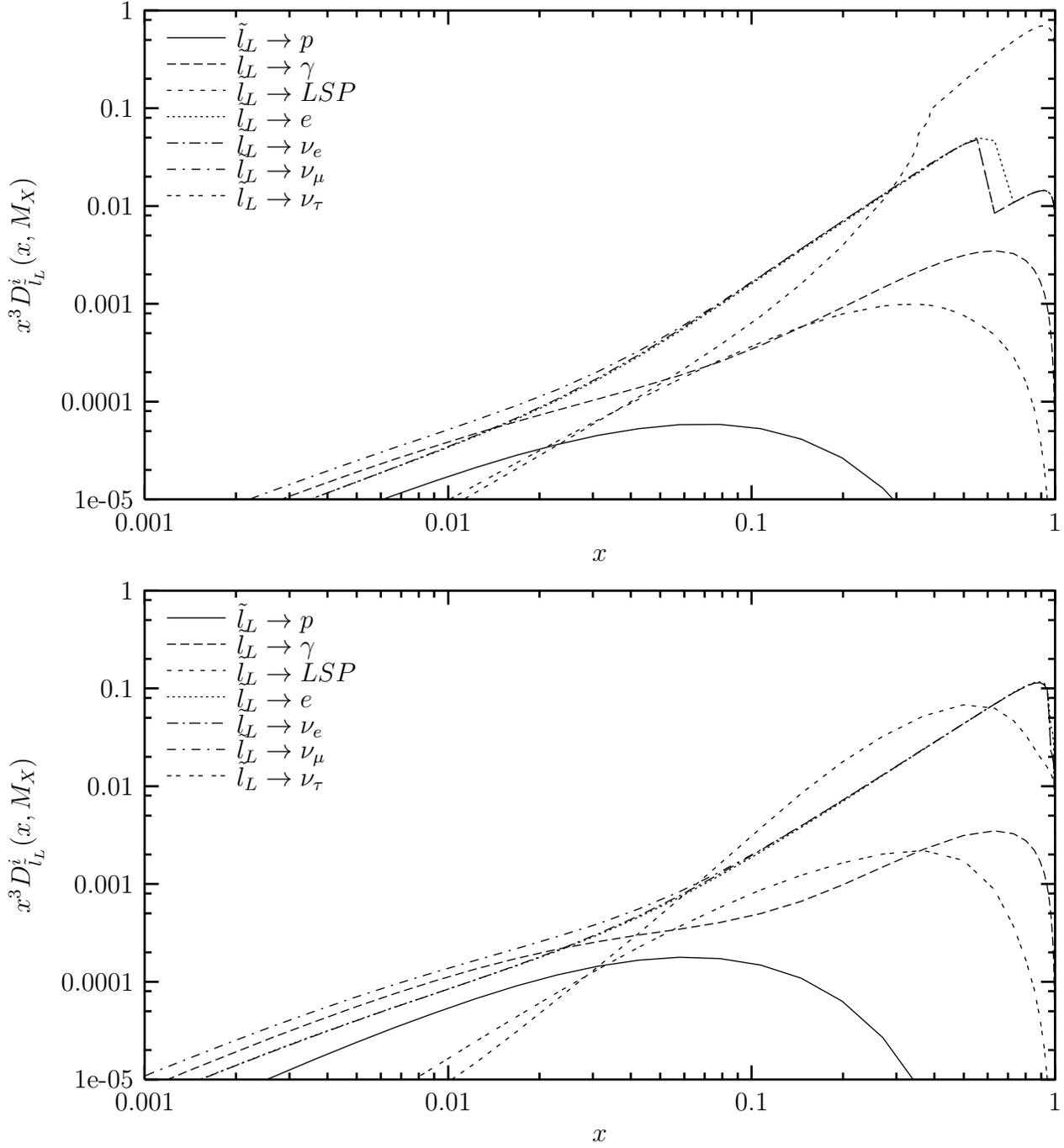

\input{Figures/Low_Ga_eL_.tex}
\input{Figures/Low_Hi_eL_.tex}
\caption{FFs into the final stable particles for an initial
$SU(2)$ doublet slepton of the first or second generation $\tilde l_L$,
for the gaugino (top) and higgsino (bottom) set of parameters.}
\label{GH_eL_}
\end{figure}

If the initial particle is strongly interacting, the rapid evolution
of the shower ensures that the generalized FFs (\ref{boundary_MSSM})
describing the evolution between $M_{\rm SUSY}$ and $M_X$ essentially
vanish at $x \simeq 1$, i.e. all spectra are smooth. In contrast, if
the initial particle $I$ has only weak interactions, a significant
$\delta-$peak will remain at $x = 1$ in the generalized FF $\tilde
D_I^I$. If $I$ is a superparticle or Higgs boson, the decays of $I$
can therefore lead to sharp edges in the final FFs. This is
illustrated in Fig.~\ref{GH_eL_}, which shows the FFs for an initial
first or second generation $SU(2)$ doublet sleptons $\tilde l_L$. The
parameters of the gaugino set are chosen such that $\tilde l_L$
sleptons can only decay into $l \ +$ LSP. The decays of the $\tilde
l_L$ which survive at $x=1$ therefore lead to edges in the FFs into
$e, \, \nu_e$ and $\nu_\mu$; recall that $\tilde l_L$ is an equal
mixture of $\tilde e_L, \, \tilde \nu_e, \, \tilde \mu_L$ and $\tilde
\nu_\mu$. The edge in the FF into $e$ occurs at a somewhat larger
value of $x$ than those in the FFs into $\nu_{e,\mu}$, since after
$SU(2)$ symmetry breaking the charged members of the slepton doublets
are a little heavier than the neutral ones; the decay $\tilde e_L
\rightarrow e \tilde \chi_1^0$ therefore deposits more energy in the
electron than $\tilde \nu_e \rightarrow \nu_e \tilde \chi_1^0$
deposits in the neutrino. However, in both cases the bulk of the
energy goes into the LSP, which is rather close in mass to the
slepton. This is quite different for the higgsino set, where sleptons
are much heavier than all $\tilde \chi$ states. As a result, almost
the entire slepton energy can go into the decay lepton, leading to FFs
into $e, \, \nu_e$ and $\nu_\mu$ that are peaked very near $x=1$
(after multiplying with $x^3$). Furthermore, since most sleptons now
first decay into heavier $\tilde \chi$ states rather than directly
into $\tilde \chi_1^0$, the FF into the LSP is much softer than for
the gaugino set. Finally, the effect of the longer decay chains of
SUSY particles on the overall multiplicity now amounts to about a
factor of 2, and is thus much more pronounced than for initial
squarks; this can be explained by the reduced importance of the shower
evolution in case of only weakly interacting primaries.

\begin{figure}
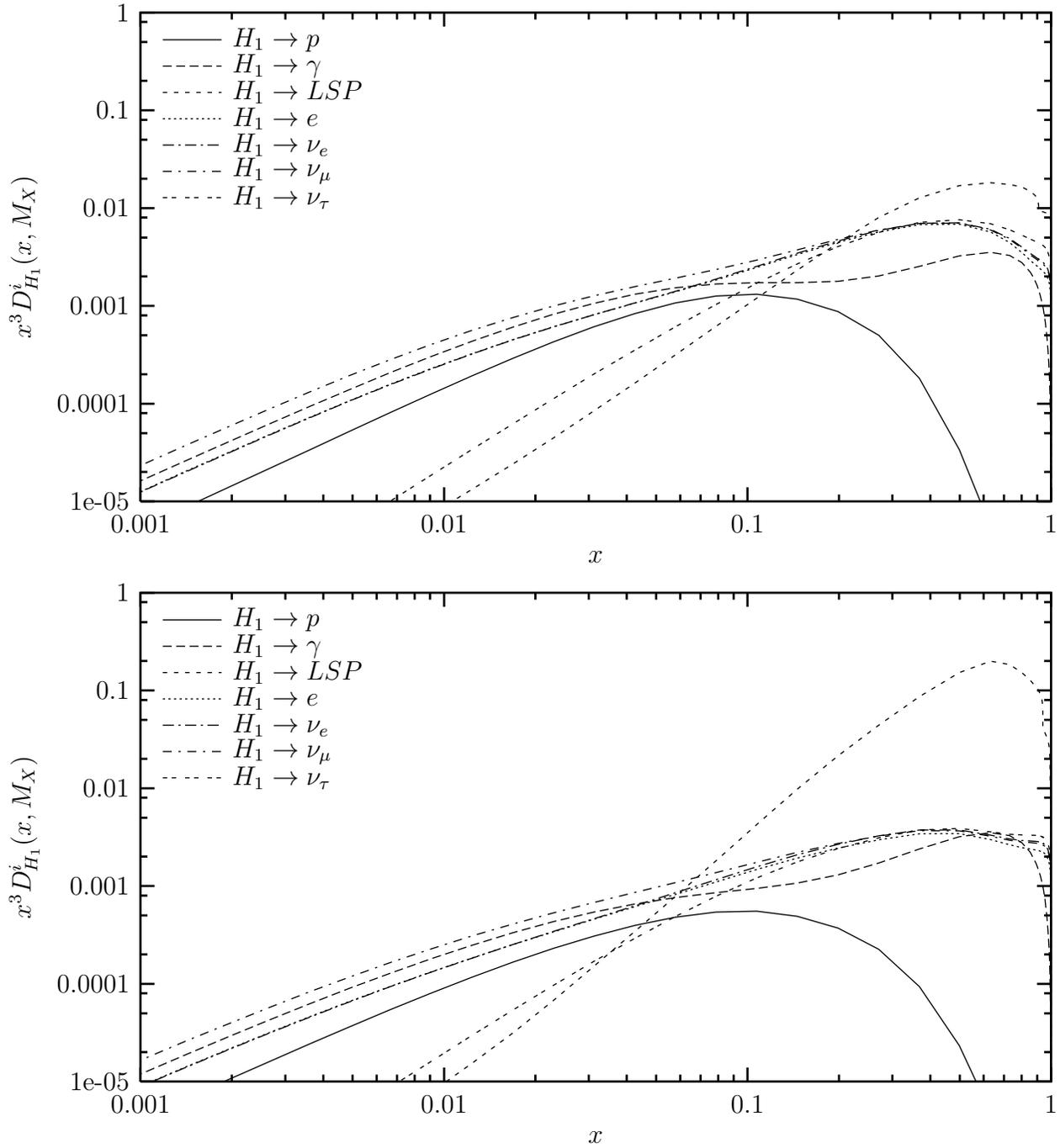

\input{Figures/Low_Ga_H1.tex}
\input{Figures/Low_Hi_H1.tex}
\caption{FFs into the final stable particles for an initial $H_1$
Higgs doublet, for the gaugino (top) and higgsino (bottom) set of
parameters.}
\label{GH_H1}
\end{figure}

Fig.~\ref{GH_H1} shows that in case of an initial $H_1$ Higgs doublet,
the role of the two parameter sets is in some sense reversed. Recall
that we chose $\tan\beta > 1$ and $m_A \gg M_Z$. In that case the
heavy Higgs bosons mostly consist of various components of the $H_1$
doublet, with only small admixture of $H_2$; see eq.(\ref{smtrafo}) in
Appendix~\ref{app:transfo}. As usual with only weakly interacting
primaries, the generalized FF $D_{H_1}^{H_1}$ remains sizable at $x=1$
even at scale $M_X$. In the higgsino set, the dominant decay modes of
the heavy Higgs bosons involve a gaugino and a higgsino, leading to a
large FF into the LSP in this case. Since in the gaugino set the mass
of the higgsino--like $\tilde \chi$ states is very close to the mass
of the heavy Higgs bosons, these supersymmetric decay modes are closed
for the heavy Higgs bosons in this case, which instead predominantly
decay into top quarks, with decays into $b$ quarks and $\tau$ leptons
also playing some role. The fragmentation and decay products of these
heavy quarks lead to a significantly larger FF into protons in the
gaugino region; semi--leptonic $t$ and $b$ decays as well as the
$\tau$ decays also lead to enhanced FFs into electrons and neutrinos
for the gaugino set. Finally, the hadronic showers initiated by the
decay products of the top quarks as well as by the $b$ quarks produced
directly in the decays of Higgs bosons raise the total multiplicity
for the gaugino set to a value which is slightly larger than that for
the higgsino set.

\clearpage

\clearpage

\begin{figure}
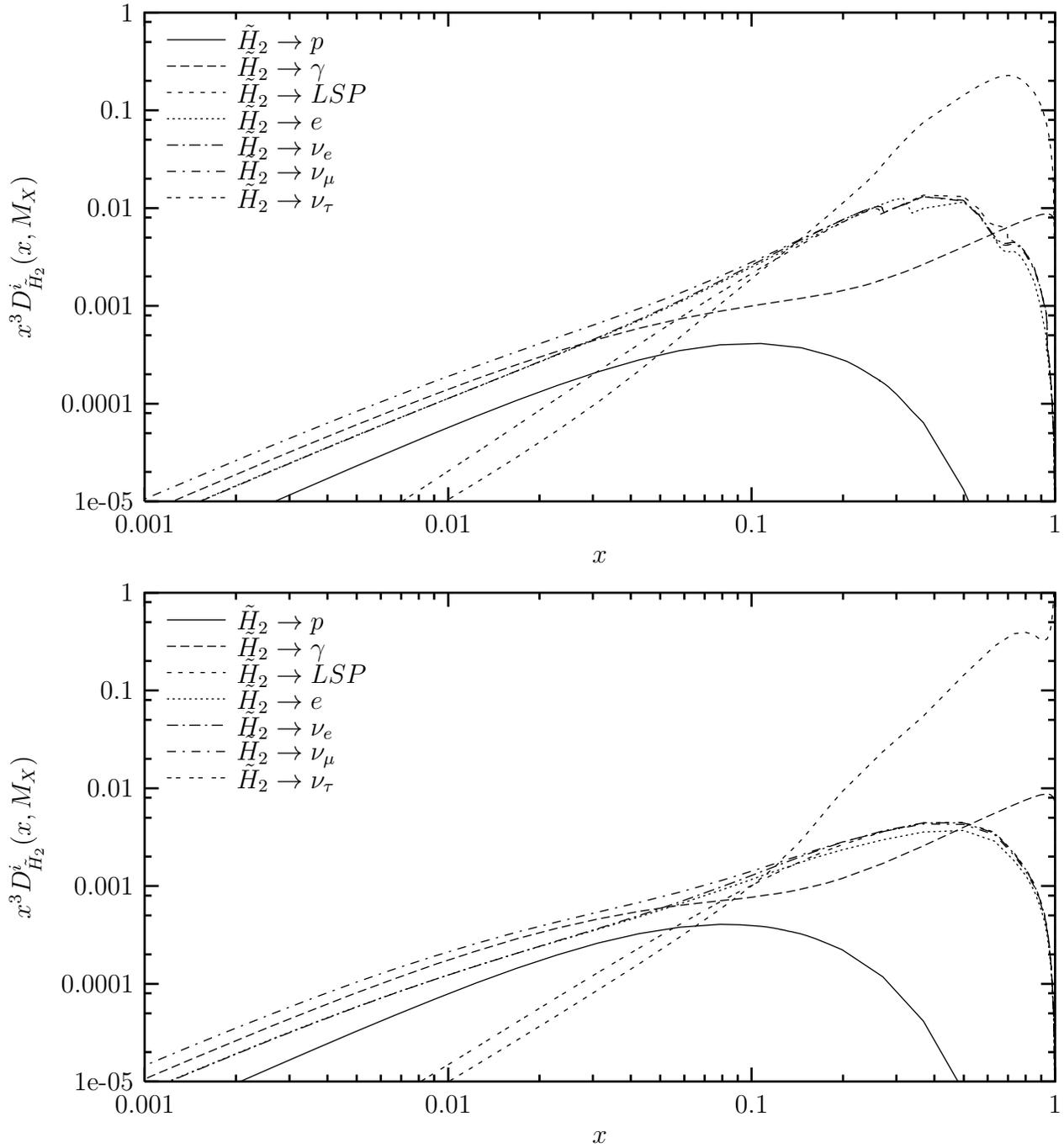

\input{Figures/Low_Ga_H2_.tex}
\input{Figures/Low_Hi_H2_.tex}
\caption{FFs into the final stable particles for an initial $\tilde H_2$
higgsino doublet, for the gaugino (top) and higgsino (bottom) set of
parameters.}
\label{GH_H2_}
\end{figure}

As final example we compare the FFs of an initial $\tilde H_2$
higgsino doublet in Fig.~\ref{GH_H2_}. Here we again find a larger FF
into the LSP for the higgsino set, including a peak at $x=1$. In this
case this is simply a reflection of the large $\tilde H_2^0$ component
of the LSP. On the other hand, in case of the gaugino set $\tilde H_2$
projects almost exclusively into the heavier $\tilde \chi$ states,
which have many two--body decay modes into sleptons and leptons. This
explains the relative enhancement at large $x$ of the FFs into leptons
that we observe for the gaugino set, as well as the structures in
these FFs. On the other hand, the longer sparticle decay chains again
imply a somewhat larger overall multiplicity for the higgsino
set. These decays of heavy sparticles are important here since the
large top Yukawa coupling of $\tilde H_2$ initiates a significant
parton shower in this case, where numerous superparticles are
produced. This is quite different for an initial $\tilde H_1$ at small
$\tan\beta$ (not shown), where we find a {\em smaller} overall
multiplicity for the higgsino set, since the number of produced
superparticles remains small, and the initial particle $\tilde H_1$
has a longer decay chain for the gaugino set.

Altogether we see that the SUSY spectrum can change the final FFs, and
thus the final spectra of $X$ decay products, significantly. Generally
this effect is stronger for an initial superparticle or heavy Higgs
boson than for an SM particle, and stronger for only weakly
interacting particles than for those with strong interactions.
However, with the exception of the FFs into the LSP, the variation is
usually not more than a factor of two, and often much less. The
dependence of the $X$ decay spectra on SUSY parameters can therefore
be significant for detailed quantitative analyses, but this dependence
is always weaker than the dependence on the primary $X$ decay mode(s).

\clearpage

\subsection{Coherence effects at small $x$: the MLLA solution}
\label{sec:MLLA}

So far we have used a simple power law extrapolation of the hadronic
(non--perturbative) FFs at small $x$. This was necessary since the
original input FFs of ref.\cite{Poetter} are valid only for $x \geq
0.1$. As noted earlier, we expect our treatment to give a reasonable
description at least for a range of $x$ below 0.1. However, at very
small $x$, color coherence effects should become important
\cite{Basics_of_QCD}. These lead to a flattening of the FFs, giving a
plateau in $x D(x)$ at $x_{\rm plateau} \sim \sqrt{Q_{\rm had}/M_X}
\sim 10^{-8}$ for $M_X = 10^{16}$ GeV. One occasionally needs the FFs
at such very small $x$. For example, the neutrino flux from $X$ decays
begins to dominate the atmospheric neutrino background at $E \sim
10^5$ GeV \cite{neutrinos, bdhh1}, corresponding to $x \sim 10^{-11}$
for our standard choice $M_X \sim 10^{16}$ GeV. In this subsection we
therefore describe a simple method to model color coherence effects in
our FFs.

This is done with the help of the so--called limiting spectrum derived
in the modified leading log approximation. The key difference to the
usual leading log approximation described by the DGLAP equations is
that QCD branching processes are ordered not towards smaller
virtualities of the particles in the shower, but towards smaller
emission angles of the emitted gluons; note that gluon radiation off
gluons is the by far most common radiation process in a QCD
shower. This angular ordering is due to color coherence, which in the
conventional scheme begins to make itself felt only in NLO (where the
emission of two gluons in one step is treated explicitly). It changes
the kinematics of the parton shower significantly. In particular, the
requirement that emitted gluons still have sufficient energy to form
hadrons strongly affects the FFs at small $x$. For sufficiently high
initial shower scale and sufficiently small $x$ the MLLA evolution
equations can be solved explicitly in terms of a one--dimensional
integral \cite{Basics_of_QCD}. This essentially yields the modified FF
describing the perturbative gluon to gluon fragmentation, $\tilde D_g^g$
in the language of eq.(\ref{split}). In order to make contact with
experiment, one makes the additional assumption that the FFs into
hadrons coincide with $\tilde D_g^g$, up to an unknown constant; this
goes under the name of ``local parton--hadron duality'' (LPHD)
\cite{LPHD}. Here we use the fit of this ``limiting spectrum'' in
terms of a distorted Gaussian \cite{distorted_gaussian}, which
(curiously enough) seems to describe LEP data on hadronic FFs somewhat
better than the ``exact'' MLLA prediction does. It is given by
\beq \label{gaussian}
F_i(\xi,\tau) \equiv xD_i(x,Q) =
\frac {\bar{n}_i} {\sigma \sqrt{2\pi} } \exp {\left[ \frac {1} {8} k +
\frac{1}{2} s \delta - \frac{1}{4} (2+k) \delta^2 + \frac{1}{6} s
\delta^3 + \frac{1}{24} k \delta^4 \right]},
\eeq
where $\bar{n}_i$ is the average multiplicity. The other quantities
appearing in eq.(\ref{gaussian}) are defined as follows:
\beqa \label{gauss_coef}
\tau &=& \log{\frac{Q}{\Lambda}} \,,\nonumber\\
\xi &=& \log{\frac{1}{x}} \,,\nonumber\\
\bar{\xi} &=& \frac{1}{2} \tau \left( 1 + \frac {\rho} {24}
\sqrt{\frac{48}{\beta \tau} } \right) + {\cal O}(1)\,,\nonumber\\
\sigma &=& \langle (\xi - \bar{\xi})^2 \rangle^{1/2} =
\sqrt{ \frac{1}{3} } \left( \frac{\beta}{48} \right)^{1/4} \tau^{3/4}
\left( 1 - \frac{1}{64} \sqrt{ \frac{48\beta} {\tau} } \right) + {\cal
O} (\tau^{-1/4}) \,,\nonumber\\
\delta &=& \frac{ \xi -\bar{\xi}} {\sigma} \,,\nonumber\\
s &=& \frac{ \langle (\xi - \bar{\xi})^3 \rangle }{\sigma^3} =
-\frac {\rho}{16} \sqrt{ \frac{3} {\tau} } \left( \frac {48}
{\beta\tau} \right)^{1/4} + {\cal O}(\tau^{-5/4}) \,,\nonumber\\
k &=& \frac{ \langle (\xi - \bar{\xi})^4 \rangle }{\sigma^4} =
- \frac {27} {5\tau} \left( \sqrt{ \frac{1}{48} \beta \tau } - \frac
{1}{24} \beta \right) + {\cal O}(\tau^{-3/2}) \,,
\eeqa
where $\beta$ is the coefficient in the one--loop beta--function of
QCD and $\rho = 11 + 2 N_f/27$, $N_f$ being the number of active
flavors. Eqs.(\ref{gaussian}) and (\ref{gauss_coef}) have been derived
in the SM, where $\beta = 11 - 2 N_f/3$. Following
ref.\cite{limiting_spectrum} we assume that it remains valid in the
MSSM, with $\beta = 3$ above the SUSY threshold $M_{\rm SUSY}$ and
$\rho = 11 + 8/9$. Note that we do not attempt to model the transition
from the full MSSM to standard QCD here; indeed, we do not know of an
easy way to do this, since the limiting spectrum cannot be written as
a convolution of two other spectra. On the other hand, the position
$\bar\xi$ of the plateau depends only on $\sqrt{\beta}$, and only via
the second term, which is suppressed by a factor $\sqrt{\tau} \sim
6.5$, whereas the parameters $\sigma$ and $s$ describing the behavior
in the vicinity of the maximum depend in leading order in $\tau$ only
on $\beta^{1/4}$. Finally, the coefficient $\rho$ is very similar in
the SM and MSSM. We therefore expect the error we make by ignoring the
transition from MSSM to SM to be smaller than the inherent accuracy of
eq.(\ref{gaussian}).

When comparing MLLA predictions with experiments, the overall
normalization $\bar n_i$ (which depends on energy) is usually taken
from data. We cannot follow this approach here, since no data with $Q
\sim M_X$ are available. Moreover, usually MLLA predictions are
compared with inclusive spectra of all (charged) particles. We need
separate predictions for various kinds of hadrons, and are therefore
forced to make the assumption that all these FFs have the same
$x-$dependence at small $x$. This is perhaps not so unreasonable; we
saw above that the DGLAP evolution predicts such a universal
$x-$dependence at small $x$. We then match these analytic solutions
(\ref{gaussian}), (\ref{gauss_coef}) with the hadronic FFs $D_I^h$ we
obtained from DGLAP evolution and our input FFs at values $x_0^h$,
where for each hadron species $h$ the matching point $x_0^h$ and the
normalization $\bar n_h$ are chosen such that the FF and its first
derivative are continuous; we typically find $x_0 \sim 10^{-4}$. Note
that this matching no longer allows to respect energy conservation
exactly. However, since the MLLA solution begins to deviate from the
original FFs only at $x \sim 10^{-7}$, the additional ``energy
losses'' are negligible.

\begin{figure}
\input{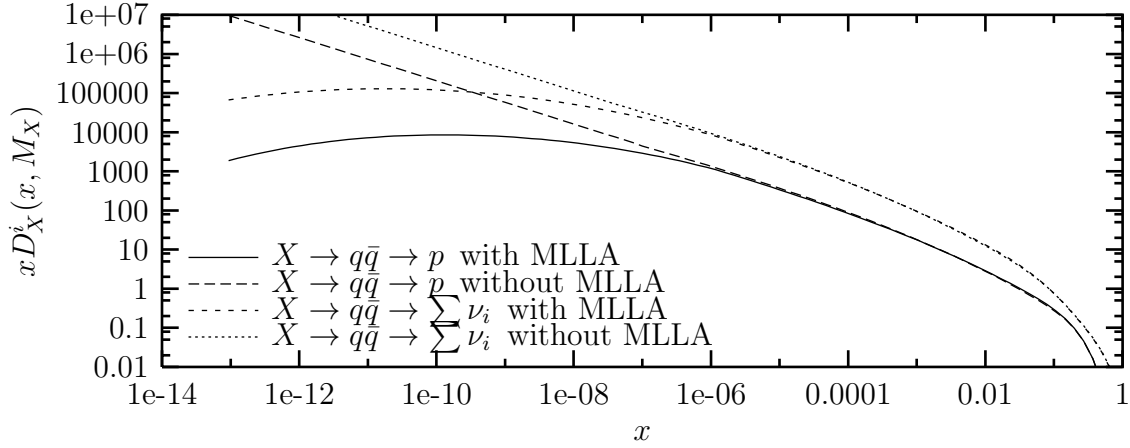}
\caption{Comparison between the MLLA solution and our results without
coherence effects, for the final proton and neutrino spectra. We
assume that $X$ undergoes two--body decay into $q_L \bar q_L$.}
\label{MLLA}
\end{figure}

Some results of our MLLA treatment are shown in Fig.~\ref{MLLA}. Here
the ``non--MLLA'' curves have been obtained by extrapolating our
numerical results described earlier, which extend ``only'' to $x =
10^{-7}$, by using simple power--law fits. We see that at $x \sim
10^{-11}$ the FFs are suppressed by about two orders of magnitude, but
the effect diminishes quickly at larger values of $x$. Note that the
FFs into protons and into neutrinos have slightly different shapes in
the small$-x$ region. By assumption the FFs have the same shape for
all {\em hadrons}; however, in going from the spectrum of pions and
kaons to the neutrino spectrum, several additional convolutions are
required, which shift the peak of the distribution to even smaller
values of $x$. This figure also shows that the MLLA predictions
closely tracks the non--MLLA solution for $x$ values that are several
orders of magnitude smaller than the matching point $x_0$; this
illustrates the advantage of requiring both the FF and its first
derivative to be continuous at $x_0$.

\section{Summary and Conclusions}
\label{sec:conclusion}
\setcounter{footnote}{1}

In this chapter, we presented a detailed analysis of the decay of a SH
particle, including all physical features which are supposed to play a
role in such decay (in our current understanding of the physics at
ultra-high energies), and using up to date results from SUSY
simulations and QCD experimental data. In particular, we included all
couplings of the MSSM in the perturbative partonic cascade above
$M_{\rm SUSY}$, and fully implemented the SUSY decay cascade; we are
able ensure energy conservation to a numerical accuracy of better than
1\%, as compared to up to several \% in ref.\cite{bd1}. Moreover, we
showed that the dependence of our results on the necessary
extrapolation of the measured FFs towards small $x$ is negligible. We
also included leading higher--order QCD corrections at very small $x$
using the MLLA approximation for taking into account color coherence
effects; this approximation is in good agreement with data from
particle colliders. These effects become significant for $x \leq
10^{-7}$, decreasing the predicted fluxes at $x \sim 10^{-11}$ by
about two orders of magnitude.

Furthermore, we showed that varying SUSY parameters can have some
impact on our results, affecting the shapes of the FFs at $x \geq
0.01$ and in some cases also the total multiplicity; however, the
dependence on the SUSY spectrum is much milder than the dependence on
the primary $X$ decay mode(s). Qualitatively the photon and LSP fluxes
are the most important ones at large $x$ if the primary is a strongly
interacting (s)particle; if the primary has only weak interactions,
the lepton fluxes can also be very large at large $x$. The proton flux
is always subdominant in this region. In contrast, the shapes of most
FFs at small $x$ can be predicted almost uniquely. This leads to the
following ordering of the fluxes at $x < 0.01$: the largest flux is of
muon neutrinos, followed by photons, $\nu_e$ and electrons, and
finally protons. The ratios of these fluxes become almost independent
of $x$ in this region, the proton flux being about a factor of five
smaller than the $\nu_\mu$ flux. On the other hand, the two smallest
fluxes at small $x$, of LSPs and finally $\nu_\tau$, do depend
sensitively on various currently unknown parameters. Generically they
rise less rapidly with decreasing $x$ than the other fluxes do;
already at $x \sim 10^{-3}$, the $\nu_\tau$ and LSP flux are usually
about one order of magnitude below the proton flux.

Finally, in the appendices we give additional details of our
description of the complete cascade. In particular,
Appendix~\ref{app:SFs} contains the first complete set of leading
order splitting functions for the MSSM, including all gauge as well as
third generation Yukawa interactions. A ``catalog'' containing an
almost complete set of FFs for a given set of parameters is given in
Appendix~\ref{app:curves}.

This work presents the to date most accurate and complete description
of the spectra at source of stable particles resulting from the decay
of a superheavy $X$ particle. These spectra are needed for all
quantitative tests of the ``top--down'' explanation of the most
energetic cosmic ray events. Of course, in order to be able to compare
with fluxes measured on or near Earth, effects due to the propagation
through the galactic, and perhaps extragalactic, medium
\cite{reviewSigl} have to be included, which depend on the
distribution of $X$ particles throughout the Universe; that is the
program of chapter~\ref{chap:applications}. On the other hand, our
description of $X$ decays is model--independent in the sense that it
allows to incorporate any primary $X$ decay mode. Indeed, it could
with very little modification also be used to describe the evolution
of very energetic jets produced through some other mechanism (e.g. the
annihilation of very massive stable particles), as long as the initial
virtuality of the produced particles is comparable to their energy.

Turning to the original problem of ultra--high energy cosmic rays
(UHECRs), the biggest obstacle towards a test of generic top--down
models is the strong dependence of the predicted decay spectra on the
primary decay mode. Most previous investigations assumed that $X$
decays into a pair of quarks, but we are not aware of any compelling
argument why this should be the dominant decay mode. On the other
hand, data may already rule out some classes of top--down models. For
example, it seems likely that few, if any, UHECR are photons
\cite{nophot}. In the context of top--down models, this leaves protons
as only choice. Our results then seem to disfavor models where $X$
decays primarily into particles with only weak interactions, since
this implies a large ratio of the photon to proton flux at large $x$.
However, this argument may not apply if $M_X \geq 10^{13}$ GeV, since
then all events seen so far are at $x \leq 0.01$, where the ratio of
photon to proton fluxes is essentially independent of the primary $X$
decay modes. Moreover, the photon flux may be diminished more
efficiently between source and detector than the proton flux. Searches
for very energetic neutrinos might therefore lead to somewhat more
robust tests of top--down models (see \cite{neutrinos, bdhh1} and
section~\ref{sec:neutrinos} of chapter~\ref{chap:applications}); as
noted earlier, the predicted neutrino flux should begin to exceed the
background from atmospheric neutrinos at very small values of $x$.
Nevertheless, the need to normalize the expected flux to the observed
flux of UHECR events, and hence to the proton and perhaps photon flux
at much larger $x$, re--introduces a large model dependence even in
this case \cite{bdhh1}. Moreover, other proposed explanations of the
UHECR also predict sizable neutrino fluxes at very high energy, e.g.
due to the GZK process itself. The failure to observe such neutrinos
could therefore exclude top--down models (given sufficiently large
detectors), but a positive signal may not be sufficient to distinguish
them from generic ``bottom--up'' models. This discrimination might be
achieved by searching for the predicted flux of very energetic LSPs,
since the LSP flux in bottom--up models is undetectably small;
however, this test will require very large detectors (see \cite{bdhh2}
and section~\ref{sec:neutralinos} of chapter~\ref{chap:applications}).
We conclude that ultimately the test of this idea will probably
require a combined analysis of different signals, at quite different
energies and in different detectors. We provide one of the tools
needed to perform such an analysis, since we are able to
systematically study the fluxes of {\em all} stable particles at
source, and their correlations, for {\em all} top--down models.


\chapter{Presentation of the code SHdecay}
\label{chap:code}

I give here a detailed user guide for the program
SHdecay\footnote{SHdecay is a public code and can be downloaded from
  http://www1.physik.tu-muenchen.de/$\sim$barbot/ .}, which has been
developed for computing the final spectra of stable particles
(protons, photons, LSPs, electrons, neutrinos of the three species and
their antiparticles) arising from the decay of a super-heavy $X$
particle. It allows to compute in great detail the complete decay
cascade for any given decay mode into particles of the Minimal
Supersymmetric Standard Model (MSSM). In particular, it takes into
account all interactions of the MSSM during the perturbative cascade
(including not only QCD, or SUSY-QCD, like the previous code of this
type \cite{codeToldra}, but also the electroweak and 3rd generation
Yukawa interactions), and includes a detailed treatment of the SUSY
decay cascade (for a given set of parameters) and of the
non-perturbative hadronization process (see chapter~\ref{chap:decay}
of this thesis for details). All these features allow us to ensure
energy conservation over the whole cascade up to a numerical accuracy
of a few per mille. Yet, this program also allows to restrict the
computation to QCD or SUSY-QCD frameworks. I detail the input and
output files, describe the role of each part of the program, and
include some advice for using it best.

In this chapter, I first describe in section~\ref{sec:black_box} the
``master program'' contained in the package, which partly allows to
use the whole program as a ``black box''. In
section~\ref{sec:programs}, I present the organigram of the code and
describe all its components in detail; I also list all the options of
the master program.

\section{How to use SHdecay as a black box}
\label{sec:black_box}

Here I would like to describe how to use this program as easily as
possible, ignoring the different internal components, and considering
the whole program as a ``black box''. I just want to stress that the
price to pay is running time... Indeed, certain component programs of
this code are pretty time consuming - especially the first one
(DGLAP\_MSSM), which is solving a set of 30 integro-differential
equations over orders of magnitude in virtuality, and needs around 30
hours of running on a modern computer\footnote{For processors of 1 GHz
  and above, the running time seems to be almost independent of the
  exact frequency, and there is no gain of time with increasing
  frequencies.}. Yet, in most applications, DGLAP\_MSSM and its
``brother'' DGLAP\_QCD have to be run only once. Moreover,
DGLAP\_MSSM can be ``cut'' into smaller pieces which can be run
independently on different computers. This will require more detailed
knowledge of this program (see section~\ref{sec:programs}).

There is another point I want to insist on: although SHdecay is a
self-contained code, it requires two Input files that have to be
obtained from an other program, like the public code ISASUSY: these
two files contain all information about the SUSY spectrum (masses and
mixing angles), and the decay modes of the sparticles, top quark and
Higgses, with the associated branching ratios (BRs). In order to keep
the completeness of the furnished code, I implemented a personalized
version of ISASUSY\footnote{I used the version 7.51 of ISASUSY.} in
this package in a fully transparent way for the user. Nevertheless, if
you want to use another code giving the same information, or
even an updated version of ISASUSY, you will have to work by yourself
for obtaining the two output files (called by default ``Mixing.dat''
and ``Decay.dat'', and stored in the Isasusy directory) in the
required format. I will come back to this point in
section~\ref{sec:programs}.

\subsection{Installation of SHdecay}

SHdecay has been written in C/C++ \footnote{In fact, it is a C program
  using a few C++ tools; in any case, it is {\it not} an object
  oriented program!} for a UNIX or Linux operating system. It
certainly can be used on a computer using windows with a C++ compiler,
but in that case you won't be able to use the provided makefiles. In
the following, I am describing the procedure for using SHdecay on a
UNIX/Linux computer.

Once you have downloaded the compressed package ``SHdecay.tar.gz'',
decompress it with the command:\newline

{\bf tar -xzf SHdecay.tar.gz\newline}

\noindent
It will create a directory SHdecay and install inside all files and
subdirectories you need. Then enter the directory SHdecay and compile the
``master program'' by typing:\newline

{\bf run\_SHdecay\newline}

\noindent
You can now call the ``master program'':\newline

{\bf SHdecay.exe\newline}

\clearpage

\noindent
The following menu should appear:\newline
\bigskip
\bigskip

 *************************** SHdecay.c  ***************************

    0: Compile all.

\bigskip

    1: Run all programs.

    2: Run all programs but Isasusy.

\bigskip

    3: DGLAP\_MSSM     (DGLAP evolution for the FFs between M\_SUSY and M\_X).

    4: Isasusy         (MSSM spectrum and decay modes).
 
    5: Susy1TeV        (SUSY and SU(2)*U(1) breaking; SUSY decay cascade).

    6: DGLAP\_QCD      (Pure QCD DGLAP evolution down to Q\_had).

    7: Fragment\_maker (Non-perturbative FFs at Q\_had).

    8: Less1GeV        (Hadronization and SM decays).

    9: Xdecay          (Final FFs for a given X decay mode).

\bigskip
 *******************************************************************\newline
\bigskip

You first have to compile all programs with the option ``0''. You'll
certainly get a few warnings that you can ignore (They arise from the
fact that you are compiling the program for the first time, and thus
it doesn't need to erase old files before compiling). Once it has been
done, call again the master program SHdecay.exe: now you can choose
the program you do want to run.

The 1st option allows to run the different programs as a whole black
box for given input files. You need two of them: 
\begin{itemize}
\item[a)] The first one is called ``Input.dat'' by default (but you
  may write your own with a different name: you will be asked for the
  name of this file at the very beginning of the run): it contains
  physical and technical parameters necessary for the run, as well as
  the name of output directories in which you will store the results.
  A default version of ``Input.dat'' is included (option ``/''), and
  all parameters have default values inside the program itself (option
  ``*''). Yet, you will need to write your own input data file in most
  cases. I present all the required input parameters in the next
  section.
\item[b)] The second input file is called ``SUSY.dat'' by default (but
  again, you can give it another name; you will be asked for it
  during the run): it contains all SUSY parameters that will be
  needed in ISASUSY, as well as the names of the two output files
  mentioned above (by default ``Mixing.dat'' and ``Decay.dat'').
\end{itemize}

This option will run successively all programs contained in SHdecay,
following the organigram given in fig.~\ref{Organigram} In this case,
the user has nothing to do but filling the two input files; yet, the
required running time will be around 36 hours on a recent computer
(see the note above) in the MSSM framework.

N.B.: The second option is the same as the first one, up to the fact
that it doesn't run the Isasusy program. It still requires the two
files ``Mixing.dat'' and ``Decay.dat'' to be present in the Isasusy
directory - no matter how you produced them. It only becomes useful
in the case you want to free yourself from ISASUSY.

The next options allow to run each code individually; for a
description, see the corresponding subsection in
section~\ref{sec:programs}.

\subsection{Parameters of the ``Input.dat'' file.}

The two first parameters concern two options of the program:
\begin{itemize}
\item[a)] ``Theory'' is an integer describing the theoretical
  framework in which the computation will be done. Four options are
  available: 1 for Minimal Supersymmetric Standard Model (MSSM), 2
  for Standard Model (SM), 3 for SUSY-QCD, and 4 for QCD alone. This
  option concerns the particles which will be included in the
  perturbative cascade at high energy, while solving the set of DGLAP
  equations. As a result, it also determines in which primary
  particles the $X$ particle is allowed to decay. But it doesn't
  affect the decays of particles with masses $\leq M_{SUSY}$ at all,
  which are governed by known physics and require the full SM
  spectrum. For example, in the QCD framework (Theory = 4), only
  quarks and gluons will be taken into account in the DGLAP equations,
  which means that we will neglect all but QCD couplings; yet the top
  quark will still decay into $b W$ and $W$ in leptonic as well as
  quark channels at $M_{SUSY}$!). By default we use the MSSM (Theory
  = 1).\newline Caution: for the moment, the SM option is not fully
  implemented.
\item[b)] ``MLLA'' is an integer describing the approximation made at
  low $x$ for the non-perturbative hadronic FFs. Two choices are
  possible:
\begin{itemize} 
\item[0:] one assumes a power law extrapolation at low $x$ for the
  final FFs (which indeed have a power law shape at the end of the non
  perturbative cascade). But this extrapolation doesn't take into
  account the saturation of the FFs at low $x$ due to the appearance
  of mass and color effects.
\item[1:] one assumes the Modified Leading Log Approximation (MLLA)
  \cite{Basics_of_QCD} with the implementation of a distorted Gaussian
  at small x \cite{distorted_gaussian}, in order to take the color
  effects into account. For details on this point, see
  section~\ref{sec:MLLA} of chapter~\ref{chap:decay}.
\end{itemize}
By default the MLLA approximation is used (MLLA = 1).
\end{itemize}

The next set of parameters describes the physical inputs of the program (all
masses, energies and virtualities are given in GeV):
\begin{itemize}
\item[a)] ``Nb\_output\_virtualities'' gives the number of different
  values for the $X$ mass you want to study. By default two final
  masses are stored (Nb\_output\_virtualities = 2).\newline Caution:
  for each $X$ mass you will get a lot of files containing partial
  results, for example $30 \times 30 = 900$ in the MSSM framework!
\item[b)] In ``Output\_virtualities\_DGLAP\_MSSM(GeV)'', you should
  specify the exact values of the virtualities at which you want to
  store the FFs. (Exactly as many values as you asked for in
  ``Nb\_output\_virtualities''!). By default, the two final
  virtualities (i.e. $X$ masses!) which are stored are: $10^{12}$ and
  $10^{16}$ GeV\footnote{Caution: As the name of this parameter
    indicates, this option {\it only concerns} the outputs of the
    first program DGLAP\_MSSM. The following programs will treat only
    {\it one} case, the one required by parameter $M_X$. See the
    technical section for the reasons of this choice.}.
\item[c)] ``$M_X$'' (in GeV) doesn't give exactly the {\it mass} of
  the X particle\footnote{In fact, the {\it real mass} of the $X$
    particle will be $2 \times M_X$.}, but the {\it initial
    virtuality} of the decay products of X, that is, the highest
  virtuality of the perturbative cascade; of course, it must be one of
  the values given in the parameter
  ``Output\_virtualities\_DGLAP\_MSSM(GeV)'' described above, at which
  the output FFs have been stored. This parameter will be used by all
  other programs following DGLAP\_MSSM. If you want to do the complete
  treatment for different $M_X$ masses, you will have to run these
  other programs as many times as necessary, with the different values
  of $M_X$. By default $M_X$ is set to $10^{16}$ GeV (the GUT scale).
\item[d)] ``N-body\_X\_decay'' must contain the value of $N$ for a N-body
  decay mode of the $X$ particle. By default we consider a 2-body decay.
\item[e)] ``X\_decay\_mode'' contains the details of the $N$-body decay
  mode you want to study. Of course, it must contain as many particles
  as asked in ``N-body\_X\_decay''; the id's of the different particles
  are given in Appendix~\ref{app:compound} of this manual. By default $X$ is
  decaying into two SU(2) doublets of the first/second generation: two left
  quark/antiquark $q_L$, with id 1.
\item[f)] The ``$M_{\rm SUSY}$'' parameter must contain the virtuality (in
  GeV) at which both SUSY and $SU(2) \otimes U(1)$ are broken; it is
  also the virtuality at which all sparticles (but the LSP), top
  quarks and heavy bosons decay. By default $M_{\rm SUSY} = 1000$ GeV.
\item[g)] $Q_{\rm had}$ gives the virtuality at which
  hadronization of the lightest quarks and gluons occurs and the
  non-perturbative fragmentation functions are convoluted with the
  perturbative ones. By default $Q_{\rm had} = 1$ GeV.
\end{itemize}

{\bf Important note}: for practical reasons, it is only possible to choose
{\it powers of ten} for all energy scales. Fortunately, such
restriction is not too constraining, because the DGLAP evolution
equations are only {\it logarithmic} in energy.\newline

The next set of input parameters (namely $X_{\rm Size}$, $X_{\rm extra
  Size}$, Part\_init, Part\_fin and $X_{\rm min}$) is essentially
technical, and I recommend to keep the default values, which have been
carefully adjusted in order to maximize the precision and minimize the
time needed for running. They will be described in more detail in the
technical sections.\newline

The next set of parameters only has a practical purpose: give their
names to the output file and directories where the results will be
stored (The output themselves will be described in the next
subsection):
\begin{itemize}
\item[1)] ``Region'' is a suffix which will be added to the name of
  certain output files. It could be used to label the set of SUSY
  parameters which has been used, or the mass of $X$, or both.
\item[2)] ``Output\_file'' gives the path and the name of the final
  output file where all the parameters of the run and some results on
  the final energy carried by each type of stable particles will be
  stored.
\item[3)] The names of the next five parameters are hopefully explicit
  enough: they describe the names of the directories where the output
  data files (containing the description of the FFs) of the different
  programs will be stored. If these directories don't exist already,
  they will be created automatically. For technical reasons, we don't
  allow to give a different output directory for each of the programs
  involved in the computation. For a first use of SHdecay as a whole,
  we advise to put all the FFs in the same directory, as it is done by
  default (``LowBeta'' being the common output directory for
  DGLAP\_MSSM, Susy1TeV, Less1GeV and X\_decay).\newline Caution: the
  output directory of Fragment\_maker contains the non-perturbative
  input FFs at low energy, which are essentially built from the
  results of \cite{Poetter} through the program called
  ``Fragment\_maker''. I advise to use a special directory for that
  purpose (by default: Fragment), because these FFs should be
  considered as {\it inputs} of SHdecay; they could be taken from
  another source if newer results become available, and thus should be
  kept independent of the rest of the code. Again we choose a
  different default output directory for the program DGLAP\_QCD,
  because it only depends on parameters $M_{\rm SUSY}$ and $Q_{\rm
    had}$, and thus need to be run only once, independently of the
  other programs, for most applications.
\end{itemize}

\subsection{Parameters of the ``SUSY.dat'' file.}

{\bf CAUTION:} even if you don't want to use the ISASUSY code, you
still have to fill partially this input file, at least with the value
of $\tan \beta$, which is required by the program
DGLAP\_MSSM\footnote{Indeed, except for $\tan \beta$ which
  determines the strength of the Yukawa couplings, the other SUSY
  parameters are not needed in SHdecay itself, but only in the ISASUSY
  program mentioned above, which provides the basic information about
  the SUSY decay cascade.} (by default $\tan \beta = 10$), and the two
last parameters giving the names of the two input files (By default
``Decay.dat'' and ``Mixing.dat''). These files have to be placed in
the Isasusy directory of SHdecay. Moreover, you should {\it always}
give a value to {\it all} parameters, even the ones which are not
used, otherwise the program won't be able to read the input files.

All SUSY parameters should be self explanatory. All masses should be
given in GeV. By default, the masses as well as the $\mu$ mass
parameter are chosen to be all 1000 GeV, and the trilinear couplings
are set to 1000. The optional values for the 2nd generation sfermions,
the gaugino and the gravitino masses, are set to $10^{20}$
GeV\footnote{Of course, that is not a physical value, but an internal
  convention for Isasusy.}. I refer to the user guide of
Isasusy\cite{Isasusy} for further information.

$\tan \beta$ is the usual parameter of SUSY theories defined by $\tan
\beta = \frac{<H_2^0>}{<H_1^0>}$, where the $<H_i^0>$ describe the
vacuum expectation values of the two Higgs fields of the MSSM. By
default, $\tan \beta = 10$.

\subsection{Output files}
 \label{subsec:output}

One inconvenience of this program is that it has to store a lot of
data files, most of them being partial results which will be needed
for the next steps of the computation. For example, the program
DGLAP\_MSSM will have to store the FFs of any (s)particle of the MSSM
into any other; this requires $30 \times 30 = 900$ files for each set
of parameters\footnote{As described in chapter~\ref{chap:decay} and
  Appendix~\ref{app:compound}, it has been assumed that certain MSSM
  particles can be treated symmetrically; this reduces the number of
  independent particles from 50 to $\sim 30$.}. These partial results
are certainly not relevant for the user who wants to use SHdecay as a
black box. So I will only describe here the final results which are
produced; all partial results will be described in the section
dedicated to the corresponding program.

In fact, there are only 2 or 3 types of relevant results:
\begin{itemize}
\item[1.] The final FFs themselves $D_i^j(x,M_X)$, of any initial
  decay product $i$ of the $X$ particle (among the 30 available
  ``particles'' of the MSSM, see Appendix~\ref{app:compound} for details) into one of
  the seven stable particle $j$ (proton, photon, electron, the three
  types of neutrinos and the LSP). They are computed at the end of the
  program called Less1GeV, and stored in the corresponding output
  directory in 30 different files (all seven FFs for a given initial
  decay product $i$ of $X$ are grouped into one file); these files are
  called generically ``fragment\_i.all\_Region'', where $i$ is the
  initial decay product of $X$ defined above, and Region is the suffix
  labeling the set of SUSY parameters which has been used. Each of
  them contains 8 columns, giving respectively: the $x$ values (in
  decreasing order from 1. to $X_{\rm min}$), and the seven FFs into protons,
  $\gamma$, LSPs, $e^{-}$, $\nu_e$, $\nu_\mu$, $\nu_\tau$ respectively
  (more precisely the results correspond to the {\it sum} of the FFs
  for final particles {\it and} antiparticles).
\item[2.] For the user who wants to study a {\it precise decay mode}
  of the $X$ particle into $N$ particles of the MSSM, the relevant
  results will be given by the program Xdecay, and stored in the {\it
    same output directory} as the one given in Input.dat for Less1GeV.
  The name of the output file is ``frag\_X\_a\_b\_c(...).all\_Region'',
  where a,b,c,... are the id's of the $N$ decay products (given in
  Input.dat as ``X\_decay\_mode''; the correspondence between
  particles and id's is given in Appendix~\ref{app:compound}), and
  ``Region'' is the same labeling parameter as above. The 8 columns of
  the file are exactly the same as the ones described above.
\item[3.] The user might be interested in the amount of
  energy stored in each type of final stable particles. This
  information, as well as all the input values for the corresponding
  run of the program, is stored in an output file whose name is given
  as ``output\_file'' in Input.dat.
\end{itemize}
  
For convenience, in addition to the output files themselves, I provide
two functions, both called ``fragment\_fct'' (one in C++ and the other in
fortran 77), which allow to use any of the FFs computed in SHdecay in
another code. These functions are reading the specified input file and
computing the necessary cubic spline of the function. They are stored
in the ``Tools'' directory (``fragment\_fct.c'' for the C++ version,
and ``fragment\_fct.f'' for the fortran 77 one).

\begin{itemize}
\item[-] In C++, you will need the string type, that you can just
 include by adding:

{\bf \#include $<$string$>$}

\noindent
You then have to declare the function through the following line:

{\bf extern double fragment\_fct(double x, char* path\_file, string p\_fin);}

\noindent
You finally can call this function through the command:

{\bf fragment\_fct(x,path\_file,p\_fin)}

\noindent
Then, if ``program.c'' is the name of your program (written in the SHdecay directory), just compile it with

{\bf g++ program.c ./Tools/fragment\_fct.c ./Tools/my\_spline.c}

\item[-] In fortran 77, you just call the function with the same command:

{\bf fragment\_fct(x,path\_file,p\_fin)}

\noindent
and compile your ``program.f'' program with:

{\bf f77 program.f ./Tools/fragment\_fct.f ./Tools/spline.f}

\end{itemize}

where 

\begin{itemize}
\item[-] $x$ is the (real) value at which one wants to compute the FF,
\item[-] path\_file is a chain of characters giving the complete
  (relative) path to the file containing the data (it must be given
  between two quotes),
\item[-] p\_fin is the final particle one is interested in (to be
  chosen between ``p'' for protons, ``gam'' for photons, ``LSP'' for
  LSPs, ``e'' for electrons, ``nu\_e'', ``nu\_mu'', or ``nu\_tau'' for
  the three species of neutrinos, or possibly ``'' if the chosen file
  only contains a single FF, as it is the case for all partial
  results of the code)\footnote{It must be clear that p\_fin is just
    needed in order to distinguish between the 7 FFs which are stored
    in the same output file when coming from Less1GeV or Xdecay. For
    other files it must be set to ``''.}. It also must be given
  between two quotes.
\end{itemize}

I also provide two toy programs (``read\_fct.c'' and ``read\_fct.f'')
which are stored in the directory SH\_decay and can be used as
examples. They ask the user an input file (in fact its {\it full
 relative path}), a final particle and an $x$ value, and return the
value of the corresponding FF at $x$.

\section{Description of the different programs}
\label{sec:programs}
\setcounter{footnote}{1}

\begin{figure}[h!]
\label{Organigram}
\setlength{\unitlength}{1cm}
\includegraphics{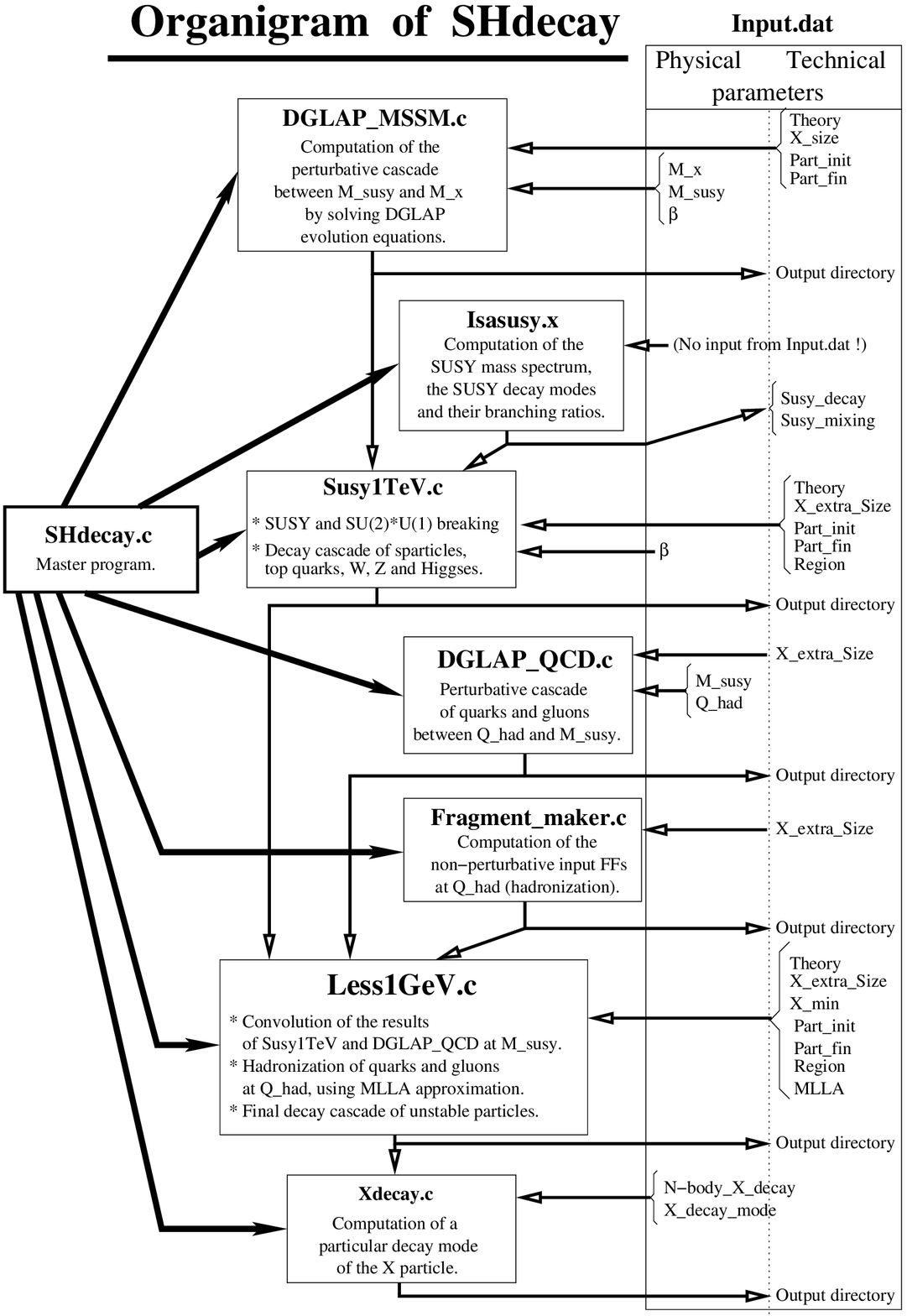}
\caption{Organigram of SHdecay with the interdependence between the
 different programs and the needed input parameters for each code.}
\end{figure}

There are mainly four successive programs treating the different parts of
the decay cascade; in order of decreasing virtuality, these are:
DGLAP\_MSSM, Susy1TeV, DGLAP\_QCD, and Less1GeV. Eventually, a
last small program called Xdecay can be run to study a particular
decay mode of the $X$ particle. We will describe in detail the role of
each of these programs, and the parameters of Input.dat they are sensitive
to. Fig~\ref{Organigram} gives a detailed organigram of the whole
code, which shows the interdependencies between the different programs
and their input parameters.

We just note here that SHdecay as a whole requires the results of other
independent codes at two different steps, namely:
\begin{itemize}
\item[1)] the SUSY mass spectrum, mixing angles, and decay modes of 
sparticles (with their branching ratios), all given by ISASUSY (a
subset of the Isajet code, written in Fortran 77).
\item[2)] the non-perturbative input fragmentation functions, 
computed (once and for all) from the results of \cite{Poetter} through a
program called Fragment\_maker (which is furnished). 
\end{itemize}
In fact, both of them are fully implemented in the body of SHdecay,
and are treated exactly the same way as the other programs. I'll
describe these two secondary procedures in more detail in the
corresponding subsections.\newline

Of course, the values of all the parameters written in Input.dat
should be kept the same for the four (or five) main programs running
successively.

All these programs have been written in C using a few C++ tools. The
compiling option of SHdecay is using the g++ compiler of gnu (given by
default on Unix and Linux OS).

I first describe all technical parameters before going into the
details of each program.

\subsection{Technical parameters}
\label{subsec:technical_param}

\begin{itemize}
\item[1)] ``$X_{\rm Size}$'' gives the number of $x$ values used to store
  the FFs on the interval [$10^{-7}$:$1-10^{-7}$]. Because of a) the
  behavior of the splitting functions at small $x$, b) the fact that we
  are beginning with ``delta functions''\footnote{modeled
    numerically by sharp gaussians centered at 1. and normalized to
    unity between 0 and 1.} at large $x$, and c) the definition of the
  convolution which is relating the low and large $x$ regions, the two
  extremities of our interval have to be modeled symmetrically with
  great accuracy, if we want the integration and (cubic spline)
  extrapolation procedures to be able to give results at the desired
  precision of $\sim 10^{-3}$. For this purpose we used a
  bi-logarithmic scale between [$10^{-7}$:0.5] and [0.5:$1-10^{-7}$],
  increasing the number of $x$ values towards the two extremities. We
  are using by default $X_{\rm Size} = 101$\footnote{Note that $X_{\rm Size}$
    {\it has to be odd}!}, i.e. 50 $x$ values on each side of the central
  value at $x = 0.5$. Note that a smaller value could lead to false
  results, while increasing $X_{\rm Size}$ is increasing greatly the running
  time needed by all programs. So I really advise the user not to
  change this value. Note finally that the smallest $x$ value $10^{-7}$
  has been chosen at the limit of the validity of the (leading order)
  DGLAP equations, before MLLA effects become strong (which happens
  at $\sqrt{\frac{Q_{had}}{M_X}} \sim 10^{-8}$ for $M_X \sim 10^{25}$
  eV and $Q_{had} \sim 1$ GeV; see \cite{bd2}). At low
  $x$, the standard LO DGLAP equations will predict a power law
  behavior\footnote{The power law can of course be extrapolated
    easily towards lower $x$, avoiding the extremely time consuming running
    of DGLAP\_MSSM on a larger $x$ interval!} (option MLLA = 0), but
  the MLLA approximation (option MLLA = 1) allows to parameterize
  some NLO effects like soft gluon emission.
\item[2)] ``$X_{\rm extra Size}$'' is a parameter which allows the
  user to increase homogeneously the overall number of $x$ values on
  the interval $[10^{-7}:1-10^{-7}]$ {\it after} the first program
  DGLAP\_MSSM (which is, once again, the most time consuming part of
  the complete code) has been completed. But it is quite {\bf
    useless}, the initial value of $X_{\rm Size}$ being large enough
  for all following programs\footnote{In fact, this is not exactly
    true, because the implementation of 2-body decays sometimes
    requires a local increase of the precision, and thus a local $x$
    array. But this is fully implemented in the programs themselves,
    and is totally hidden from the user.}. By default, $X_{\rm extra
    Size}$ is simply taken to be equal to $X_{\rm Size}$. (Of course,
  it has to be greater than (or at least equal to!) $X_{\rm Size}$).
\item[3)] ``Part\_init'' and ``Part\_fin'' describe the initial and
  final id's of an {\it interval of initial particles} for which the
  FFs have to be computed. Note that there are 30 initial
  ``compound''\footnote{Here, ``compound'' means that we regrouped
    artificially different particles following the same DGLAP
    equations. See Appendix~\ref{app:compound} for the for the
    description of these particles and their id's.} particles in the
  MSSM, and {\it all} the $30 \times 30 = 900$
  FFs from any particle to any other will be needed for the
  computation of the whole cascade. Thus the default values are
  respectively Part\_init $ = 1$ and Part\_fin $= 30$, which means
  that the program will treat successively all the 30 possible initial
  particles. Nevertheless, the treatment of an initial particle being
  fully independent of the others, any of the 3 programs DGLAP\_MSSM,
  Susy1TeV and Less1GeV can be cut into pieces to be run independently
  on different computers; for example, you can let a first computer
  run the chosen program for particles 1 to 15, and another computer
  run the {\it same} program for particles 16 to 30. These two
  parameters render this task easy and allow to save a lot of
  time.\newline {\bf Caution}: Note that each of these three programs
  has to be run over the {\it whole range of particles} before running
  the following one!
\item[4)] $X_{\rm min}$ gives the lowest value of the final x interval. As
  stated above, in the lowest x region ($[X_{\rm min}:10^{-7}]$), you can
  choose two different extrapolations of the FFs: either
  extrapolating the power law obtained from the LO DGLAP equations, or
  using the MLLA approximation for taking color coherence effects into
  account. This parameter, taken by default to be $X_{min} =
  10^{-13}$, is only used in the very last part of the computation of
  the cascade: Less1GeV.\newline
{\bf Caution}: of course, $X_{\rm min}$ has to be $\ge 10^{-7}$.

\end{itemize}

\subsection{DGLAP\_MSSM}
\label{subsec:DGLAP_MSSM}

This program treats completely the perturbative cascade above the
$M_{\rm SUSY}$ scale. Starting from input FFs at $M_{\rm SUSY}$ for each type
of primary particle $P$ ($D_P^P(x,M_{\rm SUSY}) = \delta(1-x)$ and $\forall
j \neq P$, $D_P^j(x,M_{\rm SUSY}) = 0)$, it gives the FFs of the 30
interaction eigenstates at scale $Q = M_X$: $D_P^j(x,M_X)$.

By giving the parameters of the ``Input.dat'' file, the user can
choose one of the 4 available theories, namely 1: MSSM, 2: SM, 3:
SUSY-QCD, 4: QCD alone. I point out that this complicated program is
certainly not the best one for treating a case as simple as QCD DGLAP
equations (or even SUSY-QCD), being unfortunately quite time
consuming. This program requires no external input (except
Input.dat, of course), and only needs as ``physical inputs'' the
values of $\beta$ and $M_X$, described above. The technical parameters
$X_{\rm Size}$, ``Part\_init'' and ``Part\_fin'' are used, too. As I
already mentioned before, I strongly suggest when possible to run the
program on different computers at the same time, using different
intervals of initial particles, for saving time\footnote{Again,
  the running time will depend on the computer you are using. Yet, to
  give an idea, you should foresee around one hour of running time per
  initial particle in the MSSM framework for $M_X \sim 10^{16}$ GeV on
  a modern computer (1 GHz or more, 256 Mo of RAM).}.

Finally, the user should specify the corresponding output directory,
where the output files will be stored.

Using the structure of the DGLAP evolution equations and
$\delta$-functions as input FFs at $M_{\rm SUSY}$ (practically implemented as
sharp Gaussians), this program will compute the full set of FFs from
one particle to another between $M_{\rm SUSY}$ and $M_X$. For this
purpose, we use a Runge-Kutta method with a {\it constant logarithmic}
step in virtuality for solving the system of DGLAP
equations\footnote{Unfortunately, for practical reasons, it was not
possible to choose a floating step.}. There must be an {\bf entire
number} of these steps between $M_{SUSY}$ and (any value of)
$M_X$. That's why it is only possible to use powers of 10 for these
scales. Nevertheless, as I said before, this allows already a good
accuracy.

Here we can see the interest of the variable
``Nb\_output\_virtualities'' and the corresponding array of virtuality
values ``Output\_virtualities\_DGLAP\_MSSM'': thanks to the fact that
this program is computing the FFs from $M_{SUSY}$ to $M_X$ through a
given number of Runge-Kutta steps, all intermediate virtualities used
by the Runge-Kutta program are available as possible outputs; it
allows to get the FFs at intermediate virtualities, which are equivalent
to lower $X$ masses $M_X$. As stated above, the step used for
Runge-Kutta is a constant logarithmic step, exactly one order of
magnitude each. So practically, the user who wants to study a GUT $X$
particle with mass $M_X \sim 10^{25}$ eV can get the results for any
other (power of 10) $X$ mass ($10^{21}$, $10^{22}$, $10^{23}$ eV,...)
between $M_{SUSY}$ and $M_X$. The two variables cited above allow to
put these partial results in output files that will be usable later
on. Note again that only {\it this} program will use the array of values for
$M_X$. The following ones will simply use one of these values, the one
given in the parameter $M_X$ itself.

The output is presented in $30 \times 30 = 900$ files giving the FFs
of any particle into any other with $X_{\rm Size}$ values of $x$ in
the first column and the corresponding values of the FF in the second
one. These files are called generically ``fragment\_(M\_X)eV\_p1.p2''
for the FF of particle p1 into particle p2, where (M\_X) contains the
mass of the $X$ particle at which the FF was computed. Note that,
according to the form of DGLAP equations for a generic FF
$D_{p1}^{p2}(x,Q²)$, the iterator part $\in[$Part\_init,Part\_fin$]$ of
the program DGLAP\_MSSM (and evidently its ``brother'' DGLAP\_QCD)
runs over the ``final'' particles p2. On the contrary, the equivalent
iterators in Susy1TeV and Less1GeV run over particles p1. That's why
it is {\it essential} to run the different programs successively,
after the complete end of the preceding one!.

Finally, it is worth noting that the output of this program only
depends on very few parameters: the ``theory'' chosen as framework
for the computation, the $M_{\rm SUSY}$ scale at which the
perturbative cascade is ending, and the SUSY parameter $\tan \beta$.
Yet, we have seen in section~\ref{subsec:SUSYdep} that these two
parameters have very little influence on the final
results\footnote{Indeed, the evolution being only logarithmic in
  virtuality, and running over many orders of magnitude until $M_X$,
  the exact value of $M_{SUSY}$ (say, between 200 GeV and 1 TeV)
  doesn't really matter. Similarly, the $\tan \beta$ parameter only
  affects the Yukawa interactions, which are almost negligible, except
  in some rare cases for the third generation of quarks and leptons.}.
Thus I strongly recommend to let run this program only {\it once}, for
all $X$ masses you want to study, and to carefully keep these partial
results for later use, for studying the influence of other parameters
appearing in the following programs.
\clearpage

\subsection{How to use Isasusy.x}
\label{subsec:Isasusy}

The ISASUSY program, written in fortran 77, is a subset of the whole
code called ISAJET \footnote{Isajet/Isasusy is a public code and can
 be downloaded from http://www.phy.bnl.gov/~isajet/}. I refer to the
user manual of ISASUSY \cite{Isasusy} for information about how to use
this program. But, as mentioned above, I fully implemented a
personalized version of this code, which is available through the
master program (option 7). For a given set of SUSY parameters
specified by the user in SUSY.dat, it computes the complete SUSY
spectrum (masses and mixing angles of all the sparticles, stored in
the ``Mixing.dat'' file), and the allowed decay modes with the
corresponding branching ratios (stored in ``Decay.dat''). Both files
will be stored in the Isasusy directory of SHdecay, and their names
have to be given in the ``SUSY.dat'' input file.

Of course, you can get these files from any other available code
providing the same information as ISASUSY, as long as you adapt the
output format of this code in order to get the one required by SHdecay
(see the model files provided in the Isasusy directory for information
about the required format). The furnished version of Isasusy is the
one included in Isajet 7.51. If you want to use an updated version of
Isasusy, you probably just need to replace the two files called
``aldata.f'' and ``libisajet.a'' in the Isasusy repertory of SHdecay
(but hopefully {\it not} ssrun.f and the Makefile, which I have
adapted).Yet, I obviously cannot ensure that this operation will
work...

\subsection{Susy1TeV}
\label{subsec: Susy1TeV}

This program takes the results of DGLAP\_MSSM given at the ({\it
 unique}) $M_X$ value specified in Input.dat and deals with the
breaking of SUSY and $SU(2) \otimes U(1)$, the supersymmetric decay
cascade and the decays of the top quarks, the Higgs, $W$ and $Z$
bosons. The muons and taus existing at this step are decayed too. We
considered only 2- and 3-body decays for which we computed the
relevant phase space (see Appendix~\ref{app:decays} for details),
using the branching ratios and the mass spectrum given by ISASUSY. For
any detail on these procedures, we refer to \cite{bd2}.

The input directory has to be the one where the outputs of DGLAP\_MSSM
have been stored (the user doesn't have to specify it). On the other
hand, the output directory can be different, in order to {\it
 distinguish between different parameters}. For example, the mass of
the $X$ particle, which is especially specified in the names of the
output files of DGLAP\_MSSM (as being the final virtuality of the
perturbative cascade) is {\it no more specified} in the outputs of
Susy1TeV\footnote{I made this choice for avoiding to lengthy file
  names; yet, you still can use the ``Region'' parameter as a reminder
  of the $X$ mass you used (see below).}; thus it can be useful to
define different output directories for different $X$ masses.
Moreover, the large number of output files is easier to handle when
stored in different directories.

The program will need the two output files given by Isasusy.x (or any
other program, see above); the two files have to be located in the
directory ``Isasusy'', and their names have to be given in the two
corresponding parameters of SUSY.dat: ``Decay'' and ``Mixing''. Here
the parameter ``Region'' also becomes useful. If necessary, the
extension of the $x$ array to ``$X_{\rm extra Size}$'' values instead
of ``$X_{\rm Size}$'' will occur in this program, too\footnote{Though,
  as I already mentioned, this option is not of very big use.}.

The output files contain the FFs of the 30 initial particles
(interaction eigenstates) into the remaining SM mass eigenstates after
the decays, namely the quarks u, d, s, c, b and gluons, the electrons,
neutrinos and the LSP. All of them will have a suffix ``\_1TeV'' to
distinguish them from the outputs of other programs and the
``Region''-suffix, e.g. labeling the set of SUSY parameters you used
during the run, or the $X$ mass you used (or both!).

\subsection{DGLAP\_QCD}
\label{subsec:DGLAP_QCD}

This program is a simplified copy of DGLAP\_MSSM. It computes the pure
QCD perturbative partonic cascade for quarks\footnote{Only 5 quarks
  are considered here, namely $u,d,s,c,b$, the top quarks having been
  decayed at scale $M_{\rm SUSY}$.} and gluons (so only 6 particles)
for a virtuality decreasing from $M_{\rm SUSY}$ to $Q_{\rm had} =
{\rm max}(m_{\rm quark},1$ GeV). This program is {\it not using} any previous
result from other ones, and only depends on $M_{\rm SUSY}$ and $Q_{\rm had}$,
which are not very sensitive parameters, as stated above. We thus
recommend to define their values once and for all (say, keep the default
values $M_{\rm SUSY} = 1 $ TeV and $Q_{\rm had} = 1$ GeV), and to run
DGLAP\_QCD {\it only once}. This possibility of sparing running time is
the reason why the necessary convolution between the results of this
program and the FFs given by the previous one (Susy1TeV) was
implemented in Less1GeV, in order to keep DGLAP\_QCD fully
independent.

The $6 \times 6 = 36$ output files, called generically ``fragment\_p1.p2'' -
where p1 and p2 are initial and final partons \{u,d,s,c,b,g\} - will
be stored in the corresponding directory given in Input.dat. We
recommend to use a dedicated directory, for the reason stated above:
these results are almost parameter independent and can be used for
different runs of Susy1TeV and Less1GeV.

\subsection{Fragment\_maker}
\label{subsec:Fragment}

This program is certainly the weakest part of our treatment, because
of the lack of knowledge concerning the non-perturbative FFs at very
low $x$. We used the results of \cite{Poetter} for this purpose, which
are based on LEP data. Unfortunately they are only valid for $x \geq
0.1$. The reason is that at LEP energies, it is necessary to consider
mass effects at small $x$, which can be described by the so-called
``MLLA plateau'' (see section~\ref{sec:MLLA}). Such effects can be
taken into account during the computation of the hadronization itself,
in Less1GeV. In Fragment\_maker, we just keep the FFs given in
\cite{Poetter} up to $x = 0.1$ and extrapolate them at small $x$, by
requiring continuity and the overall conservation of energy.

We finally obtain a set of input functions for light quarks (including
the $b$) and for gluons which conserve energy and agree with known
data.

This program is fully independent of the others, and is just used to
``prepare'' the non-perturbative input FFs at low energy needed in
Less1GeV. It doesn't depend on any parameter, and can be run once and for
all. Once again, we recommend to use a dedicated directory for storing the
output Files of this program; the default value is a directory called
``Fragment''.

\subsection{Less1GeV}
\label{subsec:Less1GeV}

This program first computes the convolution between the results of
Susy1TeV (describing the evolution of the FFs between $M_X$ and
$M_{\rm SUSY}$ after SUSY, top, $W$, $Z$ and Higgs decays) and the ones of
DGLAP\_QCD (describing the further evolution of the partonic part of
these FFs between $M_{\rm SUSY}$ and $Q_{\rm had}$). It further deals with the
hadronization of quarks and gluons, using external input FFs (The
results of the Fragment\_maker program described above) which have to
be convoluted with the previous results. It finally deals with the
decays of the last unstable particles. The 2- and 3-body decays are
treated exactly as in Susy1TeV.

The results are once more given in terms of FFs of any initial
(interaction eigenstate) particle (between the 30 available in the
MSSM, see Appendix~\ref{app:compound}) into the final (physical)
stable ones, namely the protons, electrons, photons, three species of
neutrinos, and LSPs. To simplify the storing and further use of these
(final) results, we grouped all the results corresponding to one
initial particle in one file generically called
``fragment\_p1.all\_Region'', where p1 is the initial particle and
Region the suffix labeling the set of SUSY parameters. Each file
contains seven FFs: the first column gives the values of $x$ (from
$1-10^{-7}$ to $X_{\rm min}$, in decreasing order), and the next
columns give successively the FFs of p1 into protons, photons, LSPs,
electrons, $\nu_e$, $\nu_\mu$, $\nu_\tau$.

\subsection{Xdecay}
\label{subsec:Xdecay}

This last program allows to study a special decay mode of the X
particle, by computing a last convolution between the results obtained
in Less1GeV and the phase space of the given decay mode. The number of
decay products and their nature (through the associated id, see
Appendix~\ref{app:compound}) have to be specified in Input.dat. 

If a decay mode for the $X$ particle has been specified in the two
parameters ``N-body\_X\_decay'' and ``X\_decay\_mode'' (respectively
the number $N$ of products and the id's associated to each product -
see Appendix~\ref{app:compound}), a last convolution with the $N$-body
decay energy spectrum will be computed and the results will be
directly given in terms of the FFs of the $X$ particle into the stable
final ones. The $N$-body energy spectrum we used is the one given in
\cite{Sarkar:2001}. If $\rho_N(z)$ is the probability density of
obtaining a decay product of energy $E$ carrying the energy fraction
$x = 2E/M_X$ of the decaying particle, we have:

\beqa
\label{e9}
&& \rho_2(x) = \delta(1-x)\,,\nonumber\\
&& \rho_N(x) = (N-1)(N-2)x(1-x)^{N-3}\,, N\geqslant 3\,.\nonumber\\
\eeqa

This program has been separated from Less1GeV to allow the user to
obtain very quickly any decay mode he wants to study. The final result
is stored in the same directory as the results of Less1GeV. It is
generically called ``frag\_X\_a\_b\_c.all\_Region'' (see
section~\ref{subsec:output}) and has the same format as the one
described above for the results of Less1GeV.

\section{Conclusion}

This chapter describes in some detail how to use the code SHdecay,
which has been designed for computing the most general decay spectra
of any super-heavy particle in the framework of the MSSM. I hope that
it will be of some use for other researchers. As I mentioned before,
the code as well as a version of this user-guide are available on the
web site of our group, under the
address: ''www1.physik.tu-muenchen.de/$\sim$barbot/'', and I will be pleased
to answer any question you have about it. Of course, any remark or
suggestion is welcome, too.


\chapter{Applications of SHdecay: phenomenology}
\label{chap:applications}

\section{Introduction}

Since the first discovery of cosmic rays (CRs) in 1912 by Victor
Hess\cite{Hess}, their energy spectrum has been measured over more
than 12 decades of magnitude. Although there still
are number of discussions going on in this field of research, our
general understanding of the typical features of this spectrum has
greatly improved. Especially, these CR particles are now known to
consist primarily of protons, helium, carbon, nitrogen and other heavy
ions up to iron.

\begin{figure}
\postscript{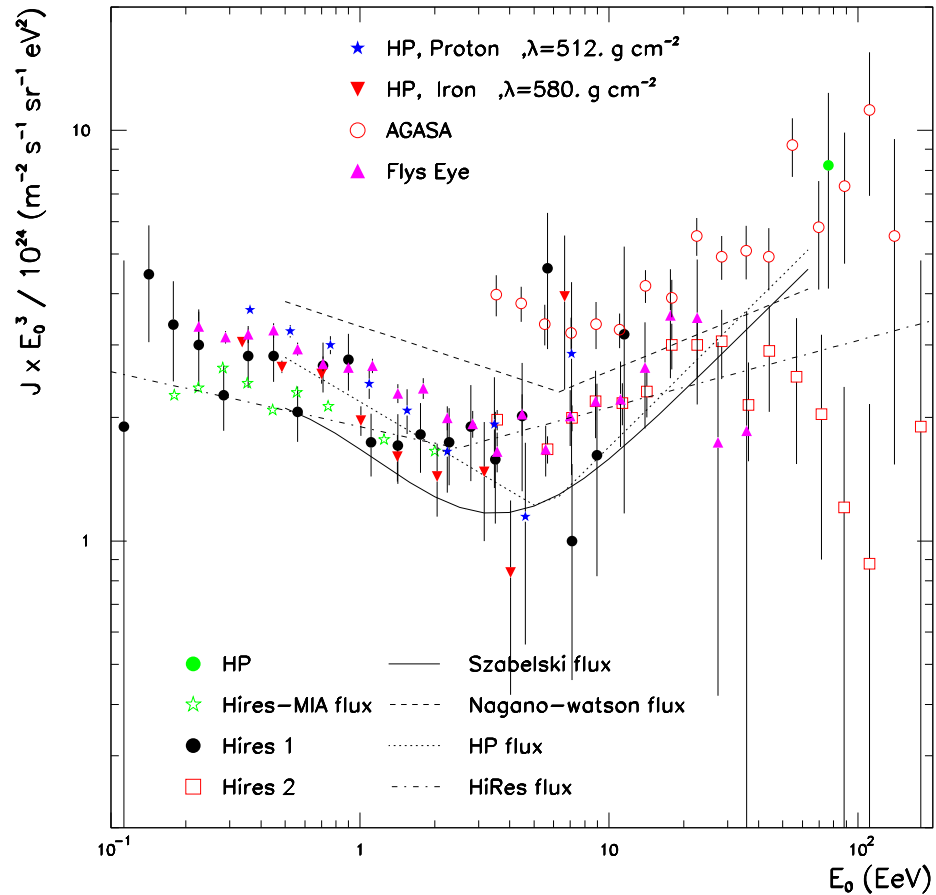}{0.80}
\caption{A composite energy spectrum including recently reanalyzed Haverah 
  Park data assuming proton and iron primaries (the parameter
  $\lambda$ measures the attenuation length of the density of charged
  particles at 600~m from the shower core), stereo Fly's Eye data,
  monocular HiRes data from both eyes up to $60^\circ$, and hybrid
  HiRes--MIA data. Published in \protect{Ref.~\cite{Watson:2001}}.}
\label{f2}
\end{figure}

Above $10^{14}$ eV, the flux becomes so low that only ground-based
experiments with large aperture and long exposure times can hope to
acquire a significant number of events. Such experiments exploit the
atmosphere as a giant calorimeter. The incident cosmic radiation
interacts with the atomic nuclei of air molecules and produces
extensive air showers (EAS) which spread out over large areas.
Already in 1938, Pierre Auger concluded from the size of EAS that the
spectrum extends up to and perhaps beyond
$10^{15}$~eV~\cite{Auger:1938,Auger:1939}. Nowadays substantial
progress has been made in measuring the extraordinarily low flux
($\sim 1$ event km$^{-2}$ yr$^{-1}$) above $10^{19}$ eV. Continuously
running experiments using both arrays of particle detectors on the
ground like AGASA and fluorescence detectors which track the cascade
through the atmosphere like HiRes, have detected events with primary
particle energies higher than $10^{20}$~eV
\cite{volcano,Suga:jh,yakutsk,haverah,agasa:1,hires,largest,newhires},
with no evidence that the highest energy recorded thus far is Nature's
upper limit (see fig.~\ref{f2}).

Such tremendous energies, well above the energies that
we can expect to reach on Earth in the current and future generations
of colliders, defy our understanding for number of reasons; in
particular, CRs of energies beyond the ankle ($\sim 10^{18}$ eV) are
expected to be of extragalactic origin, and probably coming from
distances further than the local cluster of galaxies ($\sim 20$ Mpc),
because we know no astrophysical object able to accelerate particles
enough to provide them such an energy in our vicinity. Yet, particles
carrying energies above $10^{20}$ eV traveling over cosmological
distances should lose their energy through propagation effects; for
example, a proton will interact with the cosmological microwave
background (CMB) and photoproduce pions, with an interaction length of
a few tens of Mpc, losing around 20 \% of its energy at each
interaction. Similar processes occur with nuclei, photons (through
gamma-gamma pair production over the radio background (URB)
essentially) or electrons (Inverse Compton scattering over the URB).
Thus particles with initial energy $\sim 10^{20}$ eV coming from
cosmological distances (over 50 Mpc) should reach the Earth with a
maximal energy $\sim 5.10^{19}$ eV\footnote{A notable exception are,
  of course, the neutrinos, which can travel over cosmological
  distances without loosing their energy. But the events observed on
  Earth cannot be attributed to primary neutrinos, because of the
  shape of the observed air showers.}, as it was already predicted
independently by Greisen \cite{GZK:1}, and Zatsepin and Kuzmin
\cite{GZK:2} in 1966; the expected fall off of the spectrum is
nowadays referred as the GZK cut-off. Yet, events have been registered
above this cut-off in very different experiments over the last few
decades
\cite{volcano,haverah,yakutsk,sugar,agasa:1,hires,largest,hires-mia}.
A recent review of the experimental results is also given in
\cite{reviewYoshida}.

The presence or absence of this cut-off in the CR spectrum is still a
matter of controversy: as one can see from fig.~\ref{f2}, if it seems
to be present in the updated HiRes spectrum \cite{newhires}, it is
fully absent in the AGASA one \cite{agasa:1}. But all experiments agree on the
existence of events well above this cut-off; this result is certainly
the strongest argument against the classical'' acceleration (or
``bottom-up'') theories, which are usually recognized to be the most
efficient mechanism for producing the CR spectrum at lower energies.
Indeed, the bottom--up theories exploit the electromagnetic fields
that are likely to be present in objects like gamma ray bursters
\cite{WaxmanGRB}, ``hot spots'' of radio-galaxies \cite{Biermann} or
near super--massive black holes in dormant quasars \cite{Boldt} in
order to accelerate charged particles. However, it is difficult to
find objects capable of accelerating protons to energies above
$10^{20}$ eV, partly because the product of field strength and spatial
extension of the field does not seem to be sufficiently large, and
partly because the accelerated particles can loose a fair fraction of
their energy in synchrotron radiation. Moreover, there is another
strong indication against these models: UHECRs are expected to travel
rather straight away in the universe, without being deviated by the
(inter)galactic magnetic fields. Thus they should point to there
sources within a few degrees. Yet, excepted the existence of a few
doublets and triplets in the experimental data, the observations are
compatible with an almost perfect isotropy above $4. 10^{19}$ eV
\cite{agasa_anisotropy}\footnote{Nevertheless, it should be noted that
  there are still attempts to explain the UHECRs through acceleration
  mechanisms, see for example \cite{WaxmanGRB,Biermann}.}.

These remarks lead to the development of another class of models
attempting to explain the existence of UHECRs, generically called
``top-down'' theories. Here one postulates the existence of ``new
physics'' at a very high energy scale, i.e. the existence of
super-heavy particles of masses greater than $10^{12}$ GeV, which
could decay and hence produce the observed UHECRs. Top--down models can
be motivated by a variety of arguments. For example, the recent
measurements of the cosmic microwave background and of supernova
redshifts have dramatically confirmed that our universe contains a
large fraction of cold dark matter \cite{cmb}. A top--down model in
which annihilating or decaying superheavy relic particles produce the
highest energy cosmic rays could potentially solve both of these
problems \cite{bere1,Birkel,SHDM}. Several mechanisms for the
production of such ultra--massive particles at the end of inflation
have been suggested \cite{creat}. Moreover, particles with the
required mass and lifetime are predicted to exist in certain
superstring theories \cite{wimpzillas}. Another explanation for the
required long lifetime of these particles is to confine them into
topological defects \cite{Hill}. These ideas have been lengthy
reviewed e.g. in \cite{reviewSigl,Cronin,reviewAnchordoqui,reviewSarkar}.

In this chapter, we will focus on this last class of models, in order
to see which {\it quantitative} predictions we can make within our
current knowledge, using the results presented in the previous
chapters. We will see that is it possible to make testable predictions
for both neutrino and LSP fluxes which could be observed on Earth in
the next generation of experiments, thus providing a way to reinforce
or rule out this class of models.

\section{What can we learn?}
\label{sec:learn}

As we have seen in chapter~\ref{chap:code}, the SHdecay code provides the
spectra of the stable particles at source, from the decay of a unique
X particle. In order to make predictions on the real fluxes that could
be observed on Earth in the current and future detectors, we need a
complete model for the UHECR production and propagation, which
includes:
\begin{itemize}
\item[a)] A given model of $X$ particles, specifying the nature of the $X$
  particle, or at least its decay modes (with their branching ratios)
  - and possibly its lifetime.
\item[b)] A cosmological model for $X$ particles production
  and evolution which allows to determine their current
  distribution in the universe.
\item[c)] A propagation code for all types of stable particles produced
  in the decay, taking into account the travel distances, the galactic
  and intergalactic magnetic fields, and all leading interactions
  between the UHECRs and the different backgrounds (CMB, infrared,
  radio).
\item[d)] An ``extensive air shower'' code for each kind of initial
  particles. 
\item[e)] A good knowledge of the detectors, especially their
  acceptances, exposure times and efficiencies, and a good study of
  all possible backgrounds for a given signal.
\end{itemize}

Of course, in our current understanding of these problems, the
estimations that can be done are extremely model dependent. Yet, we
can parameterize our ignorance and use the current known data to
partly get rid of this model dependency. In particular, we certainly
don't need to know all details related to the $X$ particle for pursuing
the study: we can consider a few typical decay modes
(hadronic/leptonic/bosonic with small or large number $N$ of decay
products), and different typical distributions of $X$ particles in the
universe, independently of the whole story of their production and
evolution since the very early ages of the universe. In order to get
the required density of sources, we simply can normalize our spectra
on the available data for a given type of observed particles, and use
this parameterization for making predictions on the fluxes of other
particles. In the next sections, we will follow this phenomenological
procedure in order to get ``order of magnitude'' predictions for the
neutrino and neutralino fluxes on Earth.

There is one more point that we have to address here before going
further: in order to make quantitative predictions on the neutrino and
neutralino fluxes that could be observed on Earth, we need to make an
assumption on the {\it composition} of the observed UHECR spectrum.

Established particle physics implies that UHE jets
fragment predominantly into pions and kaons, with a small admixture of
protons \cite{lep}. The mesons will eventually decay into photons or
electrons plus neutrinos. A typical QCD jet therefore produces more
photons than protons. This is true in particular at relatively low
values of $x = E_{\rm particle} / E_{\rm jet}$, but even at large $x$
the photon flux is at least as large as the proton flux in a jet
\cite{SHDM,bere2,Berezinsky:2000,Birkel,Sarkar:2001}. This seems to be
in disagreement with mounting evidence that the highest energy cosmic
rays are not photons \cite{nophot}. The observed shower profile of
the original Fly's Eye event \cite{largest}, with energy exceeding
$10^{20}$ eV, fits the assumption of a primary proton, or, possibly,
that of a nucleus. The shower profile information is sufficient to
conclude that the event is unlikely to be of photon origin
\cite{pro1}. The same conclusion is reached for the Yakutsk event that
is characterized by a large number of secondary muons, inconsistent
with a purely electromagnetic cascade initiated by a gamma ray. A
reanalysis of Haverah Park data further reinforces this conclusion
\cite{newhaverah}. Very recently AGASA published \cite{agasa:1} strong
upper limits on the $\gamma/p$ ratio. Their data are compatible with
being entirely due to protons, and strongly disfavor scenarios where
most events are photonic in origin. This conclusion is again based on
the observed number of muons in UHE showers, as well well as on the
absence of a south--north asymmetry in the events. Such an asymmetry
would be produced by $\gamma \rightarrow e^+e^-$ conversion in the
Earth's magnetic field, which becomes effective for $E_\gamma \geq
10^{19}$ eV. In light of this information, it seems likely that
protons, and not gamma rays, dominate the highest energy cosmic ray
spectrum. This strongly disfavors superheavy particles as the source
of the highest energy cosmic rays, {\em unless} UHE photons are
depleted from the cosmic ray spectrum near $10^{20}$ eV, leaving a
dominant proton component at GZK energies. In fact, the uncertainties
associated with the cascading of the jets in the universal and
galactic radio backgrounds and with the strength of intergalactic
magnetic fields leave this possibility open at least for sources at
cosmological distances \cite{reviewSigl,radio}. If most relevant
sources are located in the halo of our own galaxy, as expected for
free nonrelativistic particles \cite{Birkel}, one would have to assume
that the galactic radio background has been underestimated by about an
order of magnitude, which may not be impossible \cite{private}. With
this in mind, we will choose to normalize the proton spectrum from
top--down scenarios to the observed UHECR flux. We recognize that this
assumption may be somewhat extreme especially for local sources.
However, we believe that it is necessary for the viability of this
kind of model; and it is no more extreme than many other proposed
explanations of the post--GZK events. Moreover, as we will see in the
following sections, relaxing this assumption by allowing a possible
fraction of UHE photons in the UHECR spectrum only reduces the
predicted event rate by factors of two or three; this is less than the
inherent uncertainty of our calculation, arising from the discrepancy
in the current data and the unavoidable dependence on the unknown $X$
decay mode.

\section{UHE neutrino fluxes on Earth: detection at the corner?}
\label{sec:neutrinos}

Neutrinos are produced more numerously than protons and travel much
greater distances. Possible neutrino signals of top--down models have
therefore been suggested quite a while ago \cite{neutrinos}. Here we
perform an up--to--date analysis of these signals in various kinds of
detectors, using the most accurate available calculation of the
neutrino fluxes at source (see chapter~\ref{chap:decay} or
\cite{bd1,bd2}). Once we ``renormalize'' the observed cosmic ray flux
to protons, we generically predict observable neutrino signals in
operating experiments such as AMANDA II, RICE and AGASA. Top--down
models, if not revealed, will be severely constrained by high--energy
neutrino observations in the near future.

\subsection{Nucleons from ultra-high energy jets}

The assumption that nucleons from the decay (or annihilation) of very
massive $X$ particles are the source of the highest energy cosmic rays
normalizes the decay or annihilation rate of their sources, once the
shape of the spectrum of the produced nucleons is known. One needs
mass $M_X \geq 10^{21}$ eV in order to explain the observed UHECR
events. The presence of such very massive particles strongly
indicates the existence of superparticles with masses at or below the
TeV scale, since otherwise it would be difficult to keep the weak
energy scale ten or more orders of magnitude below $M_X$ in the
presence of quantum corrections. Moreover, we know that all gauge
interactions are of comparable strength at energies near $M_X$. These
two facts together imply that the evolution of a jet with energy $\geq
10^{21}$ eV shows some new features not present in jets produced at
current particle collider experiments.

First of all, primary $X$ decays are likely to produce approximately
equal numbers of particles and superparticles, since $M_X$ is much
larger than the scale $M_{\rm SUSY} \leq 1$ TeV of typical
superparticle masses. Even if the primary $X$ decay only produces
ordinary particles, superparticles will be produced in the subsequent
shower evolution \cite{Berezinsky:2000,Sarkar:2001}. Note also that
(at least at high energies) electroweak interactions should be
included when modeling the parton shower. Both effects taken together
imply that the jet will include many massive particles --
superparticles, electroweak gauge and Higgs bosons, and also top
quarks. The decays of these massive particles increase the overall
particle multiplicity of the jet, and also produce quite energetic
neutrinos, charged leptons and lightest supersymmetric particles
(LSPs). Eventually the quarks and gluons in the jet will hadronize into
baryons and mesons, many of which will in turn decay.

We model these jets at the point of their origin using SHdecay. The
results presented below have been obtained using a ``typical''
spectrum with superparticles in the hundreds of GeV region; the
dependence of the predicted neutrino fluxes on the spectrum of
superparticles is much weaker than that on the primary $X$ decay mode
discussed below. At virtualities below $M_{\rm SUSY}$ only ordinary
QCD interactions contribute significantly to the development of the
jet; $b$ and $c$ quarks are decoupled at their respective masses,
hadronize, and decay. At a virtuality near $1$ GeV the light quarks
and gluons hadronize, with a meson to baryon ratio of roughly thirty
to one (five to one) at small (large) $x$. All baryons will eventually
decay into protons, while the mesons (mostly pions) decay into
photons, electrons\footnote{Electrons quickly lose their energy
through synchrotron radiation, and therefore do not contribute to the
observed UHECR flux.} and neutrinos (plus their antiparticles). The
heavier charged leptons (muons and taus) also decay. The final output
of the code is the spectra of seven types of particles which are
sufficiently long--lived to reach the Earth: protons, electrons,
photons, three flavors of neutrinos, and LSPs. We assume that $X$
decays are CP--symmetric, i.e. we assume equal fluxes of particles and
antiparticles of a given species.

The first version of SHdecay (as presented in Ref.\cite{bd1}) was
based on conventional one--loop evolution equations for the relevant
fragmentation functions. These may not be reliable in the region of
very small $x$. We wish to calculate neutrino fluxes at energies down
to $\sim 10^{15}$ eV (1 PeV), which corresponds to $x \sim 10^{-6} \ 
(10^{-10})$ for $M_X = 10^{21} \ (10^{25})$ eV. At these very small
$x$ values color coherence effects are expected to suppress the shower
evolution \cite{mlla}. We tried to estimate the size of these effects by
matching our spectra computed using conventional evolution equations
to the so--called asymptotic MLLA spectra (see section~\ref{sec:MLLA}
of chapter~\ref{chap:decay}). The effect of this modification on the
neutrino event rate is relatively modest for primary jet energy near
$10^{21}$ eV, but becomes significant at $10^{25}$ eV. However, even
at this higher energy the proton flux, which we only need at $x \geq
10^{-5}$, is not affected significantly.

\begin{figure}[h!]
\centering\leavevmode
\includegraphics[width=5in]{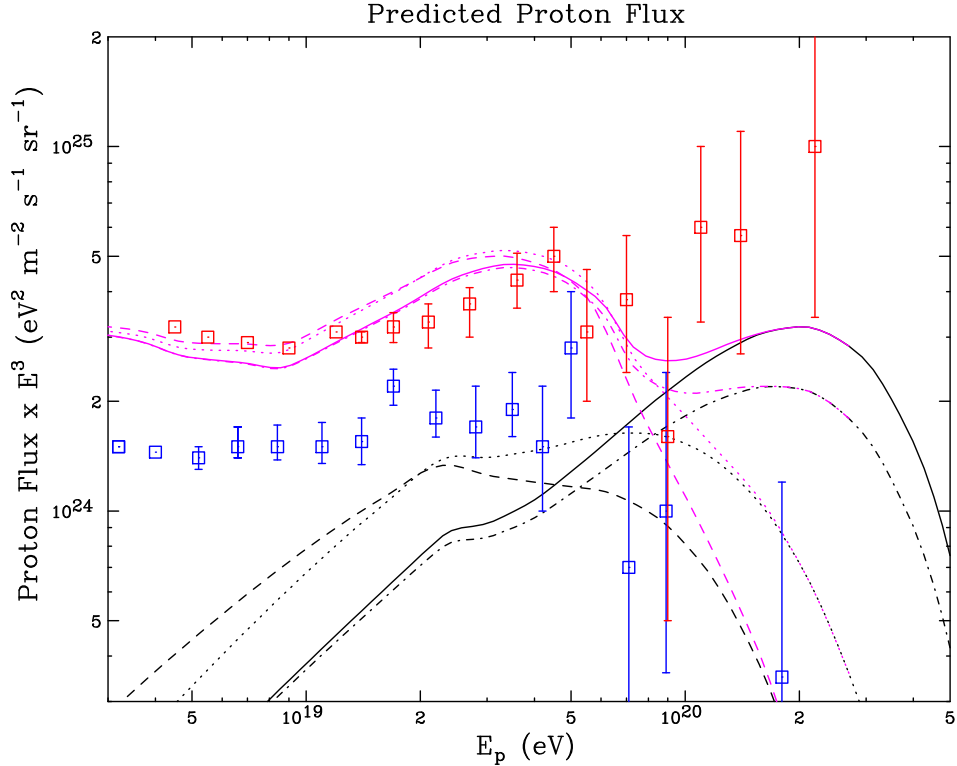}
\caption{The ultra high--energy cosmic ray flux predicted for the decay
of superheavy particles with mass $M_X = 2 \cdot 10^{21}$ eV is
compared to the HiRes (darker) and AGASA (lighter) cosmic ray data.
The distribution of jets used includes an overdensity factor of $10^5$
within 20 kpc of the galaxy. Spectra are shown for quark$+$antiquark
(solid), quark$+$squark (dot--dash), $SU(2)$ doublet lepton$+$slepton
(dots) and 5 quark$+$5 squark (dashes) initial states. Dark (lower)
lines are from top--down origin alone whereas lighter (upper) lines are
top--down plus an homogeneous extragalactic contribution as predicted
in Ref.\cite{reviewSigl}. Note that all observed super GZK events can be
explained by this mechanism.}
\label{normalisation_21}
\end{figure}

\begin{figure}[h!]
\centering\leavevmode
\includegraphics[width=5in]{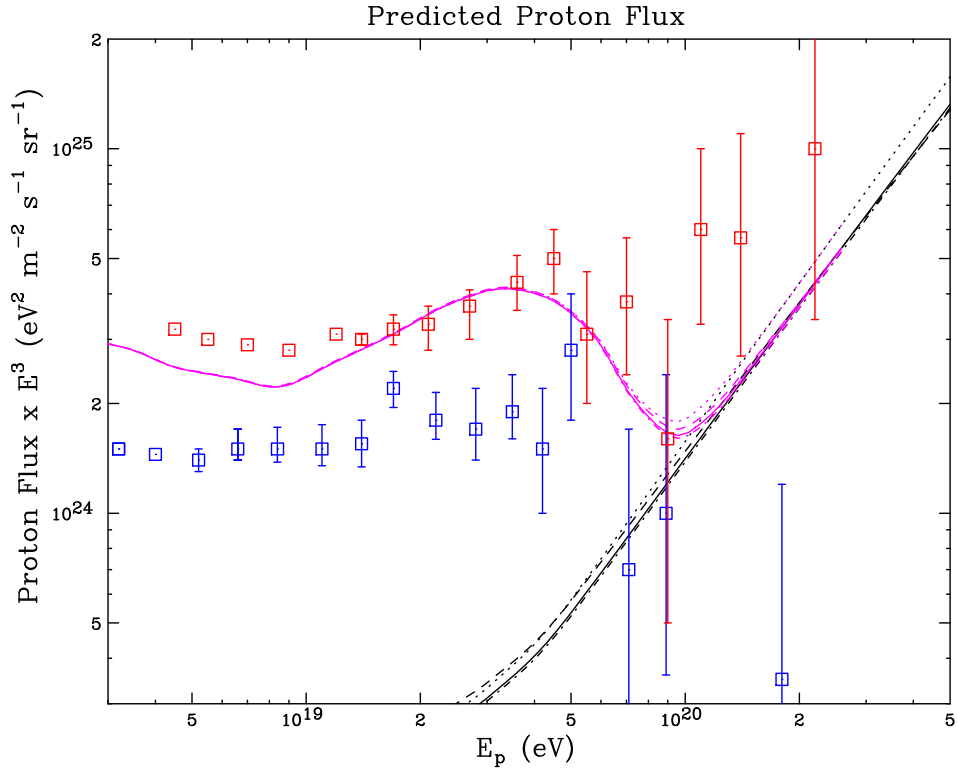}
\caption{As in figure~\ref{normalisation_21}, but using particles of mass $M_X = 2 \cdot
10^{25}$ eV.} 
\label{normalisation_25}
\end{figure}

This calculation gives us the shape of the spectra of the stable
particles at source. The spectra on Earth might differ significantly
due to propagation effects. As stated in section~\ref{sec:learn}, we will
assume that (almost) all UHE photons get absorbed. This is actually
expected to be true for a homogeneous source distribution. However,
according to current estimates of the strengths of the magnetic fields
and of the radio wave background in (the halo of) our own galaxy most
UHE photons produced in the halo of our galaxy are expected to reach
the Earth. As we already saw, this seems to be in conflict
with observation. We will therefore assume that the interaction length
of UHE photons in our galaxy has been greatly over--estimated, and
explore the consequences of this assumption for neutrino signals.

As well known, (anti)protons lose energy when traveling through the
intergalactic medium, mostly through scattering off photons of the
ubiquitous cosmic microwave background (CMB). We calculate the
observed spectrum of protons taking into account scattering off the
CMB at the $\Delta-$resonance and scattering by $e^+e^-$ pair
production; energy losses through the Hubble expansion of the Universe
are also included \cite{reviewSigl,prop}. Note that the photoproduction of
charged pions contributes to the observed neutrino flux on Earth. In
order to solve the ultra high--energy cosmic ray problem, the
(anti)proton flux must accommodate the events above the GZK cutoff.
Observations indicate on the order of a few times $10^{-27}$ events
$\rm{m}^{-2} \rm{s}^{-1} \rm{sr}^{-1} \rm{GeV}^{-1}$ in the energy
range above the GZK cutoff ($5 \times 10^{19}$ eV to $2 \times
10^{20}$ eV).

The formalism of a generic top--down scenario is sufficiently flexible
to explain the data from either the HiRes \cite{newhires} or AGASA
\cite{agasa:1} experiments. Figure~\ref{normalisation_21} compares
HiRes and AGASA data to the proton spectrum predicted for a galactic
distribution of decaying particles with mass $M_X = 2 \cdot 10^{21}\,$
eV. The drop near a few times $10^{19}$ eV is a manifestation of the
GZK cutoff. Note, however, that there are sufficient semi--local
events to explain all observed super GZK events. Similarly,
figure~\ref{normalisation_25} compares HiRes and AGASA data to the
spectrum predicted for $M_X = 2 \cdot 10^{25}$ eV, rather than $ 2
\cdot 10^{21}$ eV, decaying particles for the same distribution.
Although HiRes and AGASA data differ at face value, especially above
the GZK cutoff, top--down scenarios can accommodate all events
observed above the GZK cutoff in either experiment.

If the cosmic ray sources are not distributed with a large overdensity
in the galaxy, the resulting cosmic ray and neutrino spectrum will be
modified. For example, using a homogeneous distribution, the GZK
cutoff will again be manifest and the observed cosmic ray spectrum
will be difficult to explain\footnote{The fit would improve if we
  allowed the background to float as well. However, in such a scenario
  one expects a break in the spectrum due to the GZK effect, which is
  not seen in the AGASA data.}. A galactic overdensity of $10^3$ to
$10^4$ or more seems necessary to fit the data. The
figure~\ref{normalisation_21} shows a $10^5$ overdensity, which is the
overall overdensity of matter in our galaxy at the location of the
Sun\footnote{We assumed constant $X$ overdensity by a factor of
  $10^5$ out to a distance of 20 kpc, with homogeneous $X$
  distribution at larger distances. From galactic modeling one expects
  the Dark Matter halo to be larger, with gradually declining
  overdensity. However, all that matters for us is that in a
  ``galactic'' distribution nearly all UHE protons are produced at
  distances well below one interaction length. The actual halo profile
  does not affect our results once we normalize the proton flux to the
  observed UHECR flux.}. Note that for less extreme over-densities, the
average distance at which a proton is produced will be larger. This
implies larger energy losses, and hence a reduced proton flux on Earth
for a given number of sources. Conversely, if we fix the proton flux
to the observed flux of UHECR events, models with lower overdensity
require more sources. Since neutrino fluxes are not degraded by
propagation through the intergalactic medium, the number of neutrinos
increases proportionally to the number of sources, with additional
contributions to the neutrino flux coming from pion production on the
CMB background. Thus, the neutrino event rates and spectrum shown in
the figures reflect the most conservative choice of distributions.
Table~\ref{table:neutrinos} shows results for both homogeneous and
galactic distributions.

\subsection{Neutrinos from ultra-high energy jets}

Neutrinos, not being limited by scattering, travel up to the age of
the universe at the speed of light ($\sim$ 3000 Mpc in an Euclidean
approximation). The only nontrivial effect of neutrino propagation is
due to oscillations. In our case the propagation distance of neutrinos
amounts to many oscillation lengths, if oscillation parameters are
fixed by the currently most plausible solutions of the atmospheric and
solar neutrino deficits \cite{bahcall}. As a result, the UHE neutrino
flux on Earth is the same for all three flavors, and amounts to the
average of the fluxes of the three neutrinos flavors at source.

\begin{figure}[h!]
\centering\leavevmode
\includegraphics[width=5in]{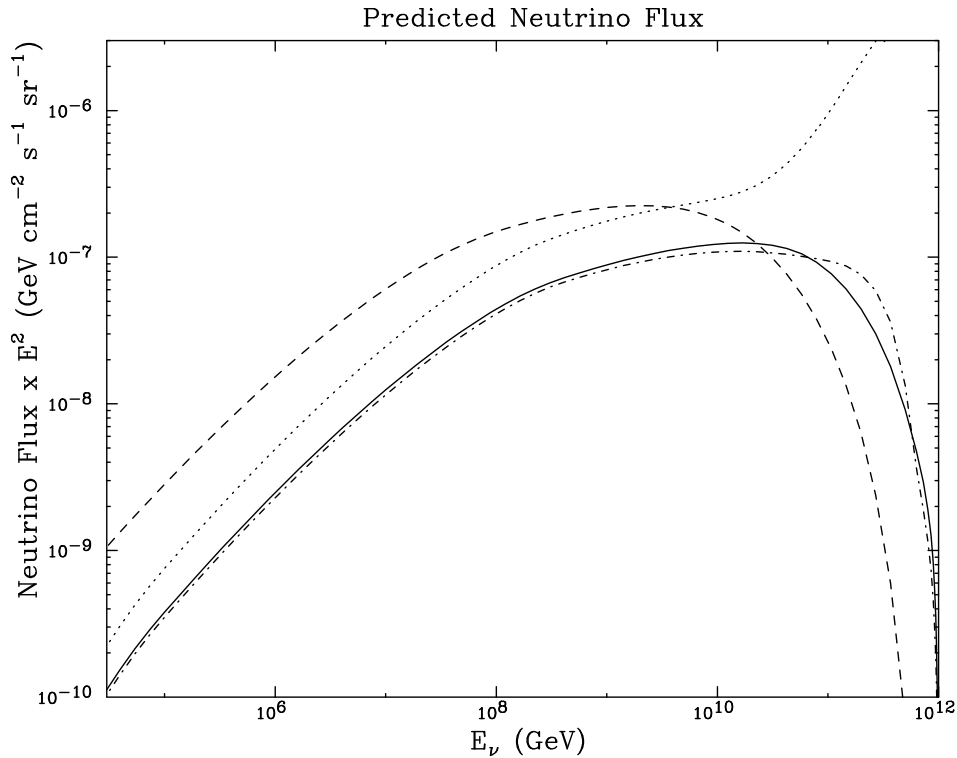}
\caption{The neutrino plus anti--neutrino flux corresponding to the
  cosmic ray spectra of figure~\ref{normalisation_21} from the decay
  of superheavy particles with mass $M_X = 2 \cdot 10^{21}$ eV.
  Spectra are shown for quark--antiquark (solid), quark--squark
  (dot--dash), lepton--slepton (dots) and 5 quark--5 squark (dashes)
  initial states.}
\label{neutrino_fluxes_21}
\end{figure}

\begin{figure}[h!]
\centering\leavevmode
\includegraphics[width=5in]{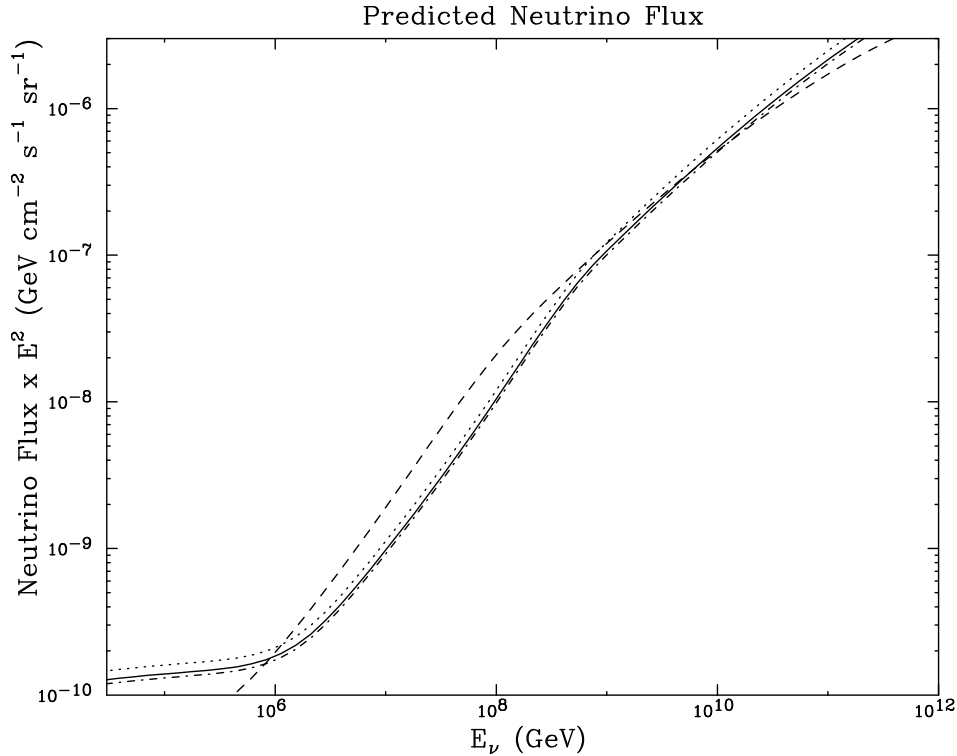}
\caption{As figure~\ref{neutrino_fluxes_21}, but corresponding to the cosmic ray spectrum of
  figure~\ref{normalisation_25} with $M_X = 2 \cdot 10^{25}$ eV.}
\label{neutrino_fluxes_25}
\end{figure}

The predicted neutrino flux is shown in
figures~\ref{neutrino_fluxes_21} and~\ref{neutrino_fluxes_25}. At
$E_\nu \ll E_{\rm jet}$ the main contribution comes from $\pi^\pm
\rightarrow \mu^\pm \nu_\mu \rightarrow e^\pm \nu_e \nu_\mu$ decays,
but at larger $E_\nu$ there can be significant contributions from the
decays of heavy (s)particles. The peak in the dotted curves at $E_\nu
= E_{\rm jet}$ results from our assumption that in this scenario $X$
decays directly into first or second generation $SU(2)$ doublet
(s)leptons, which implies that 50\% of all $X$ decays give rise to a
primary neutrino; in this case the ratio of neutrino and proton fluxes
has a maximum at high energy. On the other hand, if primary $X$ decays
are purely hadronic, the neutrino flux at the largest energy is only
slightly above the proton flux at that energy. The reason is that
neutrinos from meson decays only carry a fraction of the energy of the
meson, so a five to one meson to proton ratio at large $x$ leads to a
nearly one to one neutrino to proton ratio. We see that the neutrino
flux at the highest energy depends quite strongly on how the $X$
particles decay; there is also some dependence on the parameters of
the SUSY model \cite{bd1,bd2}. For given proton flux the neutrino flux
at smaller $x$ is much less model dependent. At very small $x$ a new
uncertainty appears due to coherence effects. These have so far only
been studied in a pure QCD parton shower; our treatment of these
effects is therefore of necessity rather crude.

\subsection{Event rates in high-energy neutrino telescopes and air shower
experiments} 

We will discuss two classes of experiments capable of observing high
energy cosmic neutrinos: neutrino telescopes and air shower
experiments.

Optical Cerenkov neutrino telescopes such as the operating AMANDA II
and next generation IceCube \cite{Icecube} are designed to observe
muon tracks from charged current interactions as well as showers which
occur in the detector. The probability of detecting a neutrino
passing through the detector from its muon track is given by
\begin{equation}
P_{\nu \rightarrow \mu}(E_\nu, \theta_{\rm zenith}) = \sigma_{\nu
N}(E_\nu)\, n_{\rm{H_2 O}} \, R_\mu(E_\mu, \theta_{\rm zenith}), 
\end{equation}
where $n_{\rm{H_2 O}}$ is the number density of nucleons in the
detector medium (water or ice), and the muon range
$R_{\mu}(E_{\mu},\theta_{\rm{zenith}})$ is the average distance
traveled by a muon of energy $E_{\mu}$ before falling below some
threshold energy (we have used 100 TeV). This quantity depends on the
zenith angle of the incoming neutrino because for a detector depth of
$\sim 2\,$km, only quasi--horizontal or upgoing events can benefit from
longer muon ranges. At the energies we are most concerned with, the
majority of muon events will be quasi--horizontal.
The number of muon events observed is then given by
\begin{equation}
N_{\rm{events}} = \int dE_{\nu}\, d\Omega \,\frac {d \phi_\nu}{dE_\nu}\, P_{\nu
\rightarrow \mu} (E_{\nu},\theta_{\rm zenith})\, A_{\rm eff}\, T, 
\end{equation}
where $T$ is the time observed and $A_{\rm{eff}}$ is the effective
area of the detector: one twentieth square kilometers for AMANDA II
and one square kilometer for IceCube.

AMANDA II and IceCube can also observe showers generated in charged or
neutral current interactions within the detector volume. The event
rate from showers is not enhanced by long muon ranges, but can be
generated by all three flavors of neutrinos and with greater cross
section (neutral + charged current). We use a shower energy threshold
of 100 TeV. The energy threshold imposed effectively removes any
background events from atmospheric neutrino events. For a review of
Optical Cerenkov neutrino telescopes see Ref.\cite{ice2}.

The operating radio Cerenkov experiment, RICE, is capable of observing
showers generated in charged current electron neutrino events. RICE's
effective volume increases with energy. At 1 PeV, RICE has an
effective volume less than one hundredth of a cubic kilometer. At
higher energies, however, it increases to about ten cubic kilometers
\cite{rice}.

Air shower experiments can also observe very high energy cosmic
neutrinos. We consider AGASA, the largest ground array currently in
operation \cite{agasa:2}, and the next generation AUGER array
\cite{auger}.

To determine that an air shower was initiated by a neutrino, rather
than a proton or other cosmic ray, we require a slant depth greater
than 4000 $\rm{g}/\rm{cm}^2$. This corresponds to a zenith angle near
75 degrees. Therefore, only quasi--horizontal air shower events can
be identified as neutrinos. Additionally, unlike showers generated in
the upper atmosphere, deeply penetrating showers provide both muon and
electromagnetic shower components which help them be differentiated
from showers with hadronic primaries. The probability of detecting and
identifying a neutrino initiated air shower is described in terms of
the array's acceptance, $A$, in units of volume times water equivalent
steradians (we sr). The detector's acceptance increases with energy.
For AGASA, the acceptance is about $0.01 \,\rm{km}^3\,\rm{we}\,{sr} $
at $10^7\,$ GeV but increases to $1.0 \,\rm{km}^3\,\rm{we}\,{sr} $ at
$10^{10}\,$ GeV and above. For AUGER, the acceptance is about $0.1
\,\rm{km}^3\,\rm{we}\,{sr} $ at $10^8\,$ GeV and reaches above
$10.0\,\rm{km}^3\,\rm{we}\,{sr} $ by $10^{10}\,$ GeV. The number of
events observed is then
\begin{equation}
N_{\rm events} = \int dE_\nu \, d\Omega \, n_{\rm{H_2 O}} \, \frac{d \phi_\nu} {dE_\nu}
\,\sigma_{\nu N}(E_\nu) \,A(E_\nu) \,T, 
\end{equation}
where $T$ is again the time observed, $n_{\rm{H_2 O}}$ is the number
density of nucleons in water and $A(E_{\nu})$ is the detector's
acceptance. AGASA presently has about five years of effective running
time between 1995 and 2000 analyzed.

The AUGER array will also be capable of observing upgoing showers
generated by tau neutrinos in the shallow earth. The rates for upgoing
tau neutrinos events are typically about an order of magnitude higher
than the rates for quasi horizontal downgoing neutrino events. For
more a detailed description of air shower acceptances and rates from
upgoing and quasi horizontal neutrino induced showers, see
Ref.\,\cite{feng}.

\begin{table}[thb]
\begin{center}
\begin{tabular}{c|c c c c c c c c c c c} 
& 
\multicolumn{2}{c }{Amanda II}&  \multicolumn{2}{c }{Agasa}  &
\multicolumn{2}{c }{Rice}  & \vline & \multicolumn{2}{c }{IceCube}
& \multicolumn{2}{c }{Auger} \\  
\hline \hline
$q \bar{q}$, $10^{21}$ eV, Gal
&~~~0.29&&~~0.030&&~~1.5&&\vline&9.6&&1.2\\
$q \tilde{q}$, $10^{21}$ eV, Gal
&~~~0.29&&~~0.028&&~~1.5&&\vline&9.5&&1.2\\
$5 \times q \tilde{q}$, $10^{21}$ eV, Gal
&~~~0.97&&~~0.065&&~~3.5&&\vline&32.1&&2.9\\
$l \tilde{l}$, $10^{21}$ eV, Gal
&~~~0.50&&~~0.079&&~~3.9&&\vline&16.1&&2.5\\
\hline  \hline 
$q \bar{q}$, $10^{25}$ eV, Gal
&~~~0.027&&~~0.0094&&~~0.38&&\vline&0.80&&0.29\\
$q \tilde{q}$, $10^{25}$ eV, Gal
&~~~0.026&&~~0.0092&&~~0.37&&\vline&0.88&&0.28\\
$5 \times q \tilde{q}$, $10^{25}$ eV, Gal
&~~~0.034&&~~0.010&&~~0.42&&\vline&1.0&&0.33\\
$l \tilde{l}$, $10^{25}$ eV, Gal
&~~~0.029&&~~0.011&&~~0.42&&\vline&0.87&&0.32\\
$q \bar{q}$, no MLLA, $10^{25}$ eV, Gal
&~~~0.041&&~~0.0099&&~~0.40&&\vline&1.2&&0.31\\
\hline  \hline 
$q \bar{q}$, $10^{21}$ eV, Hom
&~~~2.6&&~~0.27&&~~13.8&&\vline&86.0&&11.0\\
$q \tilde{q}$, $10^{21}$ eV, Hom
&~~~2.6&&~~0.25&&~~13.2&&\vline&85.1&&10.5\\
$~5 \times q \tilde{q}$, $10^{21}$ eV, Hom
&~~~8.7&&~~0.59&&~~31.5&&\vline&289.1&&25.7\\
$l \tilde{l}$, $10^{21}$ eV, Hom
&~~~4.5&&~~0.71&&~~35.3&&\vline&144.9&&22.9\\
\hline  \hline 
$q \bar{q}$, $10^{25}$ eV, Hom
&~~~0.40&&~~0.14&&~~5.7&&\vline&12.0&&4.3\\
$q \tilde{q}$, $10^{25}$ eV, Hom
&~~~0.39&&~~0.14&&~~5.6&&\vline&11.7&&4.2\\
$5 \times q \tilde{q}$, $10^{25}$ eV, Hom
&~~~0.51&&~~0.15&&~~6.3&&\vline&15.5&&5.0\\
$l \tilde{l}$, $10^{25}$ eV, Hom
&~~~0.44&&~~0.16&&~~6.4&&\vline&13.1&&4.7\\
$q \bar{q}$, no MLLA, $10^{25}$ eV, Hom
&~~~0.62&&~~0.15&&~~5.9&&\vline&18.5&&4.6\\
\end{tabular}
\caption{Neutrino events per year in top--down scenarios for several
operating and next generation experiments. For AMANDA II and IceCube,
100 TeV shower and muon energy thresholds were imposed. Events are
only calculated up to $10^{12}$ GeV as discussed in the text.}
\label{table:neutrinos} 
\end{center}
\end{table}

Table~\ref{table:neutrinos} shows the event rates expected for a
variety of models, and for several experiments. AMANDA-II, with an
effective area of $\sim$50,000 square meters can place the strongest
limits on high energy neutrino flux presently. Furthermore, AGASA,
with five years of effective observing time, has similar sensitivity.
RICE, just beginning to release results, will be capable of raising
the level to which top--down scenarios can be tested, perhaps being
capable of testing many of the models shown in the table. Even if no
events are observed with operating experiments, next generation
experiments, such as IceCube and AUGER, will be able to test all
models with adequate sensitivity.

Event rates shown in table~\ref{table:neutrinos} include only events
below $10^{12}$ GeV. Above this energy, uncertainties in the
neutrino--nucleon cross sections and in detector performance make such
calculations difficult and unreliable. Our most reasonable
extrapolations into this energy range indicate about a 20\%
enhancement to the event rate if all energies are considered for
$10^{25}$ eV jets. There is no effect for the $10^{21}$ eV jet case.

High--energy neutrino event rates have been calculated in
Ref.\,\cite{jaime} for a similar model. Their calculation used the
model of reference \cite{bere1} which normalized the ultra high--energy
cosmic ray flux to the photons and protons generated in superheavy
particle decay rather than the proton flux alone. For this reason,
their results show only two events per year in a square kilometer
neutrino telescope, a smaller rate than we predict for most
models. Another recent estimate of neutrino fluxes on Earth in
top--down models \cite{kalashev} finds broadly similar results as
our's. However, there the `MLLA' form for the fragmentation functions
was used for all energies, which (incorrectly) predicts nearly
energy--independent ratios of neutrino, photon and proton fluxes.

\subsection{Conclusions}

If a top--down scenario, such as the decay or the annihilation of
superheavy relics, is the source of the highest energy cosmic rays,
then a UHE neutrino flux should accompany the observed cosmic
ray flux. This neutrino flux will be much higher than the flux of
nucleons due to the much greater mean free path of neutrinos and
greater multiplicity of neutrinos produced in high--energy hadronic
jets.

The UHE neutrino flux generated in such a scenario can be calculated
by normalizing the flux of appropriate particles to the UHECR flux.
With mounting evidence that the highest energy cosmic rays are protons
or nuclei and not photons, we have assumed that the UHE photons are
degraded by the universal and/or galactic radio background, leaving
protons to dominate the highest energy cosmic ray flux. The neutrino
flux must then be normalized to the proton flux resulting in
significantly improved prospects for its detection.

A word about the uncertainties in our calculation might be in order.
First of all, the uncertainty of the measured UHECR flux, and in
particular the discrepancy between the HiRes and AGASA results, leads
to an overall uncertainty. On the theoretical side, the main
uncertainty probably comes from the calculation of the particle
spectra at ``small'' energies, where currently not very well
understood coherence effects can play a role. This effect is bigger
for higher primary jet energy, and can change the event rate by up to
a factor of about 2 or less (see table~\ref{table:neutrinos}).
Relaxing our assumption that {\em all} UHE photons are absorbed would
lead to a corresponding reduction of the fitted source density, and
hence of the neutrino flux. In this context it is worth mentioning
that in the scenario which seems to fit the data best, with primary
jet energy near $10^{21}$ eV and a galactic source overdensity of
about $10^5$ (see Fig.~\ref{normalisation_21} and
ref.\cite{Sarkar:2001}), including the photon flux fully would only
reduce the predicted event rate by a factor of two to three, since in
this case the flux of $10^{20}$ eV photons at source is only slightly
larger than the corresponding proton flux. This would still give a
neutrino flux in easy striking range of km$^2$ scale detectors.

This paper shows that the neutrino flux accompanying the highest
energy cosmic rays in top--down scenarios is of order of the limits
placed by operating experiments such as AMANDA II, RICE and AGASA.
Further data from these experiments, or next generation experiments
IceCube and AUGER, can test the viability of top--down scenarios which
generate the highest energy cosmic rays. If a signal is found soon,
future high statistics experiments should be able to map out the
neutrino spectrum, thereby allowing us direct experimental access to
physics at energy scales many orders of magnitude beyond the scope of
any conceivable particle collider on Earth.

\subsubsection*{Note Added}
After completion of this work HiRes published \cite{newhires} updated
spectra for both their own and the AGASA experiment. The new HiRes
spectrum (which, however, does not include the original Fly's Eye
event \cite{largest}) is somewhat below the spectrum shown in
Figs.~\ref{normalisation_21} and~\ref{normalisation_25}; it shows
clear evidence for a spectral break, as predicted by the GZK effect,
and is thus consistent with a homogeneous distribution of sources.
Ref.\cite{newhires} also contains an updated AGASA spectrum, which at
$E \sim 10^{20}$ eV is slightly higher than the one used in our fits.
The discrepancy between these two experiments, and the resulting
uncertainty of the UHECR proton flux, is thus larger than previously
anticipated. Note, however, that going from a galactic to a
homogeneous distribution of sources can over--compensate the reduced
normalization of the proton flux indicated by the HiRes spectrum, as
far as the rate for neutrinos with energy exceeding 100 TeV is
concerned.


\section{Detecting SUSY in the sky? A new window for neutralino detection}
\label{sec:neutralinos}

In models where the UHE cosmic ray problem is solved by top-down
scenarios, a significant flux of UHE neutralinos is predicted. We
calculate the number of events expected from such particles in future
experiments such as EUSO or OWL. We show that by using the Earth as a
filter, showers generated by neutralinos can be separated from
neutrino generated showers. We find that for many models, observable
rates are expected.

\subsection{Ultra-high energy fragmentation to neutralinos}

\begin{figure}[h!]
\centering\leavevmode
\includegraphics[width=5in]{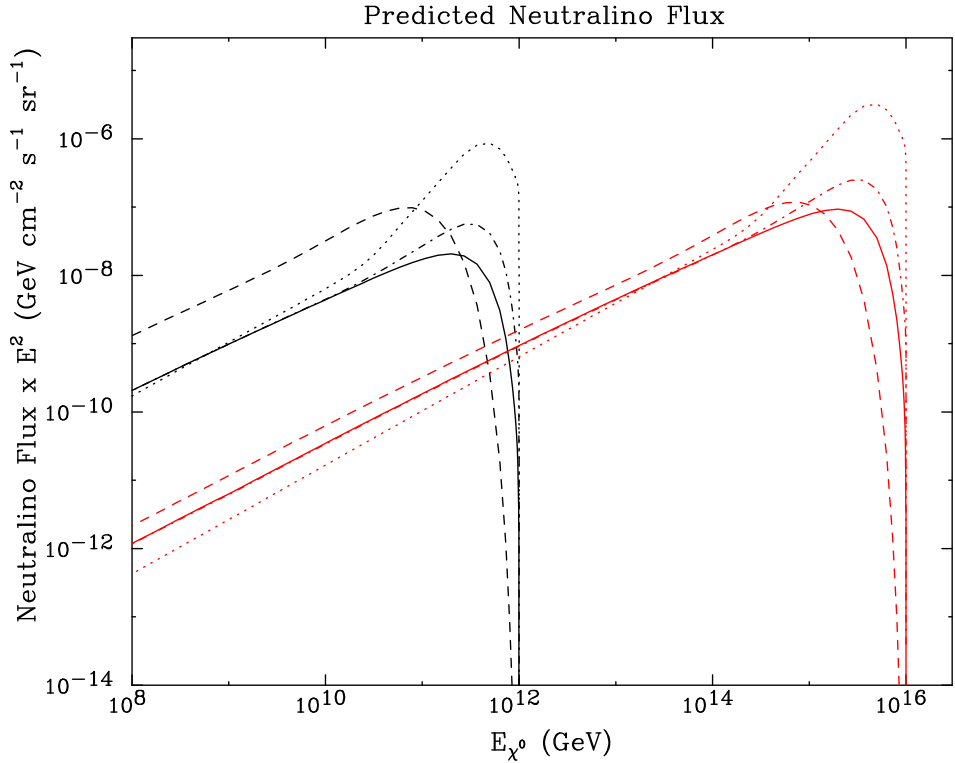}
\caption{The spectrum of neutralino LSP's predicted for the decay of
  superheavy particles with mass $M_X = 2 \cdot 10^{21}$ eV (left set
  of curves) and $M_X= 2 \cdot 10^{25}$ eV (right) normalized
  \cite{bdhh1} by the proton spectrum to the ultra-high energy cosmic
  ray flux, for a ``galactic'' distribution of sources where most
  UHECR events originate from $X$ decays in the halo of our galaxy.
  For a homogeneous distribution, the spectrum is enhanced by up to a
  factor of 15. Spectra are shown for primary $X$ decays into
  quark$+$antiquark (solid), quark$+$squark (dot-dash), $SU(2)$
  doublet lepton$+$slepton (dots) and 5 quark$+$5 squark (dashes).
  Note that for the case of $M_X = 2 \cdot 10^{21}$ eV decays, the
  spectrum peaks in the energy range most accessible to air shower
  experiments.}
\label{LSP_flux}
\end{figure}

From the general results discussed in chapter~\ref{chap:decay}, we
found that the LSP flux depends only mildly on the spectrum of
superparticles, as long as the LSP is a bino--like neutralino. Some
sample spectra are shown in Fig.~\ref{LSP_flux}.\footnote{The
  primary 10--body decay $X \rightarrow 5q 5 \tilde q$ has been
  modeled using phase space only, i.e. ignoring any possible
  dependence of the matrix element on external momenta.}

Here we have once again conservatively assumed that $X$ particles have
an overdensity of $10^5$ in the vicinity of our galaxy, as expected
\cite{Birkel} for $X$ particles that move freely under the influence
of gravity\footnote{The exact profile of the halo of $X$ particles
  does not affect our results as long as most UHECR events originate
  at distances well below one GZK interaction length.}. This minimizes
the expected neutralino flux, since, as in the preceeding section, all
scenarios are normalized by matching \cite{bdhh1} the predicted
proton spectrum to the highest energy cosmic ray observations (see
discussion in section~\ref{sec:learn}).

\subsection{Signatures of ultra-high energy neutralinos}

Ultra--relativistic neutralinos interact with quarks by $t-$channel
$Z$ and $W^{\pm}$ exchange, as well as by the exchange of squarks in
the $s-$ or $u-$channel. These interactions either directly yield an
LSP, or produce a heavier neutralino or chargino which quickly decays
to the lightest neutralino (except, perhaps, in the case of
near--degenerate masses). Either interaction generates a shower which
can be observed by air shower experiments.

\begin{figure}[h!] 
\centering\leavevmode
\includegraphics[width=4.5in]{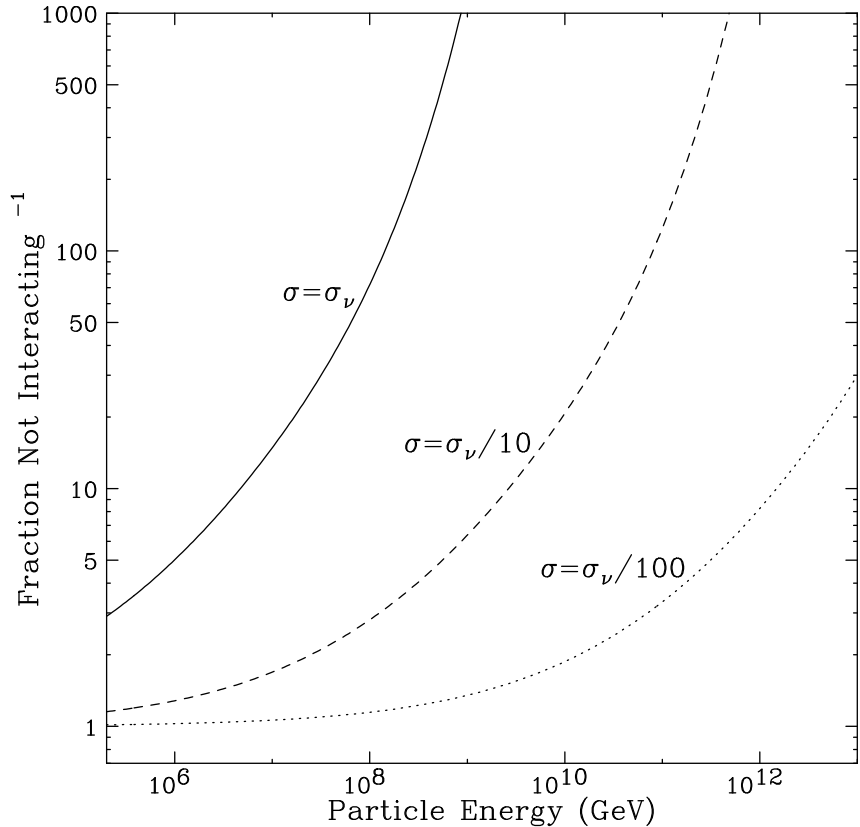}
\caption{The fraction of neutrinos or neutralinos which pass through
the Earth (integrated over zenith angle less than 85 degrees) as a
function of energy. Results are shown for particles with total cross
sections with nucleons equal to that for neutrinos as well as for
particles with cross sections ten and one hundred times smaller.
Regeneration effects are not included (see end of sec.~3).}
\label{neutralino_frac} 
\end{figure}

The background for this signal consists of showers generated by
ultra-high energy cosmic neutrinos. The neutrino interaction length
becomes comparable to the radius of the earth around $10^5$ GeV. By
$10^9$ GeV, only about one out of 1000 neutrinos passes through the
Earth without interaction (see figure~\ref{neutralino_frac}). A
neutralino, however, depending on the choice of SUSY parameters, will
have a different interaction cross section and, therefore, different
absorption properties. The size of this cross section depends
sensitively on the neutralino eigenstate, which in general is a
composition of bino, wino and neutral higgsinos. A wino-- or
higgsino--like neutralino has couplings to $W$ and/or $Z$ bosons that
resemble or even exceed those of neutrinos. In contrast, a bino--like
neutralino has very small couplings to gauge boson, because its
superpartner, the $U(1)_Y$ gauge boson, does not couple to other gauge
bosons. The couplings of bino--like neutralinos to squarks are of full
$U(1)_Y$ gauge strength, but squark searches at the Tevatron
\cite{tevsearch} tell us that first and second generation squarks must
be at least three times heavier than $W$ bosons. Note also that models
with radiative breaking of the electroweak gauge symmetry prefer the
lightest neutralino to be bino--like in most of parameter space
\cite{binomodel}. Typical parameter choices therefore predict
neutralino-nucleon cross sections one or two orders of magnitude
smaller than neutrino-nucleon cross sections \cite{SHDM}. With a
significantly smaller cross section, very high energy cosmic
neutralinos may travel through the Earth producing upgoing events at
much higher energies than neutrinos. Upgoing showers with energy above
100 PeV or so would be a smoking gun for cosmic neutralinos.

\vskip 0.5cm
 
Furthermore, by virtue of $R-$parity, neutralinos will generate less
energetic neutralinos in each interaction, thus not depleting their
number. Tau neutrinos also display this property \cite{saltzberg},
but not as dramatically. The difference comes from the fact that high
energy tau leptons lose energy in propagation whereas charginos decay
quickly enough to lose very little energy in propagation. Also, phase
space arguments indicate that a larger fraction of a decaying
chargino's energy goes into the resulting (massive) neutralinos than a
decaying tau's energy goes into the new (essentially massless) tau
neutrino. Together, these effects indicate that tau regeneration is
largely ineffective above about $10^8$ GeV. On the other hand, for
even moderately smaller neutralino cross sections, the Earth can
remain effectively transparent to cosmic neutralinos at much higher
energies.

Our calculations of tau neutrino and neutralino regeneration in the
Earth were done with a Monte Carlo simulation which, at each
interaction, calculated the energy lost in the interaction and
following propagation \cite{saltzberg}. Our treatment of $\tau$
propagation includes $e^+e^-$ pair production, photonuclear
interactions, bremsstrahlung and ionization energy losses. As stated
earlier, any unstable superparticle produced in LSP interactions is
too short--lived to lose energy prior to its decay. We estimate that
each interaction, if necessary followed by superparticle decay, will
reduce the energy of the LSP by slightly more than a factor of two;
this effect is included in our treatment of LSP regeneration. Our code
demonstrated the appearance of a `pile-up' of outgoing particles at an
energy corresponding to an interaction length equal to the size of the
Earth. For tau neutrinos, this occurs at PeV energies, but can be
considerably higher for neutralinos, due to their smaller cross
section.

\subsection{Prospects for detection in air shower experiments}

The flux of very high energy neutralinos from top-down scenarios can
be calculated assuming that this is the mechanism which generates the
highest energy cosmic rays
\cite{Hill,Berezinsky:2000,FodorKatz,Coriano:2001,Sarkar:2001,bd1,bd2}.
Given a sufficient cosmic flux, these neutralinos may be detected in
future air shower experiments. The challenge, however, is not merely
observing the showers generated in neutralino interactions but in
differentiating these cosmic neutralinos from neutrinos.

We have calculated the number of neutralino events predicted for a
variety of top-down models associated with the highest energy cosmic
rays in a future experiment such as EUSO \cite{EUSO} or OWL
\cite{OWL}. EUSO and OWL are proposed satellite experiments which
observe fluorescence in the Earth's atmosphere generated in very high
energy showers. Such experiments are expected to observe on the order
of 150,000 square kilometers of surface area on the Earth. Particles
which pass through the Earth can interact in the shallow Earth or
atmosphere generating upgoing showers observable by fluorescence or
Cerenkov radiation. Ultra-high energy showers reach a maximum near a
slant depth of 850 $\rm{g}/\rm{cm}^2$, corresponding to a depth of 8.5
meters in water. Including the effective slant depth of the lower
atmosphere extends this to $\sim 0.015 \, $ km, thus providing a water
equivalent effective volume of $\sim150,000 \times 0.015 \sim 2250 \,$
cubic kilometers, a truly enormous volume. Such an experiment will be
capable of measuring both the energy and the direction of an observed
particle. 

\begin{figure}[h!] 
\centering\leavevmode
\includegraphics[width=4.5in]{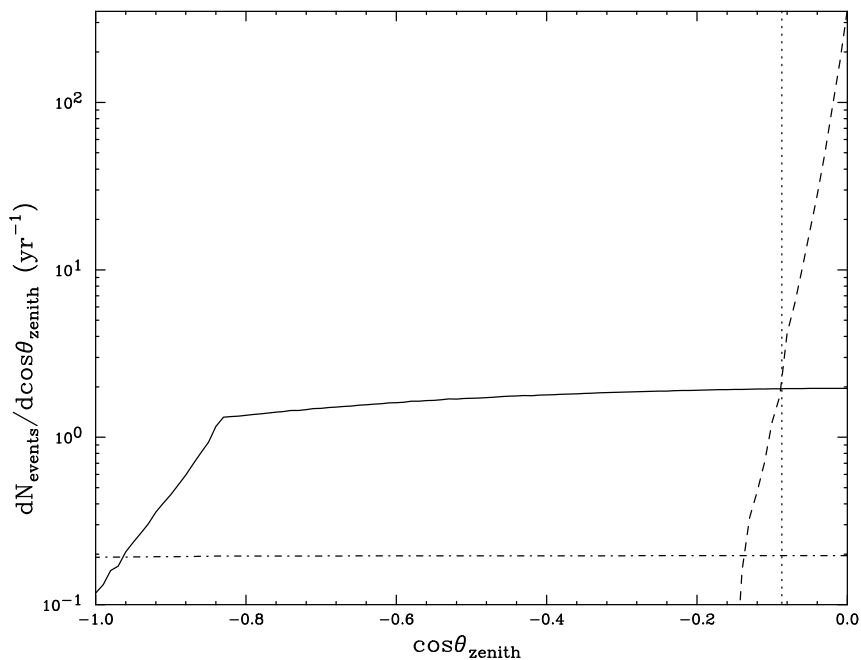}
\caption{The neutrino background (dashed) and LSP signal (solid:
  $\sigma_{\rm LSP} = \sigma_\nu/10$; dot--dashed: $\sigma_{\rm LSP} =
  \sigma_\nu /100$) at $E > 1$ EeV. Both signal and background result
  from $X \rightarrow q \bar q$ decays of $2 \cdot 10^{12}$ GeV $X$
  particles with a galactic distribution. The vertical dotted line
  indicates the angular cut of $5^\circ$ applied to the signals listed
  in table~\ref{table:neutralinos}}
\label{background} 
\end{figure}

Estimating the rate of neutrino--induced ``background'' events is
difficult at present since the neutrino flux at $E \gtrsim 10^9$ GeV is
not known. The flux of atmospheric neutrinos is completely negligible
at these energies. However, most proposed explanations of the UHECR
events also predict a significant UHE neutrino flux. We therefore use
the neutrino flux predicted by top--down models \cite{bdhh1} to
estimate the neutrino background. Fig.~\ref{background} compares signal and
background at $E \geq 1$ EeV for one such model, where we assume a
galactic distribution of $X$ particles, with primary $X \rightarrow q
\bar q$ decay and $M_X = 2 \cdot 10^{12}$ GeV. We see that signal and
background clearly have very different angular distributions even for
the larger LSP--nucleon cross section of $\sigma_\nu/10$. Regeneration
effects are included, but they cannot produce neutrino events at large
energy {\em and} large angle. Requiring the events to emerge more than
$5^\circ$ below the horizon removes almost all the background, with
little loss of signal; in the case at hand, we expect about 2 signal
events per year, compared to 0.1 background event. If the LSP--nucleon
cross section is smaller, a somewhat stronger angular cut may be
advantageous; on the other hand, at even higher energies it might be
better to use a slightly weaker cut. However, this variation of the
angular cut has negligible effect on the predicted signal rate,
compared to the uncertainty inherent in our estimates. In the
following we therefore apply a fixed angular cut of $5^\circ$ on the
signal in all cases. This cut will have to be optimized once the
angular resolution of the experiment is known. Moreover, measurements
at neutrino telescopes as well as AUGER should soon greatly improve
our knowledge of the neutrino flux at very high energies. Finally,
this figure also shows that a measurement of the angular distribution
of the signal will allow to determine the LSP scattering cross
section: for the larger cross section shown, there will be very few
vertically upgoing events. The dependence of the angular distribution
of the signal on the cross section becomes even more pronounced at
higher energies.

 
Table~\ref{table:neutralinos} shows signal event rates for two choices
of energy threshold, $E_{\chi^0} \ge$ 1 EeV and 100 EeV. We also show
results for the stronger cut on energy in order to illustrate that at
least in some cases the LSP spectrum should be measurable over a
significant range of energies. The first case shown in the table
corresponds to the situation depicted in Fig.~\ref{LSP_flux}. Of
course, the choice of a 100 EeV threshold is even more effective in
reducing the background, to the level of $10^{-3}$ events per year.
From the physics point of view an energy threshold of 100 EeV should
only be necessary in the unlikely case that the total background of
ultra--high energy neutrinos is dominated by some mechanism not
related to the observed UHECR events. Regarding the energy threshold
which can be achieved experimentally, it has been argued that for
upgoing events, the threshold could be as small as a PeV
\cite{thresh}.

\begin{table}[h!]
\begin{center}
\begin{tabular}{|c||c|c|}\hline \hline
$E_{\chi^0} \ge 1 \,\rm{EeV}$ & $\sigma_{\chi^0} =
\sigma_{\nu}/10$ & $\sigma_{\chi^0} = \sigma_{\nu}/100$  \\   
\hline \hline
$q \bar{q}$, $10^{21}$ eV, Galactic & 1.86 & 0.196 \\
$q \tilde{q}$, $10^{21}$ eV, Galactic & 2.96 & 0.306 \\
$5 \times q \tilde{q}$, $10^{21}$ eV, Galactic & 4.05 & 0.436\\
$l \tilde{l}$, $10^{21}$ eV, Galactic & 28.0 & 2.81\\
\hline  \hline 
$q \bar{q}$, $10^{25}$ eV, Galactic & 0.187 & 0.0189\\
$q \tilde{q}$, $10^{25}$ eV, Galactic & 0.213 & 0.0216\\
$5 \times q \tilde{q}$, $10^{25}$ eV, Galactic & 0.213 & 0.0216\\
$l \tilde{l}$, $10^{25}$ eV, Galactic & 0.615 & 0.0617\\
\hline  \hline 
$q \bar{q}$, $10^{21}$ eV, Homogeneous & 27.9 & 2.94 \\
$q \tilde{q}$, $10^{21}$ eV, Homogeneous & 44.4 & 4.56\\
$5 \times q \tilde{q}$, $10^{21}$ eV, Homogeneous & 60.8 & 6.54\\
$l \tilde{l}$, $10^{21}$ eV, Homogeneous & 420.0 & 42.15\\
\hline  \hline 
$q \bar{q}$, $10^{25}$ eV, Homogeneous & 2.81 & 0.284\\
$q \tilde{q}$, $10^{25}$ eV, Homogeneous & 3.20 & 0.324\\
$5 \times q \tilde{q}$, $10^{25}$ eV, Homogeneous & 3.20 & 0.324\\
$l \tilde{l}$, $10^{25}$ eV, Homogeneous & 9.23 & 0.926 \\
\hline \hline \hline
$E_{\chi^0} \ge 100\, \rm{EeV}$ & $\sigma_{\chi^0} =
\sigma_{\nu}/10$ & $\sigma_{\chi^0} = \sigma_{\nu}/100$ \\ 
\hline \hline
$q \bar{q}$, $10^{21}$ eV, Galactic & 0.0976 & 0.0344\\
$q \tilde{q}$, $10^{21}$ eV, Galactic & 0.391 & 0.122\\
$5 \times q \tilde{q}$, $10^{21}$ eV, Galactic & 0.0161 & 0.00716 \\
$l \tilde{l}$, $10^{21}$ eV, Galactic & 10.1 & 2.38\\
\hline  \hline
$q \bar{q}$, $10^{25}$ eV, Galactic & 0.0946 & 0.0143\\
$q \tilde{q}$, $10^{25}$ eV, Galactic & 0.116 & 0.0169\\
$5 \times q \tilde{q}$, $10^{25}$ eV, Galactic & 0.103 & 0.0159\\
$l \tilde{l}$, $10^{25}$ eV, Galactic & 0.435 & 0.0576\\
\hline  \hline
$q \bar{q}$, $10^{21}$ eV, Homogeneous & 1.46 & 0.516\\
$q \tilde{q}$, $10^{21}$ eV, Homogeneous & 5.87 & 1.83\\
$5 \times q \tilde{q}$, $10^{21}$ eV, Homogeneous & 0.242 & 0.107\\
$l \tilde{l}$, $10^{21}$ eV, Homogeneous & 151.5 & 35.7\\
\hline  \hline
$q \bar{q}$, $10^{25}$ eV, Homogeneous & 1.42 & 0.215 \\
$q \tilde{q}$, $10^{25}$ eV, Homogeneous & 1.74 & 0.254\\
$5 \times q \tilde{q}$, $10^{25}$ eV, Homogeneous & 1.55 & 0.239 \\
$l \tilde{l}$, $10^{25}$ eV, Homogeneous & 6.53 &0.864 \\
\hline \hline
\end{tabular}
\end{center}
\caption{Neutralino event rates per year in top-down scenarios in a
large area air shower experiment such as EUSO or OWL, with effective
volume $\simeq 2250$ cubic kilometers (water equivalent). Rates are
shown for two choices of neutralino-nucleon cross sections, two
choices of energy threshold and several top-down models. At the
energies considered, there is very little neutrino background for
upgoing events (see text).}
\label{table:neutralinos} 
\end{table}

The rates shown in table~\ref{table:neutralinos} are for a variety of
primary $X$ decay modes, and for ``galactic'' and homogeneous
distributions of $X$ particles. It seems highly unlikely that $X$
particles will indeed be distributed homogeneously, but it is
conceivable that the majority of sources contributing to the LSP flux
is at cosmological distances (e.g. if the $X$ particles are embedded
in topological defects); the homogeneous distribution is meant to be
representative for such models. Our results show that the $X$
distribution throughout the universe has significant impact on the
expected size of our signal. The models are the same as in
section~\ref{sec:neutrinos}. We note that the neutralino signal is
more sensitive to the primary $X$ decay mode than the neutrino signal
analyzed previously is. Not surprisingly, scenarios with (at least)
one superparticle in the primary decay produce a higher neutralino
flux than models where $X$ only decays into quarks. Moreover,
leptonic $X$ decays increase the predicted neutralino flux by another
order of magnitude, since in this case relatively few protons are
produced, leading to a higher source density required to explain the
observed UHECR events. On the other hand, choosing $M_X = 2 \cdot
10^{25}$ eV rather than $2 \cdot 10^{21}$ eV significantly reduces the
predicted flux. Note, however, that in this case $X$ decays can only
describe the UHECR flux above $\sim 10^{20}$ eV \cite{bdhh1}; events
at a few times $10^{19}$ eV then have to be produced by an as yet
unknown source.

As stated earlier, we normalize the LSP flux by assuming that (almost)
all UHE photons are absorbed between source and Earth. If this
evidence is ignored, i.e. if the observe UHECR spectrum is normalized
to the sum of photon and proton fluxes, the predicted LSP event rate
for models with $M_X = 2 \cdot 10^{25}$ eV would go down by about a
factor of 4. If $M_X = 2 \cdot 10^{21}$ eV, the predicted event rate
would go down by a factor of 2 to 3 for hadronic primary $X$ decays,
and by about an order of magnitude for purely leptonic primary $X$
decay. Note that this "uncertainty" in the predicted event rate from
taking the ``proton hypothesis'' (see e.g. ref. \cite{nophot})
seriously or not is comparable to the variation between different
primary $X$ decay modes. Finally, we remind the reader that the UHECR
spectra measured by AGASA and HiRes differ significantly in the
post--GZK region, leading to a corresponding uncertainty in our
predicted signal.

\subsection{Conclusions}

The cosmic neutralino flux predicted in top--down scenarios could
possibly provide an interesting test of both supersymmetry and GUT
scale particle physics. To identify any showers generated in future
experiments as being generated by cosmic neutralinos, they will need
to occur at energies and from directions at which neutrinos would be
absorbed by the Earth. We have calculated the event rates for a
variety of such models for a large area air shower experiment such as
OWL or EUSO. We find that for many scenarios, the event rate is large
enough to be observable in principle. We should mention here that our
estimates of annual event rates assume 100\% duty cycle. This is
clearly not realistic for any experiment based on optical
observations. However, planning for the kind of space--based
experiment we envision is still in its early stage; a smaller duty
cycle might be compensated by a larger area and/or a longer period of
observation. 

We believe that searching for UHE LSPs is very important, since it is
the only measurement that can {\em qualitatively} distinguish between
``top--down'' and the more conventional ``bottom--up'' explanations
for the observed UHE events: in bottom--up models superparticles can
only be produced in the collision of accelerated protons, so the UHE
LSP flux will be a tiny fraction [typically ${\cal O}(10^{-6})$ or
less] of the UHE neutrino flux, much too small to be observed in any
currently conceivable experiment. In contrast, a sizable UHE LSP flux
is a generic prediction of top--down models. Moreover, the neutralino
event rate turns out to be a far more sensitive probe of details of
the model than the flux of neutrinos with energy exceeding $\sim 1$
PeV \cite{bdhh1}. We therefore find it encouraging that the
observation of UHE LSPs along the lines suggested in this paper, while
certainly not easy, should at least be possible.



\chapter{Summary and perspectives}
\label{chap:conclusion}

The spectrum of cosmic rays (CRs) has been measured over more than 12
decades of energy. Even if our understanding of it has grown a lot in
the last few decades, many puzzles are remaining. One of them concerns
the extremity of this spectrum, at the highest energies, where
theorists were expecting a sharp cut-off to occur at energies of the
order of $5 \cdot 10^{19}$ eV: at these energies, CRs should be of
extragalactic origin, and probably coming from distances further than
the local cluster of galaxies, because we know no astrophysical object
able to accelerate particles enough to give them this energy in our
vicinity. However, particles carrying energies above $10^{20}$ eV
traveling over cosmological distances should loose their energy
through scattering; for example, a proton will interact with
the cosmological microwave background (CMB) and photoproduce pions,
with an interaction length of a few tens of Mpc, loosing around 20 \%
of its energy at each interaction. Similar processes occur with
nuclei, photons or electrons. Thus particles with initial energy
greater than $10^{20}$ eV should reach the Earth with a maximal energy
$\sim 5 \cdot 10^{19}$ eV, the so-called GZK cut-off. The fact is that
events have been registered above this cut-off in very different
experiments over the last few decades: the so-called ultra-high
energy cosmic rays (UHECR). Such an observation is almost impossible
to reconcile with any model of acceleration of charged particles in
any astrophysical object.

This lead to the development of another class of models for explaining
the existence of UHECR, namely ``top-down'' theories, which are
assuming that the observed events are generated through the decay of
some mysterious super-heavy ``$X$'' particles. The existence of such $X$
particles is e.g. predicted in GUT theories, and they can be created
rather naturally at the end of the inflation.  Among other more
``model dependent'' properties, top-down models require that these
particles should have a mass bigger than the highest energies observed
in UHECR events, $M_X > 10^{21}$ eV, and a lifetime of the order of
(or greater than) the age of the universe.  They could be trapped in
galaxies, explaining the isotropy of the data, and would constitute
semi-local sources for UHECR, avoiding the GZK problem. Moreover, if
they are abundant enough and trapped in the galaxies, they could be a
good candidate for Dark Matter.

In this thesis, I focused my attention on a detailed study of the
decay of these $X$ particles, in the framework of the Minimal
Supersymmetric Standard Model (MSSM). The main project was to write a
new public code for computing as precisely and model-independently as
possible the spectra of the stable decay products of $X$, taking into
account all relevant interactions in the parton cascade triggered by
the primary $X$ decays (at GUT energies, the electroweak coupling
constants and part of the Yukawa ones become as strong as the strong
coupling constant of QCD). The description of this code and our main
results have been published briefly in \cite{bd1} and more in detail
in \cite{bd2}.

We also wrote two other articles, in collaboration with F. Halzen and
D. Hooper, where we used our results as input data for predicting the
rates of neutrino \cite{bdhh1} and neutralino \cite{bdhh2} events
observable on Earth in different models and different kinds of future
experiments.

Still, a lot of phenomenological work remains to be done in this area.
Our results allow to study in great detail any ``top-down'' model and
give quantitative predictions for the fluxes of stable particles at
source. One interesting extension would be to connect our code to a
program that treats the propagation of photons and charged particles
through the interstellar and intergalactic medium. Inclusion of these
propagation effects is necessary for the prediction of fluxes arriving
on Earth. The treatment of the detection of neutrinos and LSPs in our
papers can also be improved.

Our code so far only includes leading order (one loop) effects. Higher
order effects at very small $x$ are estimated using the MLLA formalism.
However the application of this formalism to theory with massive
partons, like SUSY-QCD, currently rests on somewhat shaky assumptions;
this needs clarification. More generally, perturbative higher order
corrections might be significant.

Top-down models are intriguing because they allow ``direct'' access to
energies beyond the reach of any conceivable Earth-based collider.
However, to date it is not known what kinds of ``new physics''
appearing at an energy scale somewhere between $M_X$ and the
weak scale would lead to detectable changes in the fluxes of stable
particles at very high energies. Possible extensions of the (MS)SM in
that energy range include massive (gauge singlet) neutrinos,
additional gauge bosons (e.g. for a left-right symmetric gauge group),
as well as other exotic particles. Much remains to be done in this
area.

On the other hand, if the life time of $X$ particles was not long enough
to explain the current existence of UHECR, their decays in the early
times of the universe could play a role in different mechanisms like
lepto- and baryogenesis. For example, our code might be applied to
inflaton (and inflatino) decay, with minor changes.\newline

I conclude this work with the hope that the code SHdecay will be of
some use for the particle and astroparticle physics community, and
that the next few years will bring us new significant results in the
field which, with the help of the theoretical tools already
developed, could help us to finally understand what the hell is the
actual (new ?) physics hidden behind the UHECR events!...


\appendix
\chapter{Splitting functions of the MSSM}
\label{app:SFs}

\setcounter{equation}{0}
\renewcommand{\theequation}{A.\arabic{equation}}

The splitting function (SF) $P_{ji}(x)$ describes the radiation of
particle $j$ off particle $i$. Its $x-$dependence is determined by the
Lorentz structure of the corresponding vertex, while the normalization
also depends on the associated group [color and $SU(2)$] factors. If
there is no vertex relating these two particles the SF is simply 0. We
first list the functional forms we will need, together with the spins
of the particles involved in the branching process $i \rightarrow j +
k$ ($V$ for vector, $F$ for spin$-1/2$ fermion, $S$ for scalar):
\beqa
\label{def_SFs}
&& (0)\;\;\; \delta(1-x)\,,\nonumber\\
&& (1)\; i=F, j=F, k=V:\ \frac{1 + x^2} { (1-x)_+ } \,, \nonumber \\
&& (2)\; i=F, j=V, k=F:\ \frac{1 + (1-x)^2} {x} \,, \nonumber \\
&& (3)\; i=F, j=S, k=F:\  x\,, \nonumber \\
&& (4)\; i=F, j=F, k=S:\ (1-x)\,,\nonumber\\
&& (5)\; i=S, j=F, k=F:\ 1\,,\nonumber\\
&& (6)\; i=S, j=V, k=S:\ \frac {2(1-x)} {x}\,,\nonumber\\
&& (7)\; i=S, j=S, k=V:\ \frac {2x} {(1-x)_+}\,,\nonumber\\
&& (8)\; i=V, j=F, k=F:\ (1-x)^2 + x^2\,,\nonumber\\
&& (9)\; i=V, j=V, k=V:\  2 \left[ \frac{1-x}{x} + x(1-x) +
\frac{x}{(1-x)_+} \right] \,,\nonumber\\
&& (10)\; i=V, j=S, k=S:\  2x(1-x)\,. 
\eeqa
For convenience, we also define $(1') = (1)+(0)$ and $(7')=(7)+(0)$.

\clearpage

\begin{center}
\begin{table}[t]
\begin{center}
\begin{tabular}{|c||c|c|c|c|}\hline
$\alpha_S$ &\rule[-3mm]{0mm}{8mm}$ i=q $ & $\tilde{q}$ & $g$ &
$\tilde{g}$ 
\\ \hline
$j=q$ & \rule[-3mm]{0mm}{8mm} $\frac{4}{3}$ $(1^{'})$ & $\frac{4}{3}$
(5) & $\frac{N_q}{2}$ (8) & $\frac{N_q}{2}$ (4)
\\ \hline
$ \tilde{q}$ & \rule[-3mm]{0mm}{8mm} $\frac{4}{3}$ (3) & $\frac{4}{3}$
$(7^{'})$ & $\frac{N_q}{2}$ (10) & $\frac{N_q}{2}$ (3) 
\\ \hline
$g$ & \rule[-3mm]{0mm}{8mm} $\frac{4}{3}$ (2) & $\frac{4}{3}$ (6) & $ 3
\left[ (9) + \left( \frac{3}{2} - \frac{F}{6} \right) (0) \right] $ &
3 (2)
\\ \hline
$\tilde{g}$ & \rule[-3mm]{0mm}{8mm} $\frac{4}{3}$ (4) & $\frac{4}{3}$
(5) & 3 (8) & $ 3 \left[ (1) + \left( \frac{3}{2} - \frac{F}{6}
\right) (0) \right]$
\\ \hline
\end{tabular}
\caption{SUSY--QCD splitting functions $P_{ji}$, where $j$ and $i$
  determine the row and column of the table, respectively. The
  functional forms denoted by $(n), \ n=0, \dots, 10$ have been
  defined in eq.(\ref{def_SFs}), with $(1') = (1)+(0)$ and $(7') = (7)
  + (0)$. The ``multiplicity factors'' are: $N_{t_R} = N_{b_R} = 1$,
  $N_{t_L} = N_{u_R} = N_{d_R} = 2$ and $N_{q_L} = 4$. In the MSSM
  phase, i.e. for $Q > M_{\rm SUSY}$, the number of active flavors
  (quarks and squarks) is $F=6$.}
\label{alphaS}
\end{center}
\end{table}
\end{center}

The 16 SFs of SUSY--QCD listed in table~\ref{alphaS} are derived from
\cite{Jones}; in eq.(\ref{e8}) they come with a factor of the strong
coupling $\alpha_S$. Note that in ref.\cite{Jones} the chirality index
$L,R$ was always summed over; e.g. $P_{tg} = P_{t_Lg} + P_{t_R g}$,
where $t_L$ now {\em only} describes the left--handed top quark (and
not the third generation quark doublet). Since these two terms are
equal, one has $P_{t_L g} = P_{t_R g} = P_{qg}/2$. On the other hand,
our ``(s)quark'' distributions always include anti(s)quarks. This
re--introduces a factor of 2, so that for us e.g. $P_{t_R g} = P_{qg}$
of \cite{Jones}. Additional factors arise for (s)quarks of the first
and second generation. As described in Sec.~2.2, we always average
over (s)quarks and anti(s)quarks with given hypercharge of the first
two generations. This implies $P_{u_R g} = P_{d_R g} = 2 P_{qg}$ and
$P_{q_L g} = 4 P_{qg}$, where the additional factor of two in the
second expression comes from summing over the $SU(2)$ index of the
doublet $q_L$. The same factors appear in SFs describing gluon to
squark splitting as well as gluino splitting into a squark and a
quark. A complete list of these factors $N_q$ is given in the table
caption. On the other hand, in the absence of flavor--changing
interactions SFs involving quarks and squarks only always come with
factor 1 if the ``compound particles'' $u_R, q_L$ etc. are properly
normalized.

\begin{center}
\begin{table}[b!]
\begin{center}
\begin{tabular}{|c||c|c|c|c|}\hline
$g_2 = e/\sin
\theta_W$ &\rule[-3mm]{0mm}{8mm} $i=W$ & $\tilde{W}$ & $f_L$ & $\tilde
f_L$ 
\\ \hline
$j=W$ & \rule[-3mm]{0mm}{8mm} $2 \left[(9) + \left( \frac{3}{2} -
\frac{N_d}{8} \right) (0)\right] $ & 2 (2) & $\frac{3}{4}$ (2) &
$\frac{3}{4}$ (6) 
\\ \hline
$\tilde{W}$ & \rule[-3mm]{0mm}{8mm}2 (8) & 2 $\left[ (1) + \left(
\frac{3}{2} - \frac{N_d}{8} \right) \,(0) \right]$ & $ \frac{3}{4}$
(4) & $\frac{3}{4}$ (5) 
\\ \hline
$f_L$ & \rule[-3mm]{0mm}{8mm} $\frac{N_f}{2}$ (8) & $\frac{N_f}{2}$
(4) & $\frac{3}{4}$ $(1')$ & $\frac{3}{4}$ (5)
\\ \hline
$\tilde f_L$ & \rule[-3mm]{0mm}{8mm} $\frac{N_f}{2}$ (10) &
$\frac{N_f}{2}$ (3) & $\frac{3}{4}$ (3) & $\frac{3}{4}$ $(7^{'})$
\\ \hline
\end{tabular}
\caption{$SU(2)$ splitting functions $P_{ji}$, where particles $j$ and
  $i$ are associated with the row and column, respectively. The
  functional forms denoted by $(n), \ n=0, \dots, 10$ have been
  defined in eq.(\ref{def_SFs}), with $(1') = (1)+(0)$ and $(7') = (7)
  + (0)$. $N_d$ is the total number of $SU(2)$ doublets; in the MSSM,
  $N_d = 14$. $f$ stands for any matter or Higgs fermion, with $N_f$
  being the number of doublets (not counting anti--doublets) described
  by $f_L$ or $\tilde f_L$. For our ``compound'' states, these are:
  $N_{q_L} = 6$, $N_{l_L} = 2$, $N_{t_L} = 3$, $N_{\tau_L} = N_{H_1} =
  N_{H_2} = 1$.}
\label{g2}
\end{center}
\end{table}
\end{center}

The SFs stemming from electroweak interactions have similar
structures; we just need to compute the correct group and multiplicity
factors. The results are listed in tables~\ref{g2} and \ref{gY}. In
these tables we list SFs including the appropriate multiplicity
factors; a single $SU(2)$ doublet {\em without} antiparticles would
have $N_f = 1/2$. Note that there is no difference between Higgs and
$SU(2)$ doublet lepton superfields as far as gauge interactions are
concerned. Finally, due to the absence of self--interactions of $U(1)$
gauge bosons, the splitting functions $P_{BB}$ and $P_{\tilde B \tilde
B}$ are pure delta--functions, with coefficients fixed by energy
conservation, eq.(\ref{sumrule}). In all three gauge interactions we
find that the coefficient of the $\delta-$function is the same in
$P_{ff}$ and $P_{\tilde f \tilde f}$ for any matter fermion $f$, and
also in $P_{VV}$ and $P_{\tilde V \tilde V}$ for a gauge boson $V$;
this latter coefficient is $-1/2$ times the coefficient in the
$\beta-$function of the corresponding gauge coupling.

Finally, Yukawa couplings only appear in $H f_L f_R, \ \tilde h \tilde
f_L f_R$ and $\tilde h f_L \tilde f_R$ vertices. We therefore only
need functional forms (3), (4) and (5) from eq.(\ref{def_SFs}). The
coefficients can be determined from the analogous terms due to
$U(1)_Y$ interactions by replacing $ (g_Y Y_f)^2$ by $\lambda_f^2/2$,
where the extra factor of $1/2$ corrects for the factor $\sqrt{2}$
appearing in front of gaugino--fermion--sfermion vertices in the
supersymmetric Lagrangian. Since Yukawa interactions couple matter
fields with different chiral indices all diagonal SFs due to Yukawa
couplings are pure $\delta-$functions, the coefficients again being
determined by energy conservation; as before, we find equal
coefficients for diagonal SFs of a particle and its superpartner. The
resulting SFs are listed in table~\ref{Yuk}. As usual, these SFs are
multiplied with $\alpha_f/(2\pi) \equiv \lambda_f^2/(8 \pi^2)$ in the
DGLAP equations. The three interactions we consider, involving the
top, bottom and tau Yukawa couplings, can all be treated using
table~\ref{Yuk}, by identifying the matter and Higgs fields
appropriately and using the correct color factors, as explained in the
caption. The $SU(2)$ factors, which lead to the factor of 2 difference
between SFs describing radiation off $SU(2)$ singlet or doublet
(s)fermions, are the same in all three cases.\footnote{Strictly
  speaking, $H_1$ can only split into $\tau_R$ and
  $\overline{\tau_L}$, not into $\tau_L$ and $\overline{\tau_R}$,
  while the antiparticle $H_1^*$ can only split into $\tau_L$ and
  $\overline{\tau_R}$; analogous remarks hold for the other
  Yukawa--induced branching processes. However, this distinction plays
  no role for us, since we always average or sum over particle and
  antiparticle.}

\begin{center}
\begin{table}
\begin{center}
\begin{tabular}{|c||c|c|c|c|} \hline
$g_Y = e/\cos \theta_W$ & \rule[-3mm]{0mm}{8mm} $i=B$ & $\tilde{B}$ &
$f$ & $\tilde f$ 
\\ \hline
$j=B$ & \rule[-3mm]{0mm}{8mm} $-\frac{1}{2} \sum_f Y_f^2$ (0) & 0 &
$Y_f^2$ (2) & $Y_f^2$ (6)
\\ \hline
$\tilde{B}$ & \rule[-3mm]{0mm}{8mm} 0 & $-\frac{1}{2} \sum_f Y_f^2$ 
(0) & $Y_f^2$ (4) & $Y_f^2$ (5) 
\\ \hline
$f$ & \rule[-3mm]{0mm}{8mm} $ n_f Y_f^2 $ (8) & $ n_f Y_f^2$ (4) &
$Y_f^2$ (1') & $Y_f^2$ (5) 
\\ \hline
$\tilde f$ & \rule[-3mm]{0mm}{8mm}$ n_f Y_f^2$ (10) & $ n_f Y_f^2$ (3)
& $Y_f^2$ (3) & $Y_f^2$ (7') 
\\ \hline
\end{tabular}
\caption{$U(1)_Y$ splitting functions $P_{ji}$, where particles $j$
and $i$ are associated with the row and column, respectively. The
functional forms denoted by $(n), \ n=0, \dots, 10$ have been defined
in eq.(\ref{def_SFs}), with $(1') = (1)+(0)$ and $(7') = (7) + (0)$. The
sum of squared hypercharges of all particles $\sum_f
Y_f^2 = 11$ in the MSSM. $f$ stands for any matter or Higgs fermion
with hypercharge $Y_f$, while $n_f$ is the number of degrees of
freedom (not counting anti--particles) described by $f$ or $\tilde
f$. For our ``compound'' states, these are: $Y_{q_L}^2 = 1/36, n_{q_L}
= 12$; $Y^2_{u_R} = 4/9, n_{u_R} = 6$; $Y^2_{d_R} = 1/9, n_{d_R} =
6$; $Y^2_{l_L} = 1/4, n_{l_L} = 4$; $Y^2_{l_R} = 1, n_{l_R} = 2$;
$Y^2_{t_L} = 1/36, n_{t_L} = 6$; $Y^2_{t_R} = 4/9, n_{t_R} = 3$;
$Y^2_{b_R} = 1/9, n_{b_R} = 3$; $Y^2_{\tau_L} = Y^2_{H_1} = Y^2_{H_2}
= 1/4, n_{\tau_L} = n_{H_1} = n_{H_2} = 2$; $Y^2_{\tau_R} = 1,
n_{\tau_R} = 1$.}
\label{gY}
\end{center}
\end{table}
\end{center}

\begin{center}
\begin{table}
\begin{center}
\begin{tabular}{|c||c|c|c|c|c|c|}\hline
$\lambda_f$ & $i=H$ & $\tilde H$ & $f_L$ & $\tilde f_L$ & $f_R$ &
$\tilde f_R$
\\ \hline
$j=H$ & \rule[-3mm]{0mm}{8mm} $-\frac{N_c}{2}$ (0) & 0 & $\frac{1}{2}$
(3) & 0 & (3) & 0 \\ \hline
$\tilde H$ & \rule[-3mm]{0mm}{8mm} 0 &
$-\frac{N_c}{2}$ (0) & $\frac{1}{2}$ (4) & $\frac{1}{2}$ (5) & (4) & (5)
\\ \hline
$f_L$ & \rule[-3mm]{0mm}{8mm} $\frac{N_c}{2}$ (5) & $\frac{N_c}{2}$
(4) & $-\frac{1}{2}$ (0) & 0 & (4) & (5)
\\ \hline
$\tilde f_L$ & \rule[-3mm]{0mm}{8mm} 0 & $\frac{N_c}{2}$ (3) & 0 &
$-\frac{1}{2}$ (0) & (3) & 0
\\ \hline
$f_R$ & \rule[-3mm]{0mm}{8mm} $\frac{N_c}{2}$ (5) & $\frac{N_c}{2}$
(4) & $\frac{1}{2}$ (4) & $\frac{1}{2}$ (5) & $-1$ (0) & 0
\\ \hline
$\tilde f_R$ & \rule[-3mm]{0mm}{8mm} 0 & $\frac{N_c}{2}$ (3) 
& $\frac{1}{2}$ (3) & 0 & 0 & $-1$ (0)
\\ \hline
\end{tabular}
\caption{Splitting functions $P_{ji}$ originating from Yukawa
  interactions, where particles $j$ and $i$ are associated with the
  row and column, respectively. The functional forms denoted by $(n),
  \ n=0, 3, 4, 5$ have been defined in eq.(\ref{def_SFs}). Since we
  only include Yukawa interactions for the third generation, we only
  have to consider three cases. For the top Yukawa coupling, $f_L =
  t_L, \, f_R = t_R, \, H = H_2$ and number of colors $N_c = 3$; for
  the bottom Yukawa coupling, $f_L = t_L, \, f_R = b_R, \, H = H_1$
  and $N_c = 3$; finally, for the tau Yukawa coupling, $f_L = \tau_L,
  \, f_R = \tau_R, \ H = H_1$ and $N_c = 1$.}
\label{Yuk}
\end{center}
\end{table}
\end{center}

\setcounter{equation}{0}
\renewcommand{\theequation}{B.\arabic{equation}}

\chapter{Unitary transformations between current and mass
eigenstates in the MSSM} 
\label{app:transfo}

In this Appendix we describe the unitary transformations occurring
during the SUSY and $SU(2) \otimes U(1)$ breaking, where the quarks,
leptons, weak gauge bosons and Higgs bosons as well as all
superparticles acquire their masses \cite{reviewMartin}. The superscript $b$
denotes the mass eigenstates of the broken theory. The fields in the
unbroken theory are the same as those described in Sec.~2.2. For
example, $q_L$ stands for all left--handed quarks and antiquarks of
the two first generations, i.e. the $SU(2)$ doublets $(u_L,d_L)$,
$(c_L,s_L)$ and their antiparticles $(\bar{u}_L,\bar{d}_L)$,
$(\bar{c}_L, \bar{s}_L)$, and thus describes eight degrees of freedom
(times three, if color is counted separately). Similarly, $l_L$ stands
for both $SU(2)$ doublets $(e_L,\nu_e)$ and $(\mu_L,\nu_\mu)$ and
their antiparticles $(\bar{e}_L,\bar{\nu}_e)$ and
$(\bar{\mu}_L,\bar{\nu}_\mu)$. On the other hand, $u^b$ only describes
$u-$quarks and their antiparticles, but includes both chirality
states, and thus describes four degrees of freedom (not counting
color). Recall that the transformation between mass and current
eigenstates in eq.(\ref{trafo}) only affects the {\em upper} index of
the (generalized) FFs. In the given context $q_L$ therefore stands for
the sum, not the average, of its ``constituent fields'', as discussed
in Sec.~2.2. Recall finally that massive gauge bosons ``eat''
Goldstone modes from the Higgs sector. These considerations lead to
the following transformations for SM fields and Higgs bosons:
\beqa \label{smtrafo}
u^b &=& c^b = \frac{1}{4}\, q_L + \frac{1}{2}\, u_R\,,\nonumber\\
d^b &=& s^b = \frac{1}{4}\, q_L + \frac{1}{2}\, d_R\,,\nonumber\\
b^b &=& \frac{1}{2}\, t_L + b_R\,,\nonumber\\
t^b &=& \frac{1}{2}\, t_L + t_R\,,\nonumber\\\nonumber\\
e^b &=& \mu^b = \frac{1}{4}\,l_L + \frac{1}{2}\,e_R\,,\nonumber\\
\tau^b &=& \frac{1}{2}\,\tau_L + \tau_R\,,\nonumber\\
\nu^b_e &=& \nu^b_{\mu} = \frac{1}{4}\,l_L\,,\nonumber\\
\nu^b_{\tau} &=& \frac{1}{2}\,\tau_L\,,\nonumber\\
g^b &=& g\,,\nonumber\\
W^b &:=& W^+ + W^- = 2 \left( \frac{1}{3} W + \cos^2{\beta} \frac {H_1}{4} +
\sin^2{\beta} \frac {H_2}{4} \right) \, , \nonumber \\
Z^b &=& \sin^2(\theta_W) \,B + \cos^2(\theta_W) \, \frac {W}{3} +
\cos^2{\beta} \frac {H_1}{4} + \sin^2{\beta} \frac {H_2}{4} \, ,
\nonumber \\
\gamma^b &=& \cos^2(\theta_W)\,B + \sin^2(\theta_W)\, \frac{W}{3} \, ,
\nonumber \\
h^{0\,b} &=& \sin^2{\alpha} \, \frac {H_1}{4} + \cos^2{\alpha} \,
\frac {H_2} {4} \, , \nonumber \\
H^{0\,b} &=& \cos^2{\alpha} \, \frac {H_1}{4} + \sin^2{\alpha} \,
\frac {H_2} {4} \, , \nonumber \\
A^{0\,b} &=& \sin^2{\beta} \, \frac {H_1}{4} + \cos^2{\beta} \, \frac
{H_2}{4} \, , \nonumber \\
H^b &:=& H^+ + H^- = 2 \left( \sin^2{\beta} \, \frac {H_1}{4} +
\cos^2{\beta} \, \frac {H_2}{4} \right) \, . 
\eeqa

\clearpage
All superparticles also acquire masses at this stage, and the
particles with identical quantum numbers mix together to give the
mass eigenstates:
\beqa \label{susytrafo}
\tilde{q}_{L/R}^b &=& \tilde{q}_{L/R} \, \, {\rm for} \, q = u,d,s,c
\, , \nonumber \\
\tilde{b}_1^b &=& \frac{1}{2} \cos^2(\theta_b) \, \tilde{t}_L +
\sin^2(\theta_b) \, \tilde{b}_R \, , \nonumber \\
\tilde{t}_1^b &=& \frac{1}{2} \cos^2(\theta_t) \, \tilde{t}_L +
\sin^2(\theta_t) \, \tilde{t}_R \, , \nonumber \\
\tilde{b}_2^b &=& \frac{1}{2} \sin^2(\theta_b) \, \tilde{t}_L +
\cos^2(\theta_b) \, \tilde{b}_R \, , \nonumber \\
\tilde{t}_2^b &=& \frac{1}{2} \sin^2(\theta_t) \, \tilde{t}_L +
\cos^2(\theta_t) \, \tilde{t}_R \, , \nonumber \\
\tilde{e}_L^b &=& \tilde{\mu}_L^b = \frac{1}{4} \, \tilde{l}_L \, ,
\nonumber \\
\tilde{e}_R^b &=& \tilde{\mu}_R^b = \frac{1}{2} \, \tilde{e}_R \, ,
\nonumber \\
\tilde{\tau}_1^b &=& \frac{1}{2} \, \cos^2(\theta_\tau) \, \tilde{\tau}_L +
\sin^2(\theta_\tau) \, \tilde{\tau}_R \, , \nonumber \\
\tilde{\tau}_2^b &=& \frac{1}{2} \, \sin^2(\theta_\tau) \, \tilde{\tau}_L +
\cos^2(\theta_\tau) \, \tilde{\tau}_R \, , \nonumber \\
\tilde{\nu}_e^b &=& \tilde{\nu}_\mu^b = \frac{1}{4} \, \tilde{l}_L \,
, \nonumber \\
\tilde{\nu}_\tau^b &=& \frac{1}{2} \, \tilde{\tau}_L \, , \nonumber \\
\tilde{g}^b &=& \tilde{g} \, , \nonumber \\
\tilde{\chi}_1^b &:=& \tilde{\chi}_1^+ + \tilde{\chi}_1^- =
\left[ \sin^2(\gamma_R) + \sin^2(\gamma_L) \right] \, \frac
{\tilde{W}}{3} + \cos^2(\gamma_R) \, \frac{\tilde{H}_2}{2} +
\cos^2(\gamma_L) \, \frac{\tilde{H}_1}{2} \, , \nonumber \\
\tilde{\chi}_2^b &:=& \tilde{\chi}_2^+ + \tilde{\chi}_2^- =
\left[ \cos^2(\gamma_R) + \cos^2(\gamma_L) \right] \, \frac
{\tilde{W}}{3} + \sin^2(\gamma_R) \, \frac {\tilde{H}_2}{2} +
\sin^2(\gamma_L) \, \frac {\tilde{H}_1}{2} \, , \nonumber \\
\tilde{\chi}_i^{0b} &=& 
\left| v_1^{(i)} \right|^2 \frac {\tilde H_1}{2}
+ \left| v_2^{(i)} \right|^2 \frac {\tilde H_2}{2}
+ \left| v_3^{(i)} \right|^2 \frac {\tilde W}{3}
+ \left| v_4^{(i)} \right|^2 {\tilde B}.
\eeqa
Here we have largely followed the notation of ISASUSY \cite{Isasusy}.
However, we have used the more common symbol $\tilde \chi$ for
charginos and neutralinos; in ISASUSY notation, $\tilde \chi_1^b =
\widetilde W_-, \ \tilde \chi_2^b = \widetilde W_+$, and $\tilde
\chi_i^{0b} = \widetilde Z_i$. The mixing angles $\alpha$ (in the
Higgs sector), $\theta_b, \, \theta_t, \, \theta_\tau$ (in the
sfermion sector), $\gamma_L, \, \gamma_R$ (in the chargino sector) as
well as the $v_i^{(j)}$ (in the neutralino sector) have been computed
numerically using ISASUSY.

\chapter{Two- and three-body decay spectra}
\label{app:decays}

\section{Generalities}

\setcounter{equation}{0}
\setcounter{footnote}{0}
\renewcommand{\theequation}{C.\arabic{equation}}

We want to define the decay functions (DFs) $\tilde{P}_{sS}$
describing two-- or three--body decay $S \rightarrow s$, see
eq.(\ref{decay}). These DFs can be obtained directly from the decay
spectrum of $S$ in the ultra--relativistic limit, where the energy
$E_S$ is much larger than the mass $M$ of $S$:
\beq \label{decff}
\tilde P_{sS}(z) = \frac {1} {\Gamma} \frac {d \Gamma(E_S)} {d z},
\eeq
where $z = E_s / E_S$. This spectrum can e.g. be evaluated by first
computing the double differential decay distribution $d^2 \Gamma / (d
E_s^* d \cos \theta^*)$ in the {\em rest frame} of $S$, then boosting
the four--momentum of $s$ with boost factor $\gamma = E_S / M$ at
angle $\theta^*$ relative to $\vec{p}_s$, and finally integrating over
$\cos \theta^*$ subject to the constraint that the boosted energy of
$s$ equals $E_s$. Note that eq.(\ref{decff}) implies $\int_0^1 \tilde
P_{sS}(z) dz = 1$; if $S-$decays produce $N$ identical particles $s$,
the corresponding $\tilde P_{sS}$ would thus have to be multiplied
with an extra factor of $N$, in order to correctly reproduce the total
multiplicity in the final state. Finally, momentum conservation
implies $\sum_s \int_0^1 z \tilde P_{sS}(z) = 1$.

In case of two--body decays $S \rightarrow i + j$ the energy $E_s^*$
in the rest frame of $S$ is fixed completely by the kinematics. The
boost and integration over $\cos \theta^*$ then leads to a flat
decay function:
\beq
\tilde{P}^{(2)}_{iS} (z) =
\left\{ \left[ 1 - \left( \frac {m_1+m_2} {M } \right)^2 \right]
\left[ 1 - \left( \frac {m_1-m_2} {M} \right)^2 \right]
\right\}^{-\frac{1}{2}} \Theta(z - z^{(i)}_{-})\, \Theta(z^{(i)}_{+}-z)
\eeq
for the decay product $i$ with $i = 1$ or $2$. The kinematic minimum
and maximum $z^{(i)}_{\pm}$ of $z$ are given by:
\beq \label{zlim}
z^{(i)}_{\pm} = \frac{1}{2} \left( 1 + \frac {m_i^2 - m_j^2} {M^2} \pm
\sqrt{ \left[ 1 - \left( \frac {m_1+m_2}{M} \right)^2 \right] \left[ 1
- \left( \frac {m_1-m_2}{M} \right)^2 \right]} \right).
\eeq
For example, for $m_1 \rightarrow M, \ m_2 \rightarrow 0$,
eq.(\ref{zlim}) implies $z^{(1)}_\pm \rightarrow 1, \ z^{(2)}_\pm
\rightarrow 0$, i.e. the entire energy of $S$ goes into the massive
decay product. In contrast, for $m_1 = m_2$, the energy of $S$ will on
average be shared equally between the two decay products; if $m_1 =
m_2 \rightarrow 0$, the $z^{(i)}$ can lie anywhere between zero and
one. Since $E_s^*$ is fixed in this case, our treatment of two--body
decays is exact up to possible polarization effects; we do not expect
these effects to be very important, except perhaps in case of $\tau$
decays (which, however, usually do not contribute very much to the
final spectra of stable particles).

Three--body decays lead to a nontrivial distribution of the energy of
the decay products already in the rest frame of $S$. For simplicity we
assume that at most one of the three decay products is massive; this
should be a safe approximation, except for $b \rightarrow c \tau
\nu_\tau$ decays, which have a rather small branching ratio. We then
need separate DFs for the massive and massless decay products. For the
massive decay product, with mass $m$, we find
\beq
\tilde{P}^{(3)}_{sS}(z) = N_3 \left[ 1 - z + \frac{m^2}{M^2} \left(
1 - \frac{1}{z} \right) \right]
\eeq
where $z \in [\frac{m^2}{M^2},1]$ and the normalization factor is
given by:
\beq \label{n3}
N_3 = \left[ \frac{1}{2} \left( 1 - \frac{m^4}{M^4} \right) +
\frac{m^2}{M^2} \log \frac{m^2}{M^2} \right]^{-1}. 
\eeq

If on the contrary, $s$ is one of the massless decay products, we find:
\beq
\tilde{P}^{(3)}_{sS}(z) = N_3 \left [ 1 - z - \frac{m^2}{M^2} \left( 1 +
\log \frac{M^2}{m^2} + \log(1-z) \right) \right],
\eeq
where now $z \in [0,1-\frac{m^2}{M^2}]$; the normalization factor
$N_3$ has been given in eq.(\ref{n3}).

Our treatment of three--body decays is not exact, since it ignores
dynamical effects (described by the invariant Feynman amplitude) on
the decay spectrum in the $S$ rest frame.\footnote{The calculation of
the corresponding branching ratio in ISASUSY does include these
dynamical effects; in other cases the required branching ratio can be
taken directly from experiment, e.g. for $\tau$ decays.} However,
treating these effects properly is quite nontrivial, since it would
force us to introduce many different three--body decay functions. Note
in particular that massive superparticles (charginos and neutralinos)
do generally not decay via $V-A$ interactions, unlike the $b$ and $c$
quarks and heavy $\mu$ and $\tau$ leptons in the SM. Moreover, the
Feynman amplitudes in many cases depend nontrivially on the
polarization of the decaying particle; this could only be described at
the cost of introducing many additional generalized fragmentation
functions, since we would have to keep track of left-- and
right--handed particles separately. However, experience from hadron
collider physics teaches us that including the exact decay matrix
elements is usually not very important if one is only interested in
single--particle inclusive spectra. We expect this to be true in our
case as well, since the convolution with parton distribution functions
necessary at hadron colliders is reminiscent of the convolution with
generalized FFs in our case. We finally note that longer decay chains
involving two-- and three--body decays can be treated by simply
convoluting appropriate factors of $\tilde P^{(2)}_{sS}$ and $\tilde
P^{(3)}_{sS}$. 

\section{Treatment of heavy quark decays}

The top quark being very heavy ($m_t \sim 175$ GeV $\gg m_{\rm had}
\sim 1 $ GeV), it decays before hadronizing, and can thus be included
in the decay cascade at scale $M_{\rm SUSY}$. On the other hand, the
hadronization of the $b$ and $c$ quarks has to be treated with some
care. The ``input'' fragmentation functions we used \cite{Poetter}
already include the final hadrons (nucleons, kaons and pions) produced
at the end of the decay cascade of $c-$ and $b-$flavored hadrons.
However, they do not include the leptons arising from this cascade,
which are not negligible. We therefore implemented a special treatment
for this component, using the empirical FFs proposed by Peterson et
al. \cite{Peterson} for heavy quarks as a basis for the fragmentation
of $c-$ and $b-$hadrons. To that end, we used two ``generic''
particles, a $c-$ and a $b-$hadron, with respective average masses
$\overline{m}_c = 2.1$ GeV and $\overline{m}_b = 5.3$ GeV; we also had
to renormalize the complete set of FFs for $b$'s and $c$'s. The scheme
can be described by Fig.~12. Here, $B_l(b)$ and $B_l(c)$ describe the
branching ratio of the semi--leptonic decay modes of $b-$ and
$c-$flavored hadrons, respectively [summed over all accessible pairs
($l$,$\nu_l$)].

\vspace*{25mm}
\begin{center} 
\begin{figure}[h] \label{hqdec}
\begin{picture}(-400,50)(0,-120)
\SetOffset(420,-20)
\label{heavy_decays}
\SetPFont{Helvetica}{24}
\Text(-405,0)[]{$b$}
\ArrowLine(-395,0)(-330,0) 
\Text(-360,-10)[]{Peterson} 
\Text(-300,0)[]{$b-$hadron}
\ArrowLine(-265,0)(-165,0)
\Text(-220,10)[]{semi-leptonic}
\Text(-220,-10)[]{decay $B_l(b)$}
\Text(-120,0)[]{$c-$hadron   $l$   $\nu_l$}
\Line(-130,-10)(-130,-30)
\ArrowLine(-130,-30)(-40,-30)
\Text(-85,-20)[]{semi-leptonic}
\Text(-85,-40)[]{decay $B_l(c)$}
\Text(-5,-30)[]{$s-$hadron   $l$   $\nu_l$}
\Text(-405,-80)[]{$c$}
\ArrowLine(-395,-80)(-330,-80) 
\Text(-360,-90)[]{Peterson} 
\Text(-300,-80)[]{$c-$hadron}
\ArrowLine(-265,-80)(-165,-80)
\Text(-220,-70)[]{semi-leptonic}
\Text(-220,-90)[]{decay $B_l(c)$}
\Text(-120,-80)[]{$s-$hadron   $l$   $\nu_l$}
\end{picture} 
\caption{Schematic hadronization and decay cascade for heavy
quarks $c$ and $b$. The ``$s-$hadrons'', mainly kaons, are already included
in the FFs given in \cite{Poetter}.}
\end{figure}
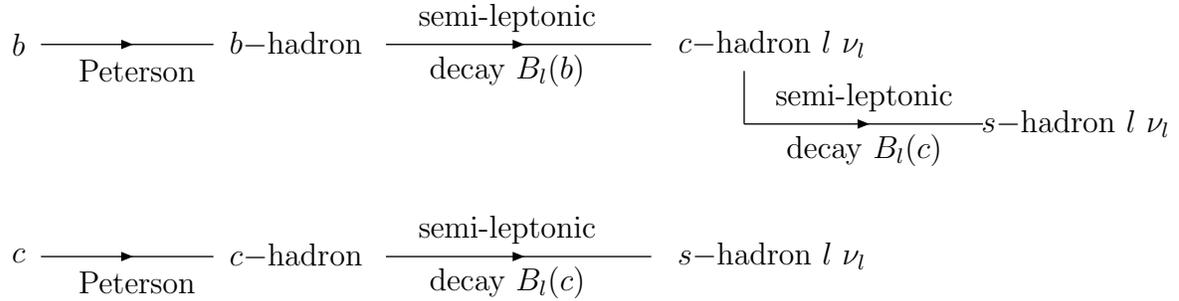
\end{center}
\vspace*{5mm}

As mentioned earlier, the leptonic $b$ and $c$ decay products have to
be included in the normalization of the FFs $D_b^h$ and $D_c^h$. To
that end, we introduce $R_c$ and $R_s$, the energy carried by the $c$
and $s$ quark in semi--leptonic $b-$ and $c-$ decays, respectively, as
well as $x_B$ and $x_D$, the energy fraction of the $b$ ($c$) quark
carried by the $b-$flavored ($c-$flavored) hadron. The latter are
given by
\beq
x_{B,D} = \int_0^1 z D_{\rm Pet}^{b,c}(z) \, dz \, ,
\eeq

where $D_{\rm Pet}^{b,c}$ is the Peterson FF \cite{Peterson} for $b$
and $c$ quarks, respectively; we took $\epsilon_c = 0.15, \,
\epsilon_b = 0.015$ for the single free parameter in these FFs. We
compute $R_c$ and $R_s$ from pure phase space, i.e. we again ignore
possible effects of the Feynman amplitudes on the three--body decay
distributions. This gives:
\beqa \label{eqR}
R_c &=& \frac {1} {\overline{m}_b \Gamma(b \rightarrow c l \nu) }
\int_{\overline{m}_c}^{E_c,{\rm max}} d E_c \, E_c \, \frac {d
\Gamma(b \rightarrow c l \nu)} {d E_c} \nonumber \\
&=& \frac {(1 - r)^3} {3 (1 - r^2) - 6 r \log r} \, ,
\eeqa
where $r = \overline{m}_c^2 / \overline{m}_b^2 = 0.157$ for our choice
of average $b-$ and $c-$hadron masses; note that $R_c \rightarrow 1/3
\ (1)$ for $r \rightarrow 0 \ (1)$. Eq.(\ref{eqR}) can also be used
for the computation of $R_s$, with $\overline{m}_c \rightarrow
\overline{m}_s \simeq 0.5$ GeV, $\overline{m}_b \rightarrow
\overline{m}_c$. The FFs of \cite{Poetter} only include the hadrons
produced in the fragmentation and decays of the $c-$ and
$b-$quarks. Their normalization, which we need to know for the
necessary extrapolation of these FFs towards small $x$ as discussed in
Appendix D, is therefore given by
\beqa 
\frac{1}{x_B} \sum_h \int_0^1 zD_b^h(z)\,dz &=&  \frac{1-x_B}{x_B}
\nonumber \\
&+& [1 - B_l(b)] \cdot [1 - B_l(c)]  \nonumber \\ 
&+& \, B_l(b) \cdot [1 - B_l(c)] \cdot R_c  \nonumber \\
&+& \, [1 - B_l(b)] \cdot B_l(c) \cdot ( R_c\cdot R_s + 1 - R_c )
\nonumber \\ 
&+& \, B_l(b) \cdot B_l(c) \cdot R_c \cdot R_s  \nonumber \\
&=& \frac{1}{x_B} - B_l(b) \cdot (1-R_c) - R_c \cdot B_l(c) \cdot (1-R_s).  
\eeqa
The right hand side can be understood as follows: the first line
describes the contribution of the light hadrons produced when the
$b-$quark hadronizes into a $b-$flavored hadron; the second line
describes purely hadronic decays; the third line describes leptonic
primary $b$ decays followed by hadronic $c$ decays (in this case only
the fraction $R_c$ of the $b-$hadrons energy goes into hadrons); the
fourth line describes the hadronic energy fraction in the case of a
hadronic primary $b$ decay followed by leptonic $c-$decays; finally,
the fifth line describes the hadronic energy fraction after leptonic
decays in both the primary and secondary decays. The same holds for
$c-$hadron decays, up to the simplifying fact that ``$s-$hadrons'' are
already fully included into the FFs of \cite{Poetter}. We get: 
\beq
\frac{1}{x_D} \sum_h \int_0^1 zD_c^h(z)\,dz =  \frac{1}{x_D} - B_l(c)
+ B_l(c) \cdot R_s.  \eeq

\setcounter{equation}{0}
\renewcommand{\theequation}{D.\arabic{equation}}

\chapter{Parameterization of the input fragmentation
  functions}
\label{app:FFs}

Here we give the input fragmentation functions (FFs) we used to
describe the hadronization of quarks and gluons, taken from
\cite{Poetter}, and the parameters of the extrapolation we made in the
small $x$ region. 

The functions taken from ref~\cite{Poetter} are given with the
functional form $N x^\alpha (1-x)^\beta$; there are given in
table~\ref{Poetter}, at the scale where quarks and hadrons hadronize,
i.e. $Q_0 = max(m_q,Q_{had})$. We used the NLO results, excepted for the
s quark, for which we had to use the LO ones, because the NLO form
didn't allow us to impose energy conservation and continuity at low x.

\begin{center}
\begin{table}
\begin{center}
\begin{tabular}{|c||c||c|c|c||c|c|}\hline
&&&&&&\\
$D_p^h(x,Q_0^2)$ & $p$ & $n$ & $\pi^\pm$ & $\pi^0$ & $K^\pm$ & $K^0$ \\ 
$= N x^\alpha (1-x)^\beta$&&&&&&\\ \hline 
$u$ & $N = 1.26$&0.63&0.448&0.224&0.178&4.96\\ 
& $\alpha = 0.0712$&0.0712&-1.48&-1.48&-0.537&0.0556\\
& $\beta = 4.13$&4.13&0.913&0.913&0.759&2.8\\ \hline  
$d$ & 0.63&1.26&0.448&0.224&4.96&0.178\\
&0.0712&0.0712&-1.48&-1.48&0.0556&-0.537\\
&4.13&4.13&0.913&0.913&2.8&0.759\\ \hline
$s$ & 4.08&4.08&22.3&11.15&0.259&0.259\\
&-0.0974&-0.0974&0.127&0.127&-0.619&-0.619\\
&4.99&4.99&6.14&6.14&0.859&0.859\\ \hline  
$c$ & 0.0825&0.0825&6.17&3.085&4.26&4.26\\
&-1.61&-1.61&-0.536&-0.536&-0.241&-0.241\\
&2.01&2.01&5.6&5.6&4.21&4.21\\ \hline  
$b$ & 24.3&24.3&0.259&0.1295&1.32&1.32\\
&0.579&0.579&-1.99&-1.99&-0.884&-0.884\\
&12.1&12.1&3.53&3.53&6.15&6.15\\ \hline  
$g$ & 1.56&1.56&3.73&1.865&0.231&0.231\\
&0.0157&0.0157&-0.742&-0.742&-1.36&-1.36\\
&3.58&3.58&2.33&2.33&1.8&1.8\\ \hline  
\end{tabular}
\caption{Input fragmentation
  functions at small $x$, with functional form $N x^\alpha
  (1-x)^\beta$, taken from \cite{Poetter} at $Q_0 = max(m_q,Q_{had})$.
  We took their NLO results for u,d,c,b and g, but the LO result for
  the s quark. See the text for further details.}
\label{Poetter}
\end{center}
\end{table}
\end{center}

As we showed in Sec.~\ref{subsec:non_pert}, the
final result at low $x$ depends very little on the chosen power law in
our parameterization
\beq \label{fpara}
f(x) = ax^{-\alpha'} + b\log{x} + c \, ,\ a > 0,
\eeq
once energy conservation has been imposed. Here we therefore only
give results for a parameterization where $\alpha'$ is taken to be
1. That is, we assume that the multiplicity due to non--perturbative
effects gets the same contribution for each decade of energy, if the
hadron's energy is small compared to that of the initial parton.

In order to obtain a unique solution with the only two constraints at
our disposal (energy conservation and continuity of the FFs), we
imposed these constraints on the sum of the FFs $\sum_h D_i^h(x,Q^2)$,
where $i$ is a given initial parton, and $h$ runs over the final
hadrons. Of course, energy will be conserved independently for each
initial parton $i$. For each $i$ we define a cut--off $x_0^i$ which
defines the transition between the functions given in \cite{Poetter}
and our extrapolation. The $x_0^i$ have to be chosen such that the
equations of energy conservation admit a solution; a necessary (but
generally not sufficient) requirement is that the integral over the
original FFs satisfy $\sum_h \int_{x_0}^1 dz z \, D_i^h(z) < 1$. Our
requirement of continuity at $x_0$ implies that the final results depend
very little on the precise values of the $x_0^i$. For simplicity we
assume that all the $D_i^h(x,Q^2_{\rm had})$ for a given $i$ have the
same shape at small $x$; recall that purely perturbative effects
ensure that this is true after DGLAP evolution, which anyway greatly
reduces the sensitivity to the input. The normalizations for the
various hadrons can then be read off directly from the results of
ref.\cite{Poetter}, once $x_0$ has been determined. The results are
presented in table.~\ref{extrapolation}, which lists the cut-off
$x_0^i$, the coefficients $a, b, c$ of the functional form
(\ref{fpara}) describing the {\em sum} $\sum_h D_i^h$ for fixed parton
$i$, and the normalization coefficients\footnote{At first sight the
relative ordering of the $N_u^K, \, N_d^K$ factors may seem
counter--intuitive. Indeed, a $u-$quark should more readily fragment
into a charged Kaon than into a neutral one, whereas the opposite
behavior is expected for $d-$quarks. Recall, however, that here we are
only interested in the behavior at small $x$. In this case
ref.\cite{Poetter} finds the opposite behavior as at large $x$,
i.e. $u-$quarks indeed seem to be more likely to produce a {\em soft}
neutral kaon than a {\em soft} charged kaon.} $N_i^h$, so that $D_i^h
= N_i^h \sum_h D_i^h$; of course, $\sum_h N_i^h = 1 \ \forall i$.

\begin{center}
\begin{table}
\begin{center}
\begin{tabular}{|c||c||c|c|c||c|c|c|c|c|c|}\hline
initial parton & $x_0$ & $a$ & $b$ & $c$ & $p$ & $n$ & $\pi^\pm$ &
$\pi^0$ & $K^\pm$ & $K^0$ \\ \hline 
$u$ & 0.27 & 4.06 & -9.74 & -14.40 & 0.05 & 0.025 & 0.38 & 0.19 & 0.05 &
0.31 \\ \hline
$d$ & 0.27 & 4.06 & -9.74 & -14.40 & 0.025 & 0.05 & 0.38 & 0.19 & 0.31 &
0.05 \\ \hline
$s$ & 0.20 & 5.74 & -18.47 & -31.42 & 0.14 & 0.14 & 0.41 & 0.21 & 0.05 &
0.05 \\ \hline
$c$ & 0.27 & 4.06 & -4.16 & -6.24 & 0.05 & 0.05 & 0.30 & 0.15 & 0.22 &
0.22 \\ \hline
$b$ & 0.20 & 5.74 & -27.81 & -49.27 & 0.08 & 0.08 & 0.35 & 0.17 & 0.16 &
0.16 \\ \hline
$g$ & 0.37 & 1.82 & -4.81 & -2.40 & 0.05 & 0.05 & 0.50 & 0.25 & 0.07 &
0.07 \\ \hline
\end{tabular}
\caption{Coefficients of the extrapolation of the input fragmentation
functions at small $x$. Column 2 gives the cut--off $x_0$ where we
switch from the FFs from \cite{Poetter} to a parameterization in the
form (\ref{fpara}). Columns 3 to 5 give the coefficients of this
parameterization, as applied to the sum $\sum_h D_i^h$. The remaining
columns give normalizations $N_i^h$, so that $D_i^h = N_i^h \, \sum_h
D_i^h$. Note that $h$ always stands for the sum of particle and
anti--particle, whenever the two are not identical; for example,
$\pi^\pm$ stands for the sum of $\pi^+$ and $\pi^-$, $K^0$ stands for
the sum of $K^0$ and $\overline{K^0}$, etc. }
\label{extrapolation}
\end{center}
\end{table}
\end{center}

\setcounter{equation}{0}
\renewcommand{\theequation}{E.\arabic{equation}}

\chapter{Description of the compound particles used in SHdecay} 
\label{app:compound}

Here I describe the 30 interaction eigenstates (or ``compound
particles'') of the MSSM which have been used as possible decay
products for the $X$ particle. As the decay is occuring well above the
breaking scales of SUSY and $SU(2) \otimes U(1)$, one has to allow a
decay into supersymmetric particles as well as SM particles, and to
distinguish between the helicities (Left or Right) of the Dirac
fermions; yet, well above the breaking scales of SUSY and $SU(2)
\otimes U(1)$, it is assumed that one doesn't need to distinguish
between the components of a given SU(2) multiplet\footnote{This is
  certainly true if $X$ is an $SU(2)$ doublet.}, in particular between
the ``up'' and ``down'' components of the SU(2) doublets. Moreover, up
to the Yukawa couplings which become relevant only for the third
generation of fermions, no difference is made between the generations,
all particles being massless above the $SU(2) \otimes U(1)$ breaking
scale. If we consider a perfect CP symmetry, one doesn't need to
distinguish between particles and antiparticles, either. In summary,
for example, the fields ($u_L$, $d_L$), ($c_L$, $s_L$), ($\bar{u}_L$,
$\bar{d}_L$), and ($\bar{c}_L$, $\bar{s}_L$) all obey exactly the same
DGLAP evolution equation and thus can be considered as a {\it single}
``particle'' which is taken to be an {\it average} over all these
fields. This ``coumpound particle'' is called $q_L$ in our
nomenclature and has id 1. We give in table~\ref{fermions} all
fermionic compound particles we used, together with the associated
superparticles, and their respective id's.

The same occurs for bosons and bosinos, where we only have to consider
the unbroken fields $B$, $W$, $g$ (for gluons), the two SU(2) Higgs doublets
of the MSSM $H_1$ (coupled to leptons and down-type quarks of the
third generation) and $H_2$ (coupled to the up-type quarks of the
third generation), and their superpartners. The well known particles
and antiparticles at lower energies are mixtures of the components
of these interaction eigenstates. We give the corresponding id's in
table~\ref{bosons}.

\begin{center}
\begin{table}[ht!]
\begin{center}
\begin{tabular}{|c|c|}\hline
compound particle &\rule[-3mm]{0mm}{8mm}id\\\hline
$q_L = \frac{1}{4}\left[\left(\frac{u_L}{d_L}\right) + \left(\frac{c_L}{s_L}\right) +
    \left(\frac{\bar{u}_L}{\bar{d}_L}\right) +
    \left(\frac{\bar{s}_L}{\bar{c}_L}\right)\right]$&1\\
$q_R = \frac{1}{4}\left[u_R + c_R + \bar{u}_R + \bar{c}_R \right]$&2\\
$d _R = \frac{1}{4}\left[d_R + s_R + \bar{d}_R + \bar{s}_R \right]$&3\\
$t_L = \frac{1}{2}\left[\left(\frac{t_L}{b_L}\right) +
  \left(\frac{\bar{t}_L}{\bar{b}_L}\right) \right]$&4\\
$t_R = \frac{1}{2}(t_R + \bar{t}_R)$&5\\
$b_R = \frac{1}{2}(b_R + \bar{b}_R)$&6\\
&\\
$\tilde{q}_L = \frac{1}{4}\left[\left(\frac{\tilde{u}_L}{\tilde{d}_L}\right) + \left(\frac{\tilde{c}_L}{\tilde{s}_L}\right) +
    \left(\frac{\tilde{\bar{u}}_L}{\tilde{\bar{d}}_L}\right) +
    \left(\frac{\tilde{\bar{s}}_L}{\tilde{\bar{c}}_L}\right)\right]$&7\\
$u_R = \frac{1}{4}\left[\tilde{u}_R + \tilde{c}_R + \tilde{\bar{u}}_R + \tilde{\bar{c}}_R \right]$&8\\
$d _R = \frac{1}{4}\left[\tilde{d}_R + \tilde{s}_R + \tilde{\bar{d}}_R + \tilde{\bar{s}}_R \right]$&9\\
$t_L = \frac{1}{2}\left[\left(\frac{\tilde{t}_L}{\tilde{b}_L}\right)+
  \left(\frac{\tilde{\bar{t}}_L}{\tilde{\bar{b}}_L}\right) \right]$&10\\
$t_R = \frac{1}{2}(\tilde{t}_R + \tilde{\bar{t}}_R)$&11\\
$b_R = \frac{1}{2}(\tilde{b}_R + \tilde{\bar{b}}_R)$&12\\
&\\
$l_L = \frac{1}{4}\left[\left(\frac{e_L}{\nu_e}\right) + \left(\frac{\mu_L}{\nu_\mu}\right) +
    \left(\frac{\bar{e}_L}{\bar{\nu}_e}\right) +
    \left(\frac{\bar{\mu}_L}{\bar{\nu}_\mu}\right)\right]$&13\\
$l_R = \frac{1}{4}\left[e_R + \mu_R + \bar{e}_R + \bar{\mu}_R
\right]$&14\\
$\tau_L = \frac{1}{2}\left[\left(\frac{\tau_L}{\nu_\tau}\right) +
  \left(\frac{\bar{\tau}_L}{\bar{\nu}_\tau}\right) \right]$&15\\
$\tau_R = \frac{1}{2}(\tau_R + \bar{\tau}_R)$&16\\
&\\
$\tilde{l}_L = \frac{1}{4}\left[\left(\frac{\tilde{e}_L}{\tilde{\nu}_e}\right) + \left(\frac{\tilde{\mu}_L}{\tilde{\nu}_\mu}\right) +
    \left(\frac{\tilde{\bar{e}}_L}{\tilde{\bar{\nu}}_e}\right) +
    \left(\frac{\tilde{\bar{\mu}}_L}{\tilde{\bar{\nu}}_\mu}\right)\right]$&17\\
$l_R = \frac{1}{4}\left[\tilde{e}_R + \tilde{\mu}_R +
  \tilde{\bar{e}}_R + \tilde{\bar{\mu}}_R \right]$&18\\
$\tau_L = \frac{1}{2}\left[\left(\frac{\tilde{\tau}_L}{\tilde{\nu}_\tau}\right) +
  \left(\frac{\tilde{\bar{\tau}}_L}{\tilde{\bar{\nu}}_\tau}\right) \right]$&19\\
$\tau_R = \frac{1}{2}(\tilde{\tau}_R + \tilde{\bar{\tau}}_R)$&20\\\hline
\end{tabular}
\caption{Definition and id's of the compound SM fermions and their superpartners in SHdecay.}
\label{fermions}
\end{center}
\end{table}
\end{center}

\begin{center}
\begin{table}[ht!]
\begin{center}
\begin{tabular}{|c|c|}\hline
compound particle &\rule[-3mm]{0mm}{8mm}id\\\hline
$W = \frac{1}{3}(W_1 + W_2 + W_3)$&21\\
$B$&22\\
$g$&23\\
$H_1 = \frac{1}{2}(H_1^1+H_1^2)$&24\\
$H_2 = \frac{1}{2}(H_2^1+H_2^2)$&25\\
&\\
$\tilde{W} = \frac{1}{3}(\tilde{W}_1 + \tilde{W}_2 + \tilde{W}_3)$&26\\
$\tilde{B}$&27\\
$\tilde{g}$&28\\
$\tilde{H}_1 = \frac{1}{2}(\tilde{H}_1^1+\tilde{H}_1^2)$&29\\
$\tilde{H}_2 = \frac{1}{2}(\tilde{H}_2^1+\tilde{H}_2^2)$&30\\\hline
\end{tabular}
\caption{Definition and id's of the compound bosonic SM particles and their superpartners in SHdecay.}
\label{bosons}
\end{center}
\end{table}
\end{center}

\chapter{Stable particle spectra for different initial
(super)particles} 
\label{app:curves}

Here we give an almost complete set of FFs for different initial
particles, for one set of SUSY parameters, with low $\tan\beta$ and
gaugino--like LSP; the dependence of these results on the SUSY
parameters has been analyzed in Sec.~\ref{subsec:SUSYdep}. We
used a ratio of Higgs vevs $\tan \beta = 3.6$, a gluino and scalar
mass scale $M_{\rm SUSY} \sim 500$ GeV, a supersymmetric Higgs mass
parameter $\mu = 500$ GeV, a CP--odd Higgs boson mass $m_A = 500$ GeV
and trilinear soft breaking parameter $A_t = 1$ TeV. As usual, we
plot $x^3 \cdot D_I^P (x,M_X)$. We take $M_X = 10^{16}$ GeV, as
appropriate for a GUT interpretation of the $X$ particle.

\clearpage


\begin{figure}
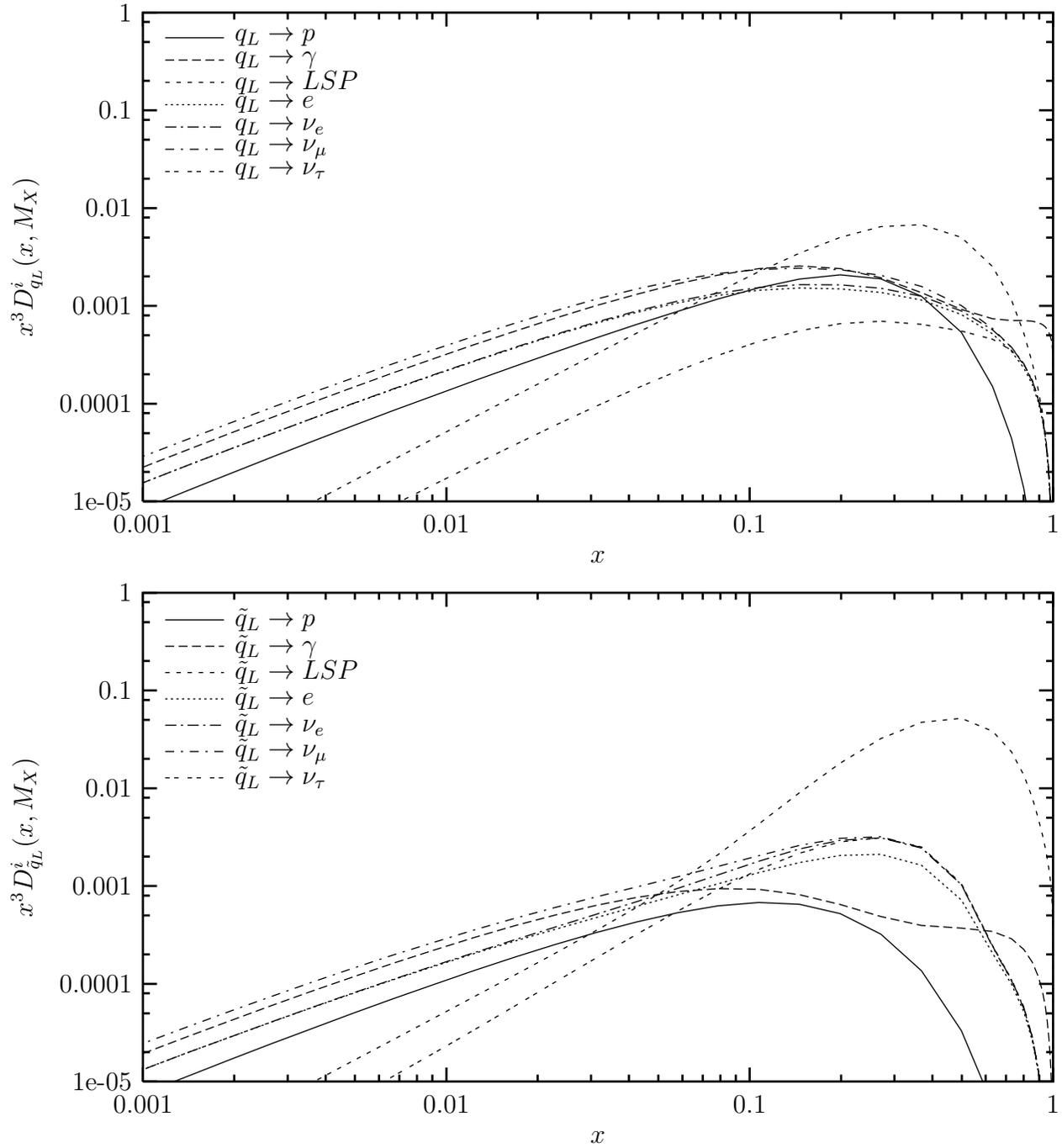

\input{Figures/Low_G_uL.tex}
\input{Figures/Low_G_uL_.tex}
\caption{Fragmentation functions of a first or second generation
  $SU(2)$ doublet quark (top) and a squark (bottom) into stable
  particles.}
\end{figure}

\begin{figure}
\input{Figures/Low_G_uR.tex}
\input{Figures/Low_G_uR_.tex}
\caption{Fragmentation functions of a first or second generation
  $SU(2)$ singlet quark (top) and a squark (bottom) into stable
  particles.}
\end{figure}

\begin{figure}
  \input{Figures/Low_G_tL.tex} \input{Figures/Low_G_tL_.tex}
\caption{Fragmentation functions of a third generation $SU(2)$ doublet
quark (top) and a squark (bottom) into stable particles.} 
\end{figure}

\begin{figure}
\input{Figures/Low_G_tR.tex}
\input{Figures/Low_G_tR_.tex}
\caption{Fragmentation functions of a third generation $SU(2)$ singlet
quark (top) and a squark (bottom) into stable particles.} 
\end{figure}

\clearpage


\begin{figure}
\input{Figures/Low_G_eL.tex}
\input{Figures/Low_G_eL_.tex}
\caption{Fragmentation functions of a first or second generation
$SU(2)$ doublet lepton (top) or slepton (bottom) into stable
particles. The structures in some of the curves in the lower frame
originate from 2--body decay kinematics.}
\end{figure}

\begin{figure}
\input{Figures/Low_G_eR.tex}
\input{Figures/Low_G_eR_.tex}
\caption{Fragmentation functions of a first or second generation
$SU(2)$ singlet lepton (top) or slepton (bottom) into stable
particles. The structures in some of the curves in the lower frame
originate from 2--body decay kinematics.}
\end{figure}

\begin{figure}
\input{Figures/Low_G_tauL.tex}
\input{Figures/Low_G_tauL_.tex}
\caption{Fragmentation functions of a third generation
$SU(2)$ doublet lepton (top) or slepton (bottom) into stable
particles. The structures in some of the curves in the lower frame
originate from 2--body decay kinematics.}
\end{figure}

\begin{figure}
\input{Figures/Low_G_tauR.tex}
\input{Figures/Low_G_tauR_.tex}
\caption{Fragmentation functions of a third generation
$SU(2)$ singlet lepton (top) or slepton (bottom) into stable
particles. The structures in some of the curves in the lower frame
originate from 2--body decay kinematics.}
\end{figure}

\clearpage


\begin{figure}
\input{Figures/Low_G_B.tex}
\input{Figures/Low_G_B_.tex}
\caption{Fragmentation functions of a $B$ boson(top) and a Bino
(bottom) into stable particles.}
\end{figure}

\begin{figure}
\input{Figures/Low_G_W.tex}
\input{Figures/Low_G_W_.tex}
\caption{Fragmentation functions of a $W$ boson (top) and a Wino
(bottom) into stable particles.}
\end{figure}

\begin{figure}
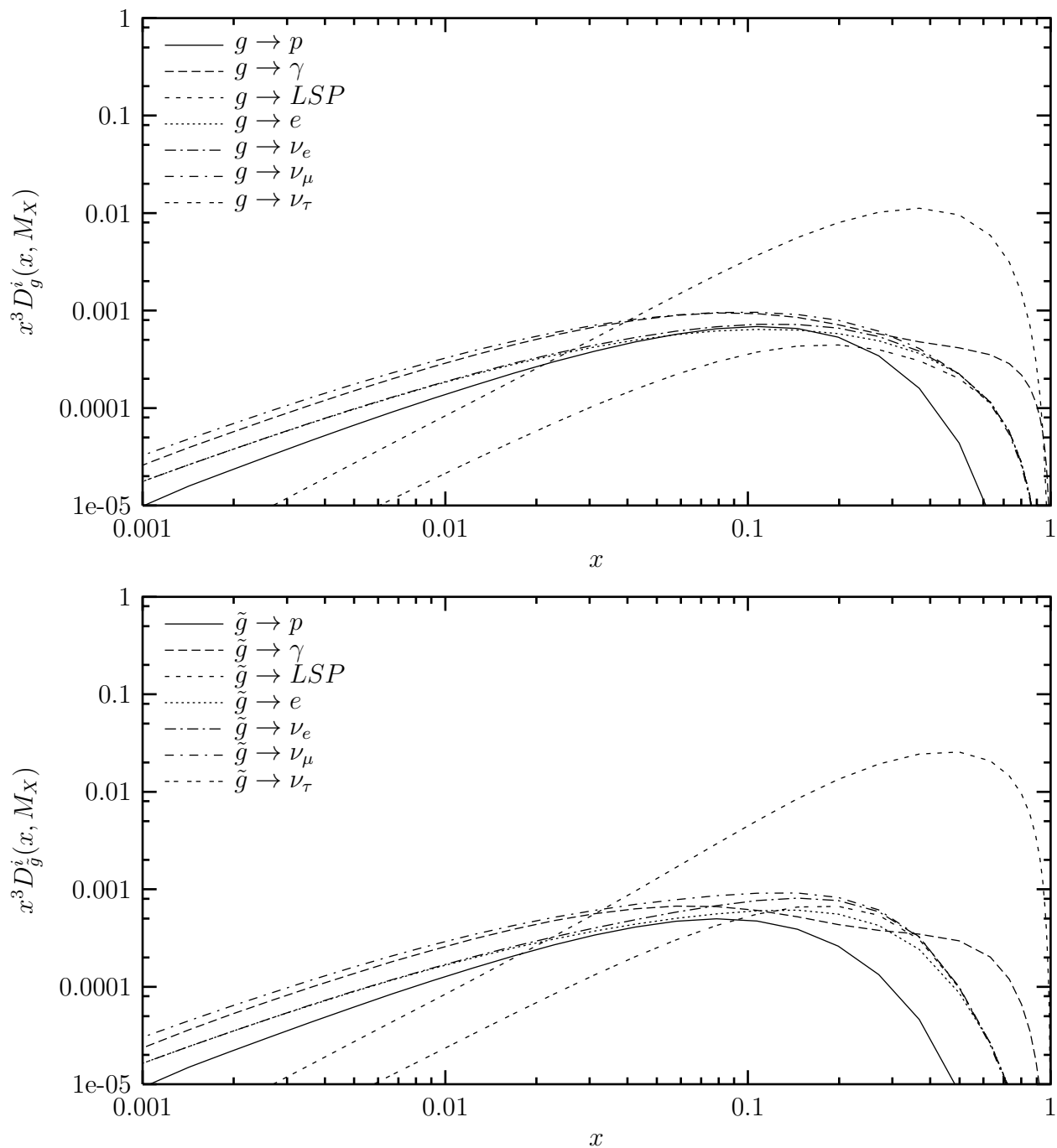

\input{Figures/Low_G_g.tex}
\input{Figures/Low_G_g_.tex}
\caption{Fragmentation functions of a gluon (top) and a gluino
(bottom) into stable particles.}
\end{figure}

\clearpage

\begin{figure}
\input{Figures/Low_G_H1.tex}
\input{Figures/Low_G_H1_.tex}
\caption{Fragmentation functions of a $H_1$ Higgs boson (top) and a
$\tilde{H}_1$ higgsino (bottom) into stable particles.}
\end{figure}

\begin{figure}
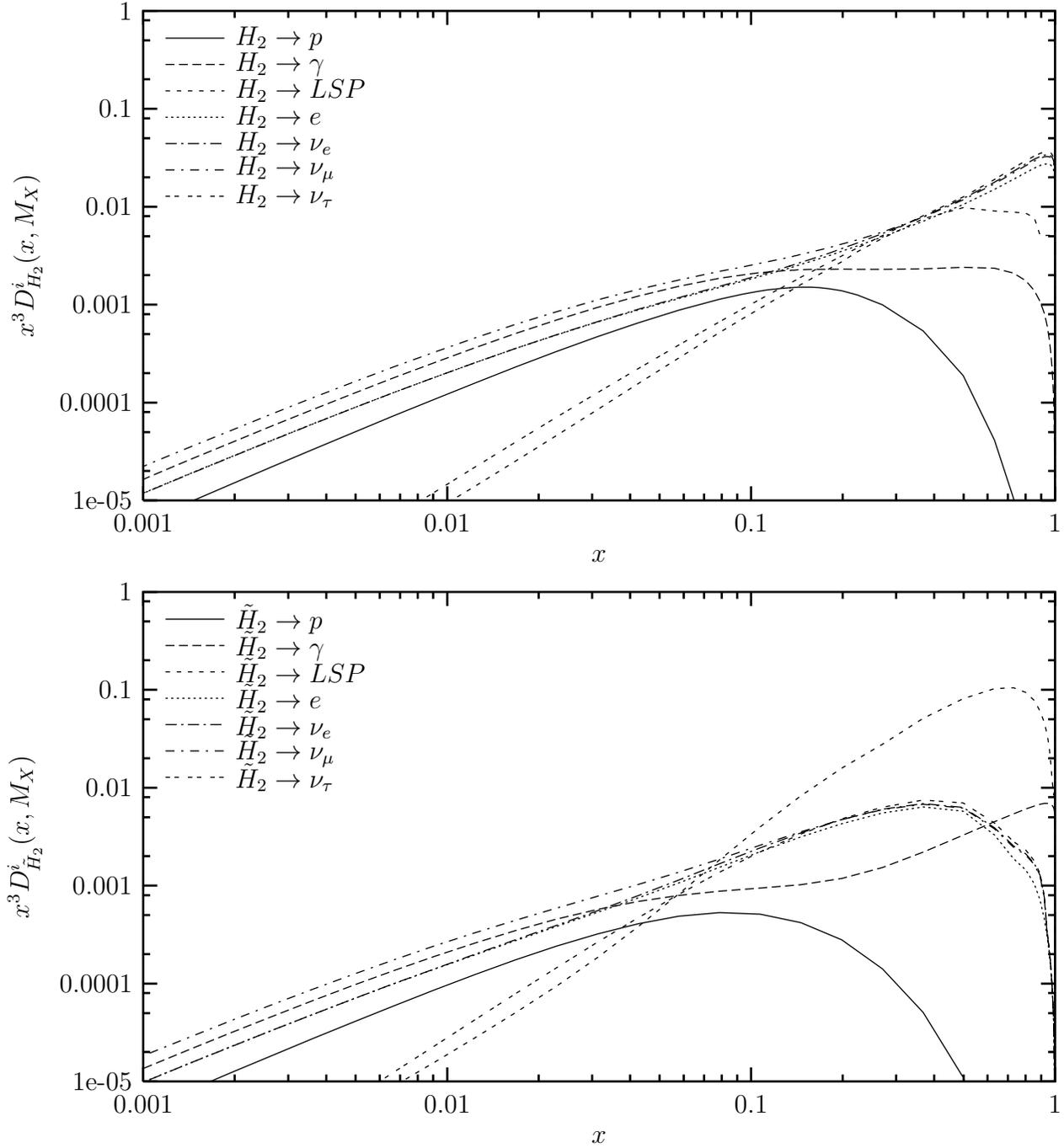

\input{Figures/Low_G_H2.tex}
\input{Figures/Low_G_H2_.tex}
\caption{Fragmentation functions of a $H_2$ Higgs boson (top) and a
$\tilde{H}_2$ higgsino (bottom) into stable particles.}
\end{figure}

\clearpage

\renewcommand{\chaptermark}[1]{\markboth{#1}{}}
\lhead[\fancyplain{}{\bf \thepage}]{\fancyplain{}{\rightmark}}
\rhead[\fancyplain{}{\rightmark}]{\fancyplain{}{\bf \thepage}}

\addcontentsline{toc}{chapter}{Bibliography}




\begin{thebibliography}{999}
  

  
\bibitem{gut}
  U. Amaldi, W. de Boer and H. F\"urstenau,\\
  ``Comparison of grand unified theories with electroweak and strong coupling constants measured at LEP,''\\
  {\em Phys. Lett.} {\bf B260} (1991) 447;\\
  P. Langacker and M. Luo,\\
  ``Implications of precision electroweak experiments for M(t), rho(0), sin**2-Theta(W) and grand unification,''\\
  {\em Phys. Rev.} {\bf D44} (1991) 817;\\
  J. Ellis, S. Kelley and D.V. Nanopoulos, \\
  ``Probing The Desert Using Gauge Coupling Unification,''\\
  {\em Phys. Lett.} {\bf B260} (1991) 131;\\
  C. Giunti, C.W. Kim and U.W. Lee,\\
  ``Running Coupling Constants And Grand Unification Models,''\\
  {\em Mod. Phys. Lett.} {\bf A6} (1991) 1745.
  
\bibitem{reviewMartin}
  S.~P.~Martin,\\
  ``A supersymmetry primer,''\\
  In Kane, G.L. (ed.): Perspectives on supersymmetry, 1-98
  (hep-ph/9709356); S.~Sarkar, {\tt
    hep-ph/0202013}; L.~Anchordoqui, T.~Paul, S.~Reucroft and
  J.~Swain, Int.\ J.\ Mod.\ Phys.\ A {\bf 18} (2003) 2229 {\tt
    hep-ph/0206072}.


\bibitem{reviewSigl}
  P.~Bhattacharjee and G.~Sigl,\\
  ``Origin and propagation of extremely high energy cosmic rays,''\\
  {\em Phys. Rept.} {\bf 327} (2000) 109, (astro-ph/9811011).
  
\bibitem{Cronin}
  J. W. Cronin, T. K. Gaisser, and S. P. Swordy,\\
  ``Cosmic Rays at the Energy Frontier,''\\
  Sci. Amer. {\bf 276}, January 44 (1997).
  
\bibitem{creat}
  D.J.H. Chung, E.W. Kolb and A. Riotto,\\
  ``Nonthermal supermassive dark matter,''\\
  Phys. Rev. Lett. {\bf 81}, 4048
  (1998), (hep-ph/9805473);\\
  D.J.H. Chung, E.W. Kolb, A. Riotto and I.I. Tkachev,\\
  ``Probing Planckian physics: Resonant production of particles during  inflation and features in the primordial power spectrum,''\\
  Phys. Rev. {\bf D62}, 043508 (2000), (hep-ph/9910437);\\
  D.J.H. Chung, P. Crotty, E.W. Kolb and A. Riotto,\\
  ``On the gravitational production of superheavy dark matter,''\\
  Phys. Rev. {\bf D64}, 043503 (2001), (hep-ph/0104100);\\
  R. Allahverdi and M. Drees,\\
  ``Production of massive stable particles in inflaton decay,''\\
  Phys.  Rev. Lett. {\bf 89}, 091302 (2002), (hep-ph/02031180), and\\
  ``Thermalization after inflation and production of massive stable
  particles,''\\
  Phys. Rev. {\bf D66}, 063513 (2002), (hep-ph/0205246).
  
\bibitem{volcano}
  J. Linsley,\\
  ``Evidence for a primary cosmic-ray particle with energy $10^{20}$
  eV''\\
  Phys. Rev. Lett. {\bf 10}, 146 (1963) and\\
  Proc.~8th {\it International Cosmic Ray Conference} 4 (1963) 295.\\
  J.~Linsley and A.~A.~Watson,\\
  ``Validity Of Scaling To 10**20-Ev And High-Energy Cosmic Ray
  Composition,''\\
  Phys.\ Rev.\ Lett.\ {\bf 46}, 459 (1981);\\
  J. Linsley, L. Scarsi, and B. Rossi,\\
  ``Extremely energetic cosmic ray event,''\\
  Phys. Rev. Lett. {\bf 6}, 485 (1961).
  H.~E.~Bergeson {\it et al.},\\
  ``Measurement Of Light Emission From Remote Cosmic Ray Air
  Showers,''\\
  Phys.\ Rev.\ Lett.\ {\bf 39}, 847 (1977).

  
\bibitem{Suga:jh}
  K.~Suga, H.~Sakuyama, S.~Kawaguchi and T.~Hara,\\
  ``Evidence For A Primary Cosmic-Ray Particle With Energy
  4x10-To-The-21 Ev,''\\
  Phys.\ Rev.\ Lett.\ {\bf 27}, 1604 (1971).

  
\bibitem{yakutsk}
  A.~V.~Glushkov {\it et al.},\\
  ``Giant Shower With E(0) = 10**20-Ev Recorded At Yakutsk
  Apparatus,''\\
  Bull. Acad. Sci. USSR;
  Phys. Ser. 55 (1991) No. 4 95-97; \\
  (Izv. Akad. Nauk SSSR, Fiz. 55 (1991) 717-719).
  
\bibitem{haverah}
  M.~A.~Lawrence, R.~J.~O.~Reid, and A.~A.~Watson,\\
  ``The Cosmic Ray Energy Spectrum Above 4x10**17-Ev As Measured By
  The Haverah Park Array,''\\
  J.~Phys.~G Nucl.~Part.~Phys. 17 (1991) 733\\
  See also {\sf http://ast.leeds.ac.uk/haverah/hav-home.html}.
  
\bibitem{agasa:1}
  N.~Hayashida {\it et al.},\\
  ``Observation of a very energetic cosmic ray well beyond the
  predicted 2.7-K cutoff in the primary energy spectrum,''\\
  Phys.\ Rev.\ Lett.\ {\bf 73}, 3491 (1994).
  N. Sakaki {\it et al.} [AGASA Collaboration],\\
  ``Cosmic Ray Energy Spectrum above $ 3 \times 10^{18}$ eV''\\
  Proc.  of 27th ICRC (Hamburg) {\bf 1}, 333 (2001).
  N.~Hayashida {\it et al.}, [AGASA collab.],\\
  ``Updated AGASA event list above 4*10**19-eV,''\\
  Astrophys.\ J.\ {\bf 522}, 225 (1999), (astro-ph/0008102).
  M.~Takeda {\it et al.}  [AGASA collab.],\\
  ``Extension of the cosmic-ray energy spectrum beyond the predicted  Greisen-Zatsepin-Kuzmin cutoff,''\\
  Phys.\ Rev.\ Lett.\  {\bf 81}, 1163 (1998), (astro-ph/9807193).\\
  See also the AGASA Homepage, {\sf
    www-akeno.icrr.u-tokyo.ac.jp/AGASA}

  
\bibitem{hires}
  C.~C.~Jui  [HiRes Collaboration],\\
  ``Results from the High Resolution Fly's Eye experiment,''\\
  in {\it 26th International Cosmic Ray Conference: Invited Rapporteur
    and Highlight Papers,} edited by B. L. Dingus, D. B. Kieda, and M.
  H.  Salamon, AIP Conf. Proc. No.516 (AIP, Melville,
  NY, 2000), p.370.\\
  D.~J.~Bird {\it et al.}  [HiRes Collaboration],\\
  ``The Cosmic Ray Energy Spectrum Observed By The Fly's Eye,''\\
  Astrophys.\ J.\ {\bf 424}, 491 (1994).\\
  See also the HIRES Homepage, www.cosmic-ray.org.
  
\bibitem{largest}
  D.~J.~Bird {\it et al.}  [HiRes Collaboration],\\
  ``Detection of a cosmic ray with measured energy well beyond the
  expected spectral cutoff due to comic microwve radiation,''\\
  Astrophys.\ J.\  {\bf 441}, 144 (1995).\\
  E.~E.~Antonov {\it et al.},\\
  ``Record Energy Of A Giant Shower,''\\
  JETP Lett.\ {\bf 69}, 650 (1999) [Pisma Zh.\ Eksp.\ Teor.\ Fiz.\ 
  {\bf 69}, 614 (1999)].
  
\bibitem{newhires}
  T. Abu--Zayyad et al., [HiRes collaboration],\\
  ``Measurement of the spectrum of UHE cosmic rays by the FADC detector
  of the HiRes experiment,''\\
  (astro-ph/0208301).
  
\bibitem{Wess}
  J. Wess and J. Bagger,\\
  ``Supersymmetry and Supergravity,''\\
  Princeton, USA: Univ. Pr. (1992) 259 p.

  
\bibitem{AP}
  G.~Altarelli and G.~Parisi,\\
  ``Asymptotic Freedom In Parton Language,''\\
  {\em Nucl. Phys.} {\bf B126} (1977) 298.
  
\bibitem{Isasusy}
  H.~Baer, F.~E. Paige, S.~D. Protopopescu, and X.~Tata,\\
  ``Simulating supersymmetry with ISAJET 7.0 / ISASUSY 1.0,''\\
  Argonne Accel.Phys.1993:0703-720 (QCD161:W588:1993) (hep-ph/9305342).\\
  See also http://www.phy.bnl.gov/~isajet/
  
\bibitem{Poetter}
  B.~A. Kniehl, G.~Kramer, and B.~P\"otter,\\
  ``Fragmentation functions for pions, kaons, and protons at
  next-to-leading order,''\\
  {\em Nucl. Phys.} {\bf B582} (2000) 514, (hep-ph/0010289).
  
\bibitem{bd1}
  C.~Barbot and M.~Drees,\\
  ``Production of ultra-energetic cosmic rays through the decay of
  super-heavy X particles,''\\
  {\em Phys. Lett.} {\bf B533} (2002) 107, (hep-ph/0202072).
  
\bibitem{bd2}
  C. Barbot and M. Drees,\\
  ``Detailed analysis of the decay spectrum of a super-heavy X
  particle,''\\
  (hep-ph/0211406), accepted for publication in Astroparticle Physics.

\bibitem{QCDrev}
  E.~Reya,\\
  `Perturbative Quantum Chromodynamics,''\\
  {\em Phys. Rept.} {\bf 69} (1981) 195.
  
\bibitem{Hill}
  C.T. Hill, D.N. Schramm and T.P. Walker,\\
  ``Ultrahigh-Energy Cosmic Rays From Superconducting Cosmic
  Strings,''\\
  {\em Phys. Rev. } {\bf D36}, (1987) 1007;\\
  P. Bhattacharjee, C.T. Hill and D.N. Schramm,\\
  ``Grand unified theories, topological defects and ultrahigh-energy cosmic  rays,''\\
  {\em Phys. Rev. Lett.} {\bf 69} (1992) 567;\\
  R.~J.~Protheroe and T.~Stanev,\\
  ``Limits on models of the ultrahigh-energy cosmic rays based on
  topological defects,''\\
  Phys.\ Rev.\ Lett.\ {\bf 77}, 3708 (1996),
  (astro-ph/9605036);\\
  G.~Sigl, S.~Lee, P.~Bhattacharjee and S.~Yoshida,\\
  ``Probing grand unified theories with cosmic ray, gamma ray and
  neutrino  astrophysics,''\\
  Phys.\ Rev.\ D {\bf 59}, 043504 (1999), (hep-ph/9809242);\\
  R.~J.~Protheroe and P.~L.~Biermann,\\
  ``A new estimate of the extragalactic radio background and
  implications...''\\
  Astropart.\ Phys.\ {\bf 6}, 45 (1996), [Erratum: ibid.\ {\bf 7}, 181
  (1996)], (astro-ph/9605119).

\bibitem{Birkel}
  M. Birkel and S. Sarkar,\\
  ``Extremely high energy cosmic rays from relic particle decays,''\\
  {\em Astropart. Phys.}  {\bf 9} (1998) 297, (hep-ph/9804285).
  
\bibitem{Berezinsky:2000}
  V.~Berezinsky and M.~Kachelriess,\\
  ``Monte Carlo simulation for jet fragmentation in SUSY-QCD,''\\
  {\em Phys. Rev.} {\bf D63} (2001) 034007, (hep-ph/0009053).

\bibitem{Rubin:1999}
N.~Rubin,\\ 
PhD thesis, http://www.stanford.edu/nrubin/Thesis.ps

\bibitem{Coriano:2001}
  C.~Coriano and A.~E. Faraggi,\\
  ``SUSY QCD and high energy cosmic rays. I: Fragmentation functions
  of SUSY QCD,''\\
  {\em Phys. Rev.} {\bf D65} (2002) 075001, (hep-ph/0106326).
  
\bibitem{Sarkar:2001}
  S.~Sarkar and R.~Toldra,\\
  ``The high energy cosmic ray spectrum from massive particle decay,''\\
  {\em Nucl. Phys.} {\bf B621} (2002) 495, (hep-ph/0108098).
  
\bibitem{FodorKatz}
  Z. Fodor and S.D. Katz,\\
  ``Grand unification signal from ultrahigh-energy cosmic rays?,''\\
  {\em Phys. Rev. Lett.} {\bf 86} (2001) 3224, (hep-ph/0008204).
  
\bibitem{ToldraLSP}
  A.~Ibarra and R.~Toldra,\\
  ``Neutralino spectrum in top-down models of UHECR,''\\
  {\em JHEP} {\bf 0206}, (2002) 006, 8hep-ph/02021119.
  
\bibitem{Berezinsky:2002}
  V. Berezinsky, M. Kachelriess and S. Ostapchenko,\\
  ``Electroweak jet cascading in the decay of superheavy particles,''\\
  {\em Phys. Rev. Lett.} {\bf 89} (2002) 171802, (hep-ph/0205218).
  
\bibitem{Jones}
  S.~K. Jones and C.~H. Llewellyn~Smith,\\
  ``Leptoproduction Of Supersymmetric Particles,''\\
  {\em Nucl. Phys.}  {\bf B217} (1983) 145.
  
\bibitem{CCC}
  M. Ciafaloni, P. Ciafaloni and D. Comelli,\\
  ``Towards collinear evolution equations in electroweak theory,''\\
  {\em Phys. Rev. Lett.} {\bf 88} (2002) 102001, (hep-ph/0111109).
  
\bibitem{susyrge}
  See e.g. K. Inoue, A. Kakuto, H. Komatsu and S. Takeshita,\\
  ``Renormalization Of Supersymmetry Breaking Parameters Revisited,''\\
  {\em Prog. Theor. Phys.} {\bf 71} (1984) 413.

  
\bibitem{Basics_of_QCD}
  Y.L. Dokshitzer, V.A. Khoze, A.H. Mueller and S.I. Troian,\\
  {\em ``Basics of Perturbative QCD''},\\
  Gif-sur-Yvette, France, Ed. Fronti\`eres (1991).
  
\bibitem{LPHD}
  Y.I. Azimov, Y.L. Dokshitzer, V.A. Khoze and S.I. Troian,\\
  ``The String Effect And QCD Coherence,''\\
  {\em Phys. Lett.} {\bf B165} (1985) 147, and\\
  ``Similarity Of Parton And Hadron Spectra In QCD Jets,''\\
  {\em Z.  Phys.} {\bf C27} (1985) 65.
  
\bibitem{Furmanski}
  W.~Furmanski and R.~Petronzio,\\
  ``Lepton - Hadron Processes Beyond Leading Order In Quantum
  Chromodynamics,''\\
  {\em Z. Phys.} {\bf C11} (1982) 293.
  
\bibitem{Peterson}
  C.~Peterson, D.~Schlatter, I.~Schmitt, and P.~M. Zerwas,\\
  ``Scaling Violations In Inclusive E+ E- Annihilation Spectra,''\\
  {\em Phys. Rev.} {\bf D27} (1983) 105.
  
\bibitem{msugra} Recent analyses are:
  J.L. Feng, K.T. Matchev and F. Wilczek,\\
  ``Neutralino dark matter in focus point supersymmetry,''\\
  {\em Phys.
    Lett.} {\bf B482}, 388 (2000), (hep-ph/0004043);\\
  R. Arnowitt, B. Dutta and Y. Santoso,\\
  ``Coannihilation effects in supergravity and D-brane models,''\\
  {\em Nucl. Phys.} {\bf B606},59 (2001), (hep-ph/0102181);\\
  J.R. Ellis et al.,\\
  ``The CMSSM parameter space at large tan(beta),''\\
  {\em Phys.  Lett.}  {\bf B510}, 236 (2001), (hep-ph/0102098);
  A. Djouadi, M. Drees and J.-L. Kneur,\\
  ``Constraints on the minimal supergravity model and prospects for
  SUSY particle production at future linear e+ e- colliders,''\\
  {\em JHEP} {\bf 0108}, 055 (2001), (hep-ph/0107316);\\
  H. Baer et al.,\\
  ``Updated constraints on the minimal supergravity model,''\\
  {\em JHEP} {\bf 0207}, 050 (2002), (hep-ph/0205325);\\
  A.B.  Lahanas, D.V. Nanopoulos and V.C. Spanos,\\
  ``Updating the constraints to CMSSM from cosmology and accelerator
  experiments,''\\
  (hep-ph/0211286).
  
\bibitem{neutrinos}
  P. Gondolo, G. Gelmini and S. Sarkar,\\
  ``Cosmic neutrinos from unstable relic particles,''\\
  {\em Nucl.
    Phys.} {\bf B392} (1993) 111, (hep-ph/9209236);\\
  F. Halzen and D. Hooper,\\
  ``High-energy neutrino astronomy: The cosmic ray connection,''\\ 
  {\em Rept. Prog. Phys.} {\bf 65} (2002) 1025, (astro-ph/0204527).
  
\bibitem{bdhh1}
  C. Barbot, M. Drees, F. Halzen and D. Hooper,\\
  ``Neutrinos associated with cosmic rays of top-down origin,''\\
  Phys.\ Lett.\ B {\bf 555}, 22 (2003) (hep-ph/0205230).
  
\bibitem{distorted_gaussian}
  C.P. Fong and B.R. Webber,\\
  ``Higher Order QCD Corrections To Hadron Energy Distributions In
  Jets,''\\
  {\em Phys. Lett.} {\bf B229} (1989) 289.
  
\bibitem{limiting_spectrum}
  V. Berezinsky and M. Kachelriess,\\
  ``Limiting SUSY-{QCD} spectrum and its application for decays of
  superheavy particles,''\\
  {\em Phys. Lett.} {\bf B434} (1998) 61, (hep-ph/9803500).
  
\bibitem{nophot}
  R.A. Vazquez et al.,\\
  {\em Astropart. Phys.} {\bf 3} (1995) 151;
  M.~Ave, J.~A.~Hinton, R.~A.~Vazquez, A.~A.~Watson and E.~Zas,\\
  ``New constraints from Haverah Park data on the photon and iron
  fluxes of UHE cosmic rays,''\\
  {\em Phys. Rev. Lett.} {\bf 85}
  (2000) 2244 (2000), (astro-ph/0007386), and\\
  ``Constraints on the ultra high energy photon flux using inclined
  showers from the Haverah Park array,''\\
  {\em Phys. Rev.} {\bf D65}
  (2002) 063007, (astro-ph/0110613);\\
  K. Shinozaki {\it et al.}, [AGASA collab.],\\
  ``Upper Limit On Gamma-Ray Flux Above 10**19-Ev Estimated By The
  Akeno Giant Air Shower Array Experiment,''\\
  Astrophys. J. {\bf 571}, L117 (2002).
 
\bibitem{bdhh2}
  C. Barbot, M. Drees, F. Halzen and D. Hooper,\\
  ``SUSY in the sky: Observing ultra-high energy cosmic neutralinos,''\\
  Accepted for publication in Phys.\ Lett.\ B (hep-ph/0207133).

  
\bibitem{codeToldra}
  R.~Toldra,\\
  ``A C++ code to solve the DGLAP equations applied to ultra high energy  cosmic rays,''\\
  Comput.\ Phys.\ Commun.\ {\bf 143}, 287 (2002) (hep-ph/0108127).



\bibitem{Hess} 
V. F. Hess,\\
Phys. Z. {\bf 13}, 1804 (1912).

\bibitem{Auger:1938}
  P. Auger, R. Maze, T. Grivet-Meyer,\\
  Comptes Rendus {\bf 206}, 1721 (1938).
  
\bibitem{Auger:1939}
  P. Auger, P. Ehrenfest, R. Maze, J. Daudin, Robley, and A. Fr\'eon,\\
  `` Extensive Cosmic Ray Showers''\\
  Rev. Mod. Phys. {\bf 11}, 288 (1939).

  
\bibitem{Watson:2001}
  A.~A.~Watson,\\
  ``Ultra high energy cosmic rays: Present status and future
  prospects,''\\
  (astro-ph/0112474).
  
\bibitem{GZK:1}
  K.~Greisen,\\
  ``End To The Cosmic Ray Spectrum?,''\\
  Phys.\ Rev.\ Lett.\ {\bf 16}, 748 (1966);
  
\bibitem{GZK:2}
  G.~T.~Zatsepin and V.~A.~Kuzmin,\\
  ``Upper Limit Of The Spectrum Of Cosmic Rays,''\\
  JETP Lett.\ {\bf 4}, 78 (1966) [Pisma Zh.\ Eksp.\ Teor.\ Fiz.\ {\bf
    4}, 114 (1966)].
  
\bibitem{sugar}
  R.~G.~Brownlee et al.,\\
  Can.~J.~Phys. 46 (1968) S259;\\
  M.~M.~Winn et al.,\\
  J.~Phys.~G 12 (1986) 653;\\
  see also {\sf http://www.physics.usyd.edu.au/hienergy/sugar.html}.
  
\bibitem{agasa_anisotropy}
  M. Teshima {\it et al.} [AGASA Collaboration],\\
  ``Anisotropy of cosmic-ray arrival direction at $10^{18}$ eV
  observed
  by AGASA''\\
  Proc. of 27th ICRC (Hamburg) {\bf 1}, 337 (2001).\\
  N.~Hayashida {\it et al.}  [AGASA Collaboration],\\
  ``The anisotropy of cosmic ray arrival directions around
  10**18-eV,''\\
  Astropart.\ Phys.\ {\bf 10}, 303 (1999) (astro-ph/9807045).\\
  M.~Takeda {\it et al.},\\
  ``Small-scale anisotropy of cosmic rays above 10**19-eV observed
  with the Akeno Giant Air Shower Array,''\\
  (astro-ph/9902239).

  
\bibitem{agasa:2}
  N.~Chiba {\it et al.},\\
  Nucl.\ Instrum.\ Meth.\ A {\bf 311}, 338 (1992);\\
  S.~Yoshida, {\it et al.} [AGASA Collaboration],\\
  {\it in Proceedings of the 27th International Cosmic Ray
    Conference}, Hamburg, Germany, 2001, Vol.3, p.1142.
  
\bibitem{hires-mia}
  A.~Borione {\it et al.},\\
  ``A Large Air Shower Array To Search For Astrophysical Sources
  Emitting Gamma Rays With Energies >= 10**14-Ev,''\\
  Nucl.\ Instrum.\ Meth.\ A {\bf 346}, 329 (1994).
  
\bibitem{reviewYoshida}
  S.~Yoshida and H.~Dai,\\
  ``The extremely high energy cosmic rays,''\\
  J.\ Phys.\ G {\bf 24}, 905 (1998) (astro-ph/9802294).
  
\bibitem{WaxmanGRB} E.~Waxman,\\
  ``Cosmological gamma-ray bursts and the highest energy cosmic
  rays,''\\
  {\em Phys. Rev. Lett.} {\bf 75} (1995) 386, (astro-ph/9505082).
  
\bibitem{Biermann}
  J.~P. Rachen and P.~L. Biermann,\\
  ``Extragalactic ultrahigh-energy cosmic rays. 1. Contribution from
  hot spots in FR-II radio galaxies,''\\
  {\em Astron. Astrophys.} {\bf 272} (1993) 161, (astro-ph/9301010).
  
\bibitem{Boldt}
  E.~Boldt and P.~Ghosh,\\
  ``Cosmic rays from remnants of quasars?,''\\
  (astro-ph/9902342).
  
\bibitem{cmb}
  C.~B.~Netterfield {\it et al.},\\
  ``A measurement by BOOMERANG of multiple peaks in the angular power
  spectrum of the cosmic microwave background,''\\
  (astro-ph/0104460);\\
  C.~Pryke, N.~W.~Halverson, E.~M.~Leitch, J.~Kovac, J.~E.~Carlstrom,
  W.~L.~Holzapfel and M.~Dragovan,\\
  ``Cosmological Parameter Extraction from the First Season of
  Observations with DASI,''\\
  Astrophys.\ J.\  {\bf 568}, 46 (2002) (astro-ph/0104490);\\
  A.~Balbi {\it et al.},\\
  ``Constraints on cosmological parameters from MAXIMA-1,''\\
  Astrophys.\ J.\ {\bf 545}, L1 (2000), (astro-ph/0005124);\\
  S.~Perlmutter {\it et al.}  [Supernova Cosmology Project
  Collaboration],\\
  ``Measurements of Omega and Lambda from 42 High-Redshift
  Supernovae,''\\
  Astrophys.\ J.\ {\bf 517}, 565 (1999), (astro-ph/9812133).
  
\bibitem{bere1}
  V. Berezinsky and M. Kachelriess,\\
  ``Ultra-high energy LSP,''\\
  Phys. Lett. {\bf B422}, 163 (1998), (hep-ph/9709485).
  
\bibitem{SHDM}
  V.~Berezinsky, M.~Kachelriess, and A.~Vilenkin,\\
  ``Ultra-high energy cosmic rays without GZK cutoff,''\\
  {\em Phys. Rev. Lett.} {\bf 79} (1997) 4302, (astro-ph/9708217);\\
  V.~A. Kuzmin and V.~A. Rubakov,\\
  ``Ultrahigh-energy cosmic rays: A window on postinflationary
  reheating epoch of the universe?,''\\
  {\em Phys. Atom. Nucl.} {\bf 61} (1998) 1028, (astro-ph/9709187);\\
  E.W. Kolb, D.J. Chung and A.  Riotto,\\
  ``WIMPzillas!,''\\
  Published in Buenos Aires 1998, Trends in
  theoretical physics 2 91-105 (hep-ph/9810361), and\\
  ``Superheavy dark matter,''\\
  {\em Phys. Rev.} {\bf D59} (1999) 023501, (hep-ph/9802238);\\
  H. Ziaeepour,\\
  ``Searching the footprint of WIMPZILLAs,''\\
  {\em Astropart. Phys.}  {\bf 16} (2001) 101, (astro-ph/0001137);\\
  V.~Berezinsky,\\
  ``Ultra high energy cosmic rays from cosmological relics,''\\
  {\em Nucl. Phys.  Proc. Suppl.}  {\bf 87} (2000) 387,
  (hep-ph/0001163).
  
\bibitem{wimpzillas}
  J. Ellis, J. Lopez and D.V. Nanopoulos,\\
  ``Confinement Of Fractional Charges Yields Integer Charged Relics In
  String Models,''\\
  {\em Phys. Lett.} {\bf B247} (1990) 257;
  S. Chang, C. Coriano and A.E. Faraggi,\\
  ``Stable superstring relics,''\\
  {\em Nucl. Phys.} {\bf B477} (1996) 65, (hep-ph/9605325);\\
  K. Benakli, J.R. Ellis and D.V. Nanopoulos,\\
  ``Natural candidates for superheavy dark matter in string and M
  theory,''\\
  {\em Phys. Rev.} {\bf D59} (1999) 047301, (hep-ph/9803333);\\
  K. Hamaguchi, Y. Nomura and T. Yanagida,\\
  ``Superheavy dark matter with discrete gauge symmetries,''\\
  {\em Phys. Rev.} {\bf D58} (1998) 103503, (hep-ph/9805346), and\\
  ``Long lived superheavy dark matter with discrete gauge
  symmetries,''\\
  {\em Phys. Rev.} {\bf D59} (1999) 063507, (hep-ph/9809426);\\
  K. Hamaguchi, K.I. Izawa, Y. Nomura and T. Yanagida,\\
  ``Long-lived superheavy particles in dynamical
  supersymmetry-breaking models in supergravity,''\\
  {\em Phys.  Rev.} {\bf D60} (1999) 125009, (hep-ph/9903207);\\
  C. Coriano, A.E. Faraggi and M. Pl\"umacher,\\
  ``Stable superstring relics and ultrahigh energy cosmic rays,''\\
  {\em Nucl. Phys.} {\bf B614} (2001) 233, (hep-ph/0107053).
  
\bibitem{reviewAnchordoqui}
  L.~Anchordoqui, T.~Paul, S.~Reucroft and J.~Swain,\\
  ``Ultrahigh energy cosmic rays: The state of the art before the
  Auger observatory,''\\
  Int.\ J.\ Mod.\ Phys.\ A {\bf 18} (2003) 2229 (hep-ph/0206072).
  
\bibitem{reviewSarkar}
  S.~Sarkar,\\
  ``Ultra-high energy cosmic rays and new physics,''\\
  (hep-ph/0202013).

  
\bibitem{lep}
  R.~Akers {\it et al.}  [OPAL Collaboration],\\
  ``Measurement of the production rates of charged hadrons in
  e+ e- annihilation at the Z0,''\\
  Z.\ Phys.\ C {\bf 63} (1994) 181;\\
  P.~Abreu {\it et al.}  [DELPHI Collaboration],\\
  ``Inclusive measurements of the K+- and p / anti-p production in
  hadronic Z0 decays,''\\
  Nucl.\ Phys.\ B {\bf 444} (1995) 3;\\
  R.~Barate {\it et al.}  [ALEPH Collaboration],\\
  ``Studies of quantum chromodynamics with the ALEPH detector,''\\
  Phys.\ Rept.\ {\bf 294}, 1 (1998).
  
\bibitem{bere2}
  V. Berezinsky, P. Blasi and A. Vilenkin,\\
  ``Signatures of topological defects,''\\
  Phys. Rev. D {\bf 58}, 103515 (1998).
  
\bibitem{pro1}
  F.~Halzen, R.~A.~Vazquez, T.~Stanev and H.~P.~Vankov,\\
  ``The highest energy cosmic ray,''\\
  Astropart.\ Phys.\ {\bf 3}, 151 (1995).
  
\bibitem{newhaverah}
  M.~Ave, J.~Knapp, J.~Lloyd-Evans, M.~Marchesini and A.~A.~Watson,\\
  ``The energy spectrum of cosmic rays above 3 x 10**17-eV as measured
  with the Haverah Park Array,''\\
  Astropart.\ Phys.\  {\bf 19}, 47 (2003) (astro-ph/0112253);\\
  
\bibitem{radio}
  G.~Sigl, S.~Lee, P.~Bhattacharjee and S.~Yoshida,\\
  ``Probing grand unified theories with cosmic ray, gamma-ray and
  neutrino astrophysics,''\\
  Phys.\ Rev.\ D {\bf 59}, 043504 (1999), (hep-ph/9809242);\\
  R.~J.~Protheroe and T.~Stanev,\\
  ``Limits on models of the ultrahigh energy cosmic rays based on
  topological defects,''\\
  Phys.\ Rev.\ Lett.\ {\bf 77}, 3708 (1996),
  [Erratum: ibid.\ {\bf 78}, 3420 (1996)], (astro-ph/9605036);\\
  R.~J.~Protheroe and P.~L.~Biermann,\\
  ``A new estimate of the extragalactic radio background and
  implications for ultra-high-energy gamma ray propagation,''\\
  Astropart.\ Phys.\ {\bf 6}, 45 (1996), [Erratum: ibid.\ {\bf 7}, 181
  (1996)], (astro-ph/9605119).

\bibitem{private}
S. Sarkar, private communication.
 
\bibitem{mlla}
  See e.g. G.~Marchesini and B.~R. Webber,\\
  ``Simulation Of QCD Jets Including Soft Gluon Interference,''\\
  Nucl. Phys.{\bf B238}, 1 (1984).
  
\bibitem{prop}
  T.~Stanev, R.~Engel, A.~Mucke, R.~J.~Protheroe and J.~P.~Rachen,\\
  ``Propagation of ultra-high energy protons in the nearby universe,''\\
  Phys.\ Rev.\ D {\bf 62}, 093005 (2000), (astro-ph/0003484).
  
\bibitem{bahcall}
  For a review, see M.C. Gonzalez--Garcia and Y. Nir,\\
  ``Developments in neutrino physics,''\\
  (hep-ph/0202058).\\
  The impact of most recent data is included in
  J. Bahcall, M.C. Gonzalez--Garcia and Carlos Pena--Garay,\\
  ``Before and after: How has the SNO neutral current measurement
  changed  things?,''\\
  JHEP {\bf 0207}, 054 (2002) (hep-ph/0204314);\\
  V.~Barger, D.~Marfatia, K.~Whisnant and B.~P.~Wood,\\
  ``Imprint of SNO neutral current data on the solar neutrino
  problem,''\\
  Phys.\ Lett.\ B {\bf 537}, 179 (2002), (hep-ph/0204253).


  
\bibitem{Icecube}
  J.~Alvarez-Muniz and F.~Halzen,\\
  ``10**20-eV cosmic ray and particle physics with IceCube,''\\
  AIP Conf.\ Proc.\ {\bf 579}, 305 (2001) (astro-ph/0102106).
  
\bibitem{ice2}
  F.~Halzen and D.~Hooper,\\
  ``High-energy neutrino astronomy: The cosmic ray connection,''\\
  Rept.\ Prog.\ Phys.\  {\bf 65}, 1025 (2002)  (astro-ph/0204527);\\
  E.~Andres {\it et al.},\\
  Nature {\bf 410}, 441 (2001).
  
\bibitem{rice}
  I.~Kravchenko {\it et al.},\\
  ``Limits on the ultra-high energy electron neutrino flux from the
  RICE experiment,''\\
  (astro-ph/0206371);\\
  I.~Kravchenko {\it et al.}  [RICE Collaboration],\\
  ``Performance and simulation of the RICE detector,''\\
  Astropart.\ Phys.\  {\bf 19}, 15 (2003)(astro-ph/0112372);\\
  S.~Seunarine,\\
  ``Status Of The Radio Ice Cerenkov Experiment (Rice),''\\
  Int.\ J.\ Mod.\ Phys.\ A {\bf 16S1C}, 1016 (2001).

\bibitem{auger}
  See homepage http://www.auger.org/admin/.
  
\bibitem{feng}
  P.~Billoir,\\
  ``Neutrino Capabilities Of The Auger Detector,''\\
  {\it Prepared for 8th International Workshop on Neutrino Telescopes, Venice, Italy, 23-26 Feb 1999};\\
  X.~Bertou, P.~Billoir, O.~Deligny, C.~Lachaud and A.~Letessier-Selvon,\\
  ``Tau neutrinos in the Auger observatory: A new window to UHECR
  sources,''\\
  Astropart.\ Phys.\  {\bf 17}, 183 (2002) (astro-ph/0104452);\\
  J.~L.~Feng, P.~Fisher, F.~Wilczek and T.~M.~Yu,\\
  ``Observability of earth-skimming ultra-high energy neutrinos,''\\
  Phys.\ Rev.\ Lett.\  {\bf 88}, 161102 (2002), (hep-ph/0105067);\\
  L.~A.~Anchordoqui, J.~L.~Feng, H.~Goldberg and A.~D.~Shapere,\\
  ``Black holes from cosmic rays: Probes of extra dimensions and new
  limits on TeV-scale gravity,''\\
  Phys.\ Rev.\ D {\bf 65}, 124027 (2002) (hep-ph/0112247).
  
\bibitem{jaime}
  J.~Alvarez--Muniz and F.~Halzen,\\
  ``10**20-eV cosmic-ray and particle physics with kilometer-scale
  neutrino telescopes,''\\
  Phys.\ Rev.\ D {\bf 63}, 037302 (2001), (astro-ph/0007329);\\
  J.~Alvarez--Muniz and F.~Halzen,\\
  ``10**20-eV cosmic ray and particle physics with IceCube,''\\
  AIP Conf.\ Proc.\  {\bf 579}, 305 (2001)  (astro-ph/0102106).
  
\bibitem{kalashev}
  O.E. Kalashev, V.A. Kuzmin, D.V. Semikoz and G. Sigl,\\
  ``Ultra-high energy neutrino fluxes and their constraints,''\\
  Phys.\ Rev.\ D {\bf 66}, 063004 (2002) (hep-ph/0205050).
  
\bibitem{tevsearch}
  S. Abachi et al., [D0 collab.]\\
  ``Search for squarks and gluinos in p anti-p collisions at S**(1/2)
  = 1.8-TeV,''\\
  Phys. Rev. Lett. {\bf 75}, 618 (1995);\\
  F. Abe et al., [CDF collab.]\\
  ``Search for gluinos and squarks at the Fermilab Tevatron
  collider,''\\
  Phys. Rev. {\bf D56}, 1357 (1997).
  
\bibitem{binomodel}
  See e.g. B.C. Allanach et al.,\\
  presented at {\it APS/DPF/DPB Summer Study on the Future of Particle
    Physics (Snowmass 2001)}, Snowmass, Colorado, 30 June - 21 July
  2001 (hep-ph/0202233).
  
\bibitem{saltzberg}
  F.~Halzen and D.~Saltzberg,\\
  ``Tau neutrino appearance with a 1000-Megaparsec baseline,''\\
  Phys.\ Rev.\ Lett.\ {\bf 81}, 4305 (1998), (hep-ph/9804354);\\
  J.~F.~Beacom, P.~Crotty and E.~W.~Kolb,\\
  ``Enhanced signal of astrophysical tau neutrinos propagating through
  earth,''\\
  Phys. Rev. {\bf D66}, 021302 (2002), (astro-ph/0111482);\\
  S.~I.~Dutta, M.~H.~Reno and I.~Sarcevic,\\
  ``Tau-neutrinos underground: Signals of nu/mu $\to$ nu/tau
  oscillations with extragalactic neutrinos,''\\
  Phys.\ Rev.\ {\bf D62}, 123001 (2000), (hep-ph/0005310).
  
\bibitem{EUSO}
  L.~Scarsi,\\
  in {\it Metepec 2000}, ``Observing ultrahigh energy cosmic rays from
    space and earth'', 113-127;\\
  O.~Catalano,\\
  ``Extreme Universe Space Observatory - Euso: An Innovative Project
  For The Detection Of Extreme Energy Cosmic Rays And Neutrinos,''\\
  Nuovo Cim.\ {\bf 24C}, 445 (2001).
  
\bibitem{OWL}
  D.~B.~Cline,\\
  prepared for the {\it Ultra High-Energy Cosmic Ray Workshop on
    Observing Giant Cosmic Ray Air Showers for $> 10^{20}$ eV
    Particles from Space}, College Park, MD, 13-15 Nov 1997;\\
  R.~E.~Streitmatter [OWL Collaboration],\\
  prepared for the {\it Ultra High-Energy Cosmic Ray Workshop on
    Observing Giant Cosmic Ray Air Showers for $> 10^{20}$ eV
    Particles from Space}, College Park, MD, 13-15 Nov 1997.
  
\bibitem{thresh}
  D.~B.~Cline and F.~W.~Stecker,\\
  ``Exploring the ultrahigh energy neutrino universe,''\\
  (astro-ph/0003459).


\end{thebibliography}
\end{document}